# Photodetectors based on 2D materials and heterostructures: Design and performance analysis

Nagaraj Nandihalli

Ames National Laboratory and Department of Materials Science and Engineering, Iowa State University, 2408 Pammel Drive, Ames, IA, 50011-1015, USA

✉ nnandiha@uwaterloo.ca, ORCID: 0000-0001-7067-5394

*Abstract*

The unprecedented demand for sophisticated, self-powered, compact, ultrafast, cost-effective, and broadband light sensors for a myriad of applications has spurred a lot of research, precipitating in a slew of studies over the last decade. Apart from the photosensing ability of an active element in the light sensor, the device architecture is crucial in terms of photoinduced charge carrier generation and separation. Since the inception of graphene and the subsequent research growth in the atomically thin 2D materials, researchers have developed and adapted different families of 2D materials and device architectures, including single element 2D, 0D/2D, 2D/2D, 1D/2D stacked structures, and so on. This review discusses the recent reports on the light-sensing properties of various 2D materials, their heterostructures, and characteristics applicable to the ultraviolet-near infrared (UV-NIR), short-wave IR (SWIR), mid-wave IR (MWIR), long-wave IR (LWIR), and terahertz (THz) spectral ranges. It highlights the novelty of the burgeoning field, the heightened activity at the boundaries of engineering and materials science, particularly in the generation of charge carriers, their separation, and extraction, and the increased understanding of the underpinning science through modern experimental approaches. Devices based on the simultaneous effects of the pyro-phototronic effect (PPE) and the localized surface plasmon resonance (LSPR) effect, the photothermoelectric effect (PTE)-assisted photodetectors (PDs), waveguide-integrated silicon-2D PDs, metal-2D-metal PDs, and organic material PDs are also examined rigorously. At the end, current challenges, and solutions to enhance the *figures of merit* of photodetectors are proposed.

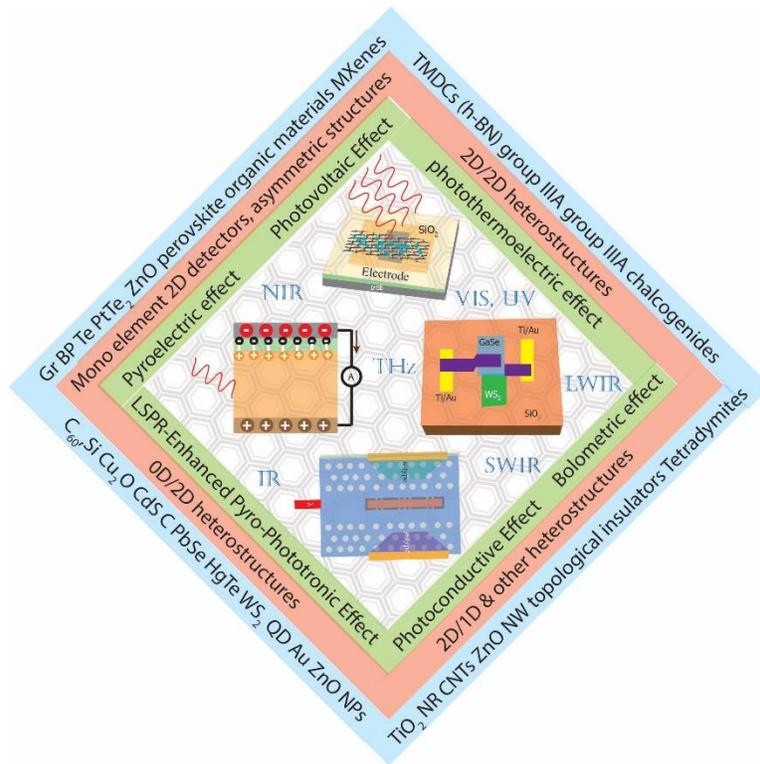

## 1 Introduction

Photodetectors are sensors designed to detect and convert incident photons into sensible electric signals. The sensing element (or active element) of photodetectors is the most important component, as it determines their overall performance. Photodetectors (PDs) have to be fast and sensitive enough to detect light from a wide range of wavelengths. This is important for a broad array of applications: aerospace,[1] communications, [2] infrared remote sensing,[3] biomedical diagnosis,[4], industrial,[5], and detection of IR for night vision applications.[6]

The light spectrum is typically classified as the ultraviolet (UV) (wavelength ($\lambda$) range: 10-400 nm), visible (Vis, $\lambda$ range: 400-750 nm), infrared (IR, $\lambda$ range: 0.75-30 µm), and terahertz (THz, $\lambda$ range: 30 µm -3 mm) bands. UV $\lambda$s are further classified into UV *A* (320-400 nm), UV *B* (290-320 nm), and UV *C* (200-290 nm). Similarly, IR $\lambda$s are classified as near-IR (NIR, 750 nm-1.1 µm), short-wave IR (SWIR, 1-3 µm), mid-wave IR (MWIR, 3-5 µm), long-wave IR (LWIR, 8-12 µm), and very long-wave IR (12-30 µm) (Fig.*1*a).

In order to detect the three prominent sub-bands, such as UV, Vis, and the NIR wavelength ($\lambda$) range, disparate sensors or materials are employed. PDs can detect light in different regimes of the light spectrum depending on the bandgap of the materials used in the active region. In a PD, the electron-hole ($e^-$-$h^+$) pair generation upon photon exposure and their subsequent collection are strongly dependent on the PD's operating mechanism and its efficiency. In general, photoelectric conversion from light to an electric signal involves four steps: i) generation of $e^-$-$h^+$ pairs upon photon absorption; (ii) separation; (iii) transportation; and (iv) extraction of $e^-$-$h^+$ pairs.

Several mainstream semiconductors with notable properties (band gaps) have been used as photosensitive materials in the fabrication of PDs (PD). For example, because of its simplicity of synthesis and, more crucially, compatibility with well-established Si-based circuit technologies, Si ($E_g$ ~1.12 eV) has been extensively used for Vis range to near-infrared (NIR)

## Framework

The essential elements of optical communications, biomedical imaging, detection of spatially resolved images, defense-related, environmental sensing, security, night-vision, gas sensing, and motion detection involve photodetectors. Infrared photodetectors account for the largest share of the detector market, which signifies the essence of these detectors. Si has made a unique place for itself in visible band photodetection by bringing processing technology to a high level of maturity. The finesse of a photodetector is dictated by its various *figures of merit*, such as its *responsivity*, *specific detectivity*, *spectral response*, *quantum efficiency*, *frequency bandwidth*, *gain*, *noise equivalent power*, *dark current*, *cutoff frequency*, and *response time* (*rise time* and *fall time*). In general, high *quantum efficiencies* and quick response times cannot be achieved simultaneously. A 2D photodetector that exhibits high *photoresponsivity*, high speed, wide spectral range, low noise, and high *specific detectivity* is promising. A common approach for harnessing the photoelectric performance of an active element is the fabrication of photodiodes and hybrid phototransistors. These devices are based on either mono element materials (graphene or other 2D structures) or heterostructures (2D/XD (X = 0, 1, 2, and 3)). Photodiodes feature a quick *response time*, a low *dark current*, a high *detectivity*, and a dismal *photogain*. In general, hybrid phototransistors have a high *photoresponsivity*. Their relative *response time*, however, is slow. Therefore, to optimize *figures of merit*, a well-calibrated approach is needed. This review appraises the performance of recently reported photodetectors applicable to the ultraviolet, visible, near-infrared, short-wave infrared, mid-infrared, and long-wave infrared bands. 2D IR room-temperature detectors are currently not ready for large-scale practical applications and cannot totally replace classic thin-film detectors. However, the rapid phase in the development of new photoactive materials and new device configurations indicates that 2D photodetectors with room-temperature operation, high performance, cost-effectiveness, reliability, and long-life are viable.

light detection. Silicon PDs have detectivities of ~$4 \times 10^{12}$ Jones (1 Jones = 1 cm Hz$^{1/2}$/W). However, when the $\lambda$ is outside the NIR range, the photoresponse range is considerably reduced.[7] Other detectors, such as the InGaAs-based membrane PD, are crucial in the infrared spectrum. At 4.2 K, InGaAs PDs generally have detectivities greater than $10^{12}$ Jones.[8] However, these InGaAs-based PDs require a complicated film growth process, a high cost, a lattice mismatching problem, a working environment with very low temperatures, a specific detection wavelength, and other issues that make measuring detection without problems impossible.[9]

In another scenario, due to its customizable bandgap spanning the 1-30 μm range, large optical coefficients that permit high quantum efficiency, and favorable inherent recombination mechanisms that lead to high working temperatures, $Hg_xCd_{1-x}Te$ has been commercially used for IR detection. Nonetheless, they require an additional cooling system (at least 77 K), limiting their applications in portable devices.[10] Due to the aforementioned challenges, it becomes necessary to explore alternative methods for enhancing PD performance. These may include the incorporation of novel optoelectronic materials, streamlining the preparation process, designing a distinctive structure, among other potential approaches. A low-cost multicolor PD system that does not require cooling and has high quantum efficiency, sensitivity, and speed over a broad spectral range (UV to the NIR) would be beneficial. Among the most investigated materials in this direction are diverse 2D materials and their heterostructures.

Ever since the inception of graphene in 2004, atomically thin materials (also known as two-dimensional (2D) materials), have garnered a lot of traction because of their unusual structures and physical properties.[11] Graphene has many salient features, including extremely high carrier mobility (>$10^5$ cm$^2$/V-s).[12] However, as will be detailed later, its zero $E_g$ presents numerous difficulties for photodetection applications. But many other 2D materials (mostly 2D semiconducting materials) have shown excellent electronic[13] and optoelectronic properties: such as high power/mass light harvesting,[14] very sensitive photon detection, and low-threshold lasing.[15]

The family of 2D materials encompasses (1) transition metal dichalcogenides (TMDCs); (2) hexagonal BN (h-BN); (3) X-ene (graphene, germanene, and silicene); (4) black phosphorus (BP, i.e., phosphorene); (5) group IIIA chalcogenides (namely GaS, GaSe) and (6) group IVA dichalcogenides (for instance $SnS_2$) and so forth.[16] Among the TMDCs ($MX_2$, $M$ = transition metal element and $X$ = chalcogen atom): $MoS_2$, $MoSe_2$, $WS_2$, $WSe_2$, $ZrS_2$, $ReS_2$ etc. are prominent (Fig.1d-e). The electronic properties of TMDCs can range from insulating to semiconducting and metallic. The direct energy band gaps and non-centrosymmetric structure of the monolayer TMDCs make them very interesting.[17]

Numerous PDs based on new 2D materials have been reported in recent years. Among these are 2D layered materials, 2D nonlayered materials, 2D-2D heterostructures, and 2D-$X$D heterostructures ($X = 0$, 1, and 3). Many exceptional properties including ultrahigh photoresponsivity, [18] swift response time, [19] broadband photodetection, [20] large photogain, [21] ultrahigh sensitivity, [22] very low dark current, [22-23] large photocurrent ON/OFF ratio, [24], and so on have been brought by PDs based on 2D materials and their stacked heterojunction. 2D materials with atomic thickness, free-dangling bonding, and mechanical workability would provide certain intriguing properties: ultrafast carrier transport speed,[25] [26] exceptional bending properties, optoelectronic properties,[27] and higher gate controllability.[28]

At present, the requirement for an exterior power source to tune the directional movement of the photogenerated carriers and therefore generate photocurrent poses a challenge to the miniaturization of the PDs. This requirement clearly increases the device's size, making it difficult to package and transport while also making it overly reliant on external power supplies, limiting its practical application in a variety of adverse complex settings. With its light weight, small size, and low energy consumption, this self-powered PD can independently realize self-powered light detection without necessitating an external power supply.[29] To make electronic devices smarter, more efficient, and autonomous, they have to be self-powered so that they can rely less on external power or inputs while still performing their functions.

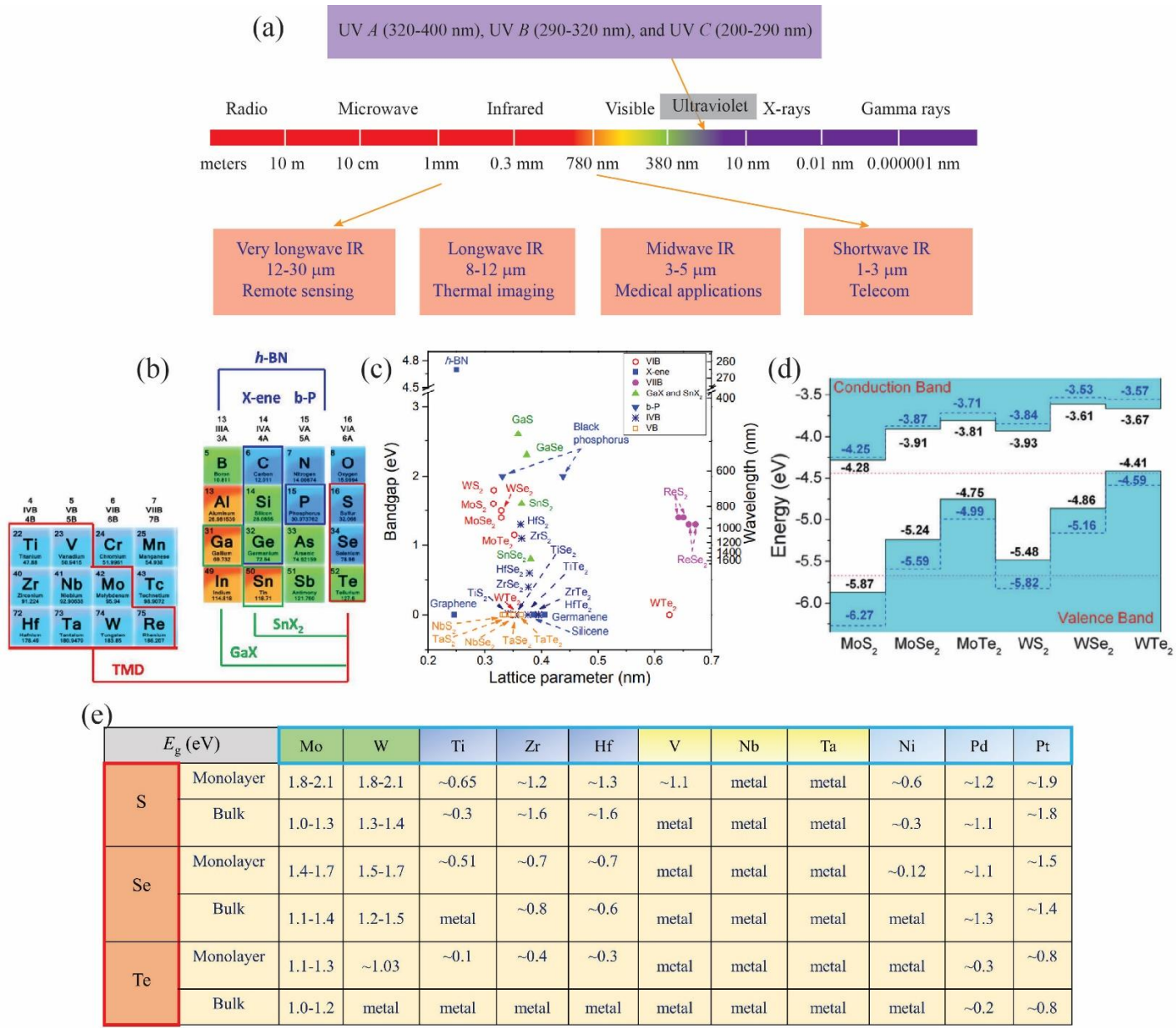

Fig.1: Electromagnetic (EM) spectrum and 2D materials. (a) EM spectrum from radio to γ-ray. Different IR regimes and their representative applications; (b) 2D materials from the periodic table; (c) Plots of calculated bandgaps vs. lattice parameters and $\lambda$s. Adapted with permission.[16] Copyright 2015, RSC. (d) The relative valence and conduction band edge of some common TMDs (monolayer); (e) Common TMDs and their $E_g$s. Adapted with permission.[30] Copyright 2015, RSC.

The classification of the 2D material family is based on the existence or absence of covalent bonds between adjacent atomic layers, resulting in two distinct categories: layered materials and nonlayered materials. Layered materials include (in addition to graphene and TMDCs semiconductors), black phosphorus (BP), Te, Sn-based materials (SnS$_2$, SnS, SnSe, SnTe), In-based materials (InSe, In$_2$Se$_3$), Bi-based oxychalcogenide materials (Bi$_2$O$_2$Se, BiOBr), and others. On the other hand, Pb-based materials (PbS, $Pb_{1-x}Sn_xSe$, and $Pb_{1-x}Sn_xTe$), cadmium chalcogenides (CdS, CdSe, CdTe), organic-inorganic hybrid perovskites, and other materials make up the nonlayered 2$D$ material family. Bottom-up methods account for the vast majority of 2D material preparation processes.

2D van der Waals (vdW) materials, which are part of the family of materials that naturally have vdW gaps between atomic layers that are chemically bonded, are one of the new candidates for these next-generation PDs.[31]. Moreover, in the absence of crystal matching limitations, two-dimensional (2D) materials have the potential to be employed in the fabrication of van der Waals heterostructures (vdWHs). Van der Waals epitaxy (vdWE) is another method for producing 2D non-layered materials. The vdWE technique relies on the presence of relatively weak van der Waals (vdW) interactions between the epitaxial layer and substrates.[32] Two conditions must be met in order for vdWE to produce 2D non-layered crystals[33]: (1) The initiation of 2D anisotropic growth in non-layered materials can be achieved by manipulating growth factors, namely temperature and pressure; (2) the substrates, such as mica and 2D layered materials like Gr, should have a passivated atomically flat surface free of dangling bonds. As a result, unlike traditional epitaxial processes, lattice matching is not essential during vdWE growth. Furthermore, the weak interaction allows the epilayers to be easily released and transferred to other substrates. The vdWE method has been successfully implemented to synthesize many 2D single-crystalline materials: elemental Te, binary semiconductors (PbS and CdS), and ternary semiconductors ($Pb_{1-x}Sn_xSe$).[33-34] These epitaxial nanosheets, with their broad planar morphology and good crystalline quality, are ideal for manufacturing optoelectronic devices such as PDs.

This article presents an in-depth review of the latest developments in photodetection utilizing two-dimensional (2D) materials, encompassing both individual materials and heterostructures. This article's features are organized around the following aspects: To begin, it briefly introduces several key parameters as well as different photodetection operation mechanisms. Then it discusses in detail the recent reports on the performance of PDs based on 2D and 2D-$X$D heterostructures ($X$ = 0, 1, 2, and 3). Since reports on graphene-based PDs are already accessible in numerous earlier publications, the discussion on this category is short.

## 2 Different mechanisms in Photodetectors

The main function of a photodetector is to transform an optical signal into a discernible electrical signal. A photoresponse can be classified based on how an electrical signal is generated when photons interact with an active element: photovoltaic effect, photoconductive effect, photogating effect, pyro-phototronic effect, bolometric effect, and photothermoelectric effect are examples. To fully comprehend the operation of PDs, it is necessary to first perceive the aforementioned key mechanisms of photodetection.

### 2.1 Photoconductive Effect

In the photoconductive effect, the conductivity increases because there are more photogenerated carriers in the semiconductor when it is exposed to laser light.

The applied external voltage would cause $e^-$-$h^+$ pairs to travel in the opposite direction, resulting in photocurrent ($I_{phot}$).[35] The band diagram of a metal/$n$-type semiconductor/metal structure is shown in **Fig.2**a. In the absence of illumination, when an external voltage is applied, a dark current ($I_{dark}$) arises from the limited free carriers. The photoinduced carriers, which might be driven by a suitable bias voltage at the source and drain electrodes, cause a considerable rise in current ($I_{illum}$, **Fig.2**b,). The net incremental photocurrent ($I_{phot}$) may indicate an improved conductivity within the same semiconductor photosensing element under two distinct scenarios, as shown in transfer curves (**Fig.2**c). As a result, $I_{phot}$ can be defined as $I_{phot} = I_{illum} - I_{dark}$.

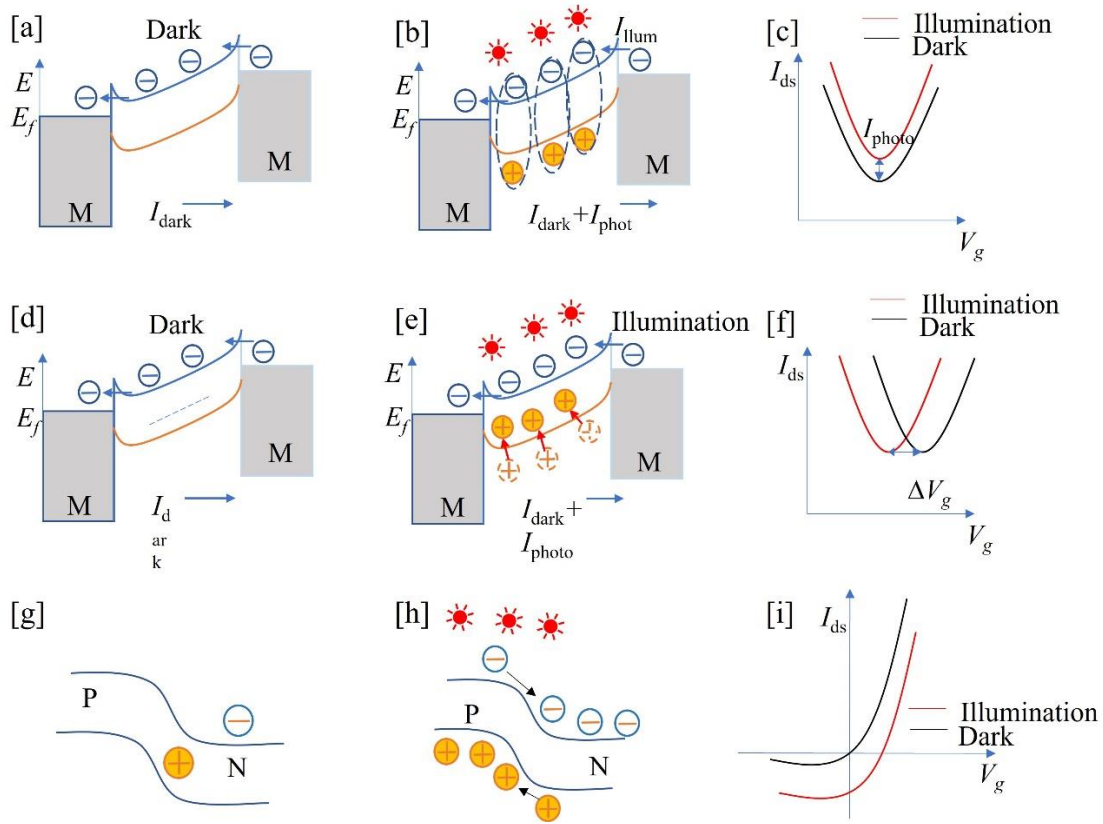

Fig.2: Mechanisms in PDs. (a,b) Photoconductive effect in a metal/semiconductor/metal structure without (a) and with (b) light illumination; (c) Transfer curves of the FET photoconductive effect in a dark state and under light illumination state; (d,e) Energy band diagram of the photogating effect in a metal/semiconductor/metal structure in dark and light states, respectively; (f) Transfer curves of a FET PD having photogating effect; (g,h) $p$-$n$ junction

band diagrams without and with light illumination, respectively; (i) *p-n* junction's current-voltage (*I-V*) characteristic curves in presence and absence of illumination.

## 2.2 Photogating Effect

The photogating effect is seen as a variant of the photoconductive effect. Many carrier traps can be found at defects or on the surface of semiconductors. The photogenerated charge carriers (either electrons or holes) are trapped and have a longer lifetime than other carriers in the channel. As a result, trapped carriers can be used to modify the conductivity of semiconductors.[36] The band diagram of a metal/semiconductor/metal structure in the absence of light is shown in Fig.2d.

Because there are many trap centers in the bandgap closed to the semiconductor's valence band, $e^--h^+$ pairs are generated by proper optical excitation in the presence of illumination (Fig.2e). If the trap states continue to retain holes, the charged traps have the potential to exert significant control over the conductivity of the channel. Furthermore, the shift in the Fermi level induces additional electrons, which modulates conductivity. As a result, the photogating effect could be used to modify channel conductivity. A horizontal left shift ($\Delta V_g$) under illumination, as shown in transfer curves (Fig.2f), reflects the photogating controllable capability and this feature is very crucial for low-dimensional materials. In contrast to conventional methods for intensifying absorption efficiency: *p-n* junctions,[37] hetero-electrodes,[2, 38] plasmonic metasurfaces,[39] optical cavities,[40] tunneling currents between parallel graphene sheets,[41] carrier multiplication,[42] and nanoribbons,[43] the photogating effect is a potential approach for increasing the responsivity of graphene PDs. The photogating effect can be activated in Gr-based PDs by photosensitizers surrounding a graphene channel that couples incident light. Electronic modifications in the photosensitizers cause large modulations in carrier concentration and graphene electrical signals. As a result, a photogating effect for an arbitrary range of $\lambda$s can be achieved by selecting a suitable photosensitizer.

## 2.3 Photovoltaic Effect (PVE)

Solar cell researchers have widely used the term "photovoltaic effect" to describe a wide range of mechanisms involving photon absorption, excitons generation, separation into free charge carriers, and collection of free charge carriers at electrodes. The PV effect is a term used in the field of atomically thin materials research to describe the separation of photo-generated carriers by a built-in electric field and this technique of using built-in electric field to separate the photogenerated carriers quickly is very crucial in photovoltaic effect.

In a *p-n* junction, for instance, the charge carrier concentration gradient would affect the relative movement of electrons and holes, leading to the development of an electric field (Fig.2g). In photovoltaic effect, under light irradiation, $e^--h$ pairs would be separated at the *p-n* junction interface by a built-in electric field and collected by electrodes (Fig.2h).[44] As-fabricated PDs are often self-powered due to their photovoltaic characteristics without any applied voltage. Fig.2i depicts the representative *I-V* characteristic curves of a *p-n* junction. Under light illumination, the phenomenal photogenerated current can be detected at zero bias voltage. Short circuit voltage and open circuit current are two essential metrics of photovoltaic devices that can also be used to assess self-powered capabilities.

## 2.4 Photothermoelectric effect (PTE)

The Seebeck effect causes a photothermoelectric voltage difference ($\Delta V_{PTE}$) when a light-induced heating creates a temperature difference ($\Delta T$) at the two ends of a photosensitive channel. The impact of the $\Delta V_{PTE}$ is determined by $\Delta V_{PTE} = \Delta T.S$, where $S$ is the Seebeck coefficient.[45] To get a good photo-thermoelectric signal from a photo-thermoelectric device with a homogeneous semiconductor, a focused laser spot must be used to light up a small area of the semiconductor and establish a $\Delta T$ between its two ends. However, when a photosensitive element is flooded with uniform illumination, the PTE within the devices cannot be noticed due to the uniform temperature distribution. PTE detectors are well suited for their ultra-broadband (IR and THz) response due to their working mechanism, which is based on the Seebeck effect. This is in contrast to photoconductors and photovoltaic detectors, whose bandgap of the active semiconductors used in such detectors restricts their spectral response. The PTE detector has several advantages, including low noise, no need for a cooling unit, and no extra bulky power source.[46] There are two processes that contribute to the PTE effect: photothermal conversion and thermoelectric effects. Photothermal conversion happens when carriers absorb energy from photons and convey it to other carriers or the crystal lattice via interactions between electron-electrons and electron-phonons, respectively. The Seebeck effect governs the second process, the thermoelectric effect, as previously explained. Because $S$ and the $\Delta T$ vary along the length of a semiconductor material, the nonuniform $S$ and the $\Delta T$ along the device channel both play important roles in photovoltaic generation.

## 2.5 Bolometric effect (BE)

In a BE, the light heating effect changes the resistivity of a light sensing material. The amount of this effect is proportionate to the temperature-dependent conductance ($G$) change of photosensitive materials, d$G$/d$T$, and the homogenous temperature increase ($\Delta T$) generated by laser heating under uniform illumination. When the d$G$/d$T > 0$, a positive photoresponse occurs, resulting in an increasing $I_{phot}$. A negative photoresponse is observed as a result of the negative gradient.[47]

## 2.6 Pyro-Phototronic Effect (PPE)

The pyro-phototronic effect couples the pyroelectric effect and photoexcitation in a pyroelectric semiconductor. This mechanism was employed to manipulate the separation, transportation, and extraction of photo-generated $e^-$-$h^+$ pairs in optoelectronic devices (Fig.3b). In this process, the $I_{phot}$ of a detector is altered, triggered by a transient temperature change ($\Delta T$) when the shined light is turned ON or OFF. In pyroelectric materials, this $\Delta T$ would create a pyroelectric polarization field, which would then couple with the initial built-in electronic field. When the orientation of the built-in electronic field and the pyroelectric polarization field are both the same, the total electric field rises. Conversely, the electric field as a whole would get smaller. The height of the barrier for charge carriers is determined by the magnitude of the overall electric field, and this barrier also affects the photogenerated $e^-$-$h$ pair separation. ZnO, CdS with a wurtzite structure, perovskites, and a few organic compounds such as crystalline rubrene have all been investigated for their potential to exhibit the pyro-phototronic effect. Because of their noncentral symmetric crystal structures, these materials exhibit a pyroelectric effect and can thus convert temperature fluctuations ($dT/dt$) into electric signals. The pyro-phototronic effect has been used to significantly improve the performance of ZnO-based PDs: their ultrafast photoresponse, high photoresponsivity, and ultrahigh detectivity.[48]

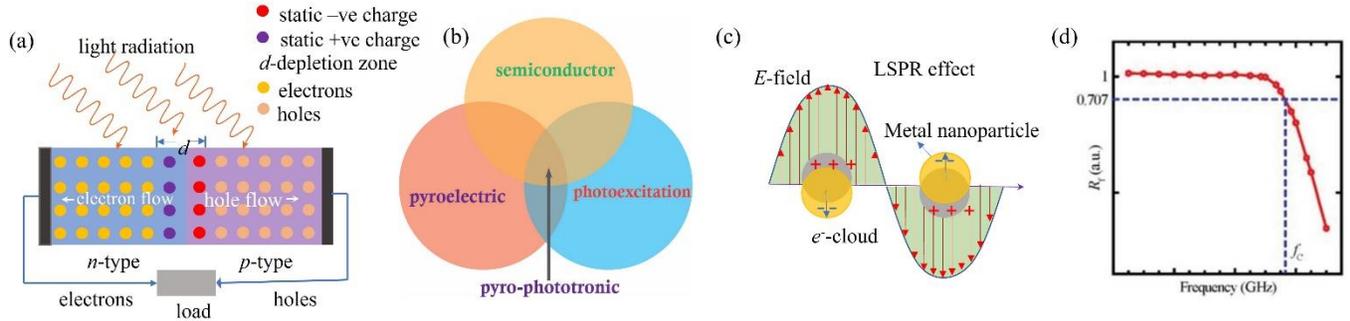

Fig.3: (a) Photovoltaic Effect; (b) Origin of pyro-phototronic effect; (c) LSPR effect; (d) Schematic presentation of the variation of responsivity with frequency. Adapted with permission.[49] © 2018 WILEY-VCH Verlag GmbH & Co. KGaA, Weinheim.

Another distinguishing feature of pyro-phototronic effect enhanced PDs is that they are self-powered with no bias. Because of their low weight, small size, and high level of energy efficiency, self-powered PDs are ideally suited for wire-free environmental sensing, chemical/biological sensing, and in-situ medical monitoring.

The pyroelectric current produced by the pyroelectric effect is given by:[50]

$I_{pyro} = P_c A \, (dT/dt)$ where $P_c$ is the pyroelectric current coefficient; $T$ is the temperature of material; $A$ is the area of electrode; $t$ is the time. Therefore, $I_{pyro} \propto dT/dt$. Recent reports on the photoresponse properties of PDs based on pyro-phototronic effect are discussed in Section 4.2.3.

### 2.7 Localized surface plasmon resonance (LSPR)

Although the LSPR effect does not come under the purview of photovoltaics, LSPR in conjunction with excitonic, pyro-phototronic, and absorption due to interband transition can be utilized to enhance the photodetection property.

LSPR is an optical phenomenon triggered by light interaction with conductive nanoparticles (NPs) smaller than the incident $\lambda$. A plasmon is defined as the mechanical oscillations of a metal's electron plasma caused by an external electric field. Traditional SPR employs a continuous gold film, whereas LSPR employs metal NPs, commonly silver or gold. LSPR has been accomplished using a variety of nanostructures, including NPs, nanorods, nanoporous sheets, and an array of nanoholes.[51] The frequency of LSPR is mostly determined by NP size, shape, material qualities, and surrounding media.[51b, 52] When a surface plasmon is excited at the surface of a metallic NP (with a size similar to the $\lambda$), free electrons are trapped and take part in collective oscillations (**Fig. 3**c).

The resonance frequency ($\omega_{LSPR}$) of the resulting LSPR is largely dependent on the carrier density ($n$) as demonstrated by the following equation:

$$\omega_{LSPR} = \frac{1}{2\pi}\sqrt{\frac{ne^2}{\varepsilon_o m_e(\varepsilon_\infty + 2\varepsilon_d)}}$$

Equation 1

where $\varepsilon_0$, $\varepsilon_\infty$, $\varepsilon_d$, and $m_e$ are the free space permittivity, high-frequency dielectric constant, dielectric constant, and effective mass of the free carrier, respectively. The photons strongly pair with the plasmon during resonant excitation, and the electromagnetic field is localized around the metal nanostructure, improving the absorption probability. The LSPR property of Au NPs is often used in different plasmonic optoelectronic devices to absorb light and make charge carriers through plasmon decay.[53] However, such uses of Au NPs are primarily restricted to light detection in the Vis and NIR ranges. Plasmonic Au NPs are also well known for their ability to generate charge carriers via interband transitions. As a result, Au NPs can be efficiently utilized for UV light absorption via interband transitions as well as by modulating the LSPR band in that region.[54] Reports on such studies will be covered in the subsequent section. The LSPR effect increases the range of possible uses for PDs by producing a strong electric field surrounding NPs by radiation and non-radiation channels, transferring the hot electrons produced to the surface, and radiating heat close to the NPs.

## 3 *Figures of merit*

The simplest approach to comprehend how a PD functions is to shine light on it (either the entire surface or a selected spot) while recording its electrical response. This approach elucidates the physical nature of the observed photoresponse and can be used to extract the important quantities that enumerate the performance of a PD, as detailed more below.

### 3.1 Photocurrent ($I_{phot}$)

The $I_{phot}$ of a PD can be defined by the expression:[45]

$$I_{phot} = I_{illum} - I_{dark} = \frac{P_{in}\mu q\eta E\tau_{photocarriers}}{Lh\nu}$$

Equation 2

where $P_{in}$ is the incident light power, $\nu$ is the frequency of incident light, $h$ is Planck constant, $\eta$ is quantum efficiency (number of generated $e^-$-$h^+$ pairs/number of incident photons), $L$ is the separation distance of source-drain electrodes, and $E$ is the strength of electric field. Before saturated absorption occurs in PDs without trap states, the $I_{ph}$ has a linear relationship with incident light power. Particularly, the exponent $\alpha$, calculated via $I_{phot} \propto P_{in}^\alpha$ is equal to 1 or $\approx 1$. A nonlinear photoresponse ($0 < \alpha < 1$), on the other hand, denotes the complicated mechanism of carrier production, trapping, and recombination within semiconductors.[55]

## 3.2 Responsivity (*R*)

$R$ is the photocurrent ($I_{phot}$) or photovoltage ($V_{ph}$) to incident power density ($P_{in}$) ratio, expressed in A/W or V/W units and can be written as:

$$R = I_{phot} \; or \; V_{ph}/P_{in}A \qquad \text{Equation 3}$$

where $A$ is the device's effective illumination area. This parameter details photoelectric conversion capability in PDs.[56]

## 3.3 External quantum efficiency (*EQE*)

$EQE$ is ratio of the number of $e^-$-$h^+$ pairs that contribute to a $I_{\text{phot}}$ to the number of photons incident on the material and can be expressed as:

$$EQE = \frac{I_{phot}}{q} \Big/ \frac{P_{in}}{h\nu} = R\frac{hc}{q\lambda} \qquad \text{Equation 4}$$

In many circumstances, the calculated $EQE$ or $G > 1$. That is, because the lifetime of the photo-generated carrier surpasses the transit time due to charge entrapment and/or mobility mismatch between the $e^-$ and $h^+$, an incoming photon induces more than one carrier.

## 3.4 Internal quantum efficiency (*IQE*)

$IQE$ = the number of $e^-$-$h^+$ pairs producing a $I_{\text{phot}}$/the number of absorbed photons.

$$IQE = \frac{I_{phot}}{q} \Big/ \frac{A_{ab}P_{in}}{h\nu} \qquad \text{Equation 5}$$

Where $A_{ab}$ is the absorbed fraction. Optical interference effects can augment photon absorption in an ultrathin photosensitive layer.[57]

## 3.5 Signal to noise ratio (*SNR*)

$SNR$ is the ratio of signal power ($P_s$) to noise power ($P_n$) in a PD and is expressed in units of decibels (dB) and can be written as:

$$SNR = P_s/P_n \qquad \text{Equation 6}$$

Low noise is advantageous for extending the lower limit of detectable signals; therefore, a high $SNR$ is required for a highly sensitive PD.

### 3.6 Noise equivalent power (*NEP*)

*NEP* is normally expressed in units of W/Hz$^{1/2}$ and is defined as the minimum illumination power required to achieve an *SNR* of 1 in a 1 Hz bandwidth. The *NEP* can be calculated by:

$$NEP = \frac{I_{in} \ or \ V_n}{R} \qquad \text{Equation 7}$$

Where $I_{in}$ and $V_n$ are the mean-square noise current and mean-square noise voltage, respectively, at the bandwidth of 1 Hz in dark in $A$/Hz$^{1/2}$ and $V$/Hz$^{1/2}$ units, respectively. $R$ is the responsivity. To realize better performance in a detector, a smaller *NEP* is required.

### 3.7 Specific detectivity (*D\**)

*D\** determines the detector's ability to detect the smallest optical singles and is expressed as:[31a]

$$D^* = \frac{(AB)^{1/2}}{NEP} \qquad \text{Equation 8}$$

Where $A$ and $B$ are the PDs' area and bandwidth, respectively. $D^*$ is expressed in units of Jones (cm Hz$^{1/2}$/W). In general, having PDs with a high $D^*$ value equates to better performance.

### 3.8 Response time ($\tau_{res}$)

$\tau_{res}$ is the ability of a PD to follow (respond) a fast-varied optical signal. It represents the PDs' reaction time when the laser source is turned on or off. In 2D PDs, ultrashort carrier lifetime ensures their high $\tau_{res}$. The $\tau_{res}$ is divided into the rise time, $\tau_r$ and the decay or fall time, $\tau_d$ or $\tau_{fall}$. Determining the amount of time needed to change from 10%/90% of current to 90%/10% of current yields the relevant $\tau_r$ and $\tau_d$ values. In general, the $\tau_r$ is defined as the $I_{phot}$ increasing to 63% and the $\tau_d$ as the $I_{phot}$ decreasing to 37% of the stable photocurrent.[58] They can also be found by fitting a single exponential function to the parts of the time-resolved photoresponse that rise and fall:[59]

$$I = I_{ph}e^{-(\frac{\tau_{res}}{t})\beta} \qquad \text{Equation 9}$$

Where $I_{ph}$ is the photocurrent when the light is turned off and $\beta$ is a relaxation mechanism exponent. The $\tau_{res}$ of layered PDs is in the micro-to millisecond range[60] and is highly dependent on trap states in the photosensor material and metal contacts.[61] The photoresponsivity of most photoconductors depends on the light modulation frequency, denoted by $R(f) = \frac{R_o}{\sqrt{1+(2\pi f\tau_{res})^2}}$. Where $R_o$ denotes photoresponsivity under static illumination. The frequency at which the photoresponsivity gets attenuated to 0.707 $R_o$ (3 dB) is called the cutoff frequency $f_C$ (also called as 3 dB bandwidth), as shown in **Fig.3**d.

### 3.9 Gain (G)

*G* describes the number of charge carriers recorded for each incident photon and expressed as:

$$G = \frac{I_{ph}}{q} \bigg/ \frac{\eta P_{in}}{h\nu} = R \frac{hc}{\eta q \lambda} \qquad \text{Equation 10}$$

Where $c$, $\lambda$, and $\eta$ are light velocity, the excitation wavelength, and quantum efficiency, respectively. When $\eta$ is ignored, then $G \approx EQE$. Because of the long lifetime ($\tau_{life}$) and short drift transit time ($\tau_{tran}$) of photogenerated carriers (electron or holes), these carriers can wander in the channel many times, resulting in photoconductive gain. The photoconductive gain due to the long-life of the carrier is given by $G = \tau_{life}/\tau_{tran}$. The $\tau_{tran}$ is dependent on the applied bias voltage ($V_{bias}$), the length of the channel ($L$), carrier mobility ($\mu$) and is expressed as $\tau_{tran} = (L^2/\mu V_{bias})$

### 3.10 On–off ratio ($\alpha$)

At a given voltage, the $\alpha$ is defined as the ratio of $I_{phot}$ (or photocurrent density) to $I_{dark}$ (or dark current density) and can be written as:[62]

$$\alpha = \frac{I_{phot}}{I_{dark}} = \frac{J_{phot}}{J_{dark}} \qquad \text{Equation 11}$$

Where, $J_{phot}$, and $J_{dark}$ are the photocurrent density and dark current density, respectively. As the preceding equation shows, boosting the $I_{phot}$ is one of the most fruitful ways to increase $\alpha$. Another efficient technique to boost $\alpha$ is to lower the $I_{dark}$, which is an important factor for high-sensitivity PDs.

### 3.11 Linear dynamic range (LDR)

The term LDR refers to the range of incident light intensity that the detector can detect and is measured in decibels (dB).[63]

$$LDR = 20 \log \frac{J_{phot}}{J_{dark}} \qquad \text{Equation 12}$$

A higher LDR means that the PD can detect the intensity of light across a larger range.

The noise caused by the stochastic character of thermal production of charge carriers in narrow band gap semiconductors is anticipated to restrict the ultimate performance of long wavelength IR (LWIR) PDs operating at high temperatures. LWIR radiation has low photon energy. As a result, detection necessitates electron transitions with minimum energies lower than photon energy. The thermal energy of charge carriers becomes close to the transition energy at near room temperature, resulting in a very fast rate of thermal creation of charge carriers. The existence of signal noise can be attributed to the stochastic character of this process. As a direct consequence of this, the long wavelength detectors, when used at temperatures very close to room temperature, generate a significant amount of background noise.[64] Cooling is the simplest and most efficient approach to reduce thermal generation. At the same time, cooling is an extremely inefficient process. The requirement for cooling is one of the most significant limitations of photodetectors, which prevents a more widespread application of infrared technology. A vacancy in the valence band is created when an electron is excited to the conduction band, thus creating an electron-hole pair. The probability of an $e^-$-$h^+$ pair being thermally created/unit time is determined by:

$$P(T) = CT^{3/2} \exp(-\frac{E_g}{2k_B T})$$
<div align="right">Equation 13</div>

Where $T$ is the absolute temperature, $E_g$ is the band gap energy, $k_B$ is the Boltzmann constant, and $C$ is a material proportionality constant. In the absence of radiation, an ideal radiation detector should be devoid of charges, and numerous charges should be present in the presence of an ionizing radiation. The decrease in temperature results in a lower total number of $e^-$-$h^+$ pairs within the crystal. Low-cost, high-performance infrared systems require infrared detectors that are both cost-effective and can operate without requiring additional cooling.

## 4 Photodetectors based on 2D materials

Over the last decade, the fabrication process, assembly, and photoresponse properties of numerous types of 2D material based PDs have been reported. These include mono-element (single material)-based, including graphene and heterostructure-based systems such as 0D/2D, 1D/2D, 2D/2D and others. These encompass layered and non-layered 2D materials alike. Architecture-wise, these structures include planar and vertically stacked ones. The subsequent section highlights some of the recent reports on such structures.

### 4.1 2D single-element-based PDs

2D materials are often classified into the UV, visible, and infrared wavebands based on their respective photoresponses to wavelengths. Among the 2D single-element-based materials, single-layer graphene and BP are the most explored. Other materials, such as 2D TMDC (UV to THz), 2D halide perovskites (for IR), oxyselenides (for IR), black arsenic phosphorus (4 and 8 μm), and 2D tellurium (tellurene, MWIR to NIR) have also been considered. In this section, GR, BP, and tellurene based monoelement PDs are covered. PDs based on some other 2D materials are covered in Section 4.2.4.

### Graphene (Gr)

Gr is a 2D carbon allotrope made up of a single layer of carbon atoms arranged in a honeycomb pattern. The incredible electrical and optical properties of Gr crystallites are the fundamental reasons for Gr being the sole focus of experimental and theoretical efforts, bypassing other 2D materials. Gr has extraordinary electrical and thermal properties, as well as unusual electronic features: Dirac fermions [65], the quantum Hall effect (QHE)[66], and an ambipolar electric field effect.[67] In Gr electrons follow long mean-free trajectories without interrupting electron-electron interactions and producing disorder. As a result, the physical structures and electrical properties of Gr differ from those of other conventional metals and semiconductors.

Near the Dirac point, Gr is gapless and linearly spread[68] that enables fast charge carrier generation by light absorption over a broad light spectrum, particularly in UV to THz spectral regimes.[31a] Gr's high $\mu$ allows ultrafast conversion of photons to electric current/voltages.[69] The photovoltaic effect, photothermoelectric effect, photoconductive effect, photogating effect, and bolometric effect are some of the basic mechanisms behind Gr PDs.[70] Gr is an appealing material for photonic and optoelectronic devices such as Gr modulators, Gr PDs, and surface plasmon enhanced PDs due to its potential combination of ultra-high $\mu$, broadband absorption, and ultra-fast luminescence.[71] Further, Gr's *IQE* is reported to be high.[69]

The fragile materials and low absorbance at nanoscale thickness make ultra-thin and flexible PDs challenging to fabricate with traditional semiconductor technology. Gr-based light sensors are transparent to light and flexible enough for future wearable electronics.

Gr can be obtained in a variety of ways. The micro-mechanical cleavage of bulk graphite is the first and most direct way.[72] This process produces extremely high-quality single- and multi-layer Gr flakes with extremely high mobility values and low defect density. Exfoliated Gr is frequently used as a starting material for functionalization or the development of hybrids and heterostructures. The exfoliation method, on the other hand, is not scalable and has a low throughput. Gr synthesis methods that are more scalable have been developed, with chemical vapor deposition (CVD) being the most promising methodology for large-area Gr fabrication.[73] CVD Gr can be generated as a single layer or as a multilayer, depending on the substrate. CVD Gr is ideally suited to intercalation in its multilayer form. The liquid-exfoliation of graphite to produce Gr dispersion in water or other solvents is another scalable approach. [74] Epitaxial growth on SiC[75] and reduction of graphene oxide (GO), a functionalized version of Gr, are two further methods of synthesis.

The most simple and direct configuration for Gr-based PDs is the metal-graphene-metal (MGrM) PD, in which the source and drain are metal electrodes in contact with the Gr. In this configuration, the photoresponse is caused by both photothermoelectric and photovoltaic effects.[76] The two electrodes may be made of two different metals (asymmetric configuration) or the same metal (symmetric configuration). They can be fabricated easily without the use of nanoscale lithography. The difference in work functions between metal contacts and Gr causes charge transfer, resulting in a shift in the Gr Fermi level in the region below the metal contacts. When moving from the metal-covered region to the metal-free channel, the Fermi level gradually returns to that of uncontacted Gr. As a result, a potential gradient extends from the end of the metal pad to the metal-free Gr channel. This non-homogeneous doping profile forms a junction along the channel. Because the back gate may alter the channel Fermi level, this can theoretically be a *p-n-*, *n-n-*, or *p-p-*junction between the Gr beneath and within the channel.

**Fig.4**a gives a schematic representation of the MGrM PD. **Fig.4**c depicts a schematic of the metal contact-induced doping profile. The generation of this junction is critical in the photodetection process because it induces an internal electric field capable of segregating the light-induced $e^- - h^+$ pairs. MGrM-PDs work over a wide $\lambda$ range since the light-matter interaction is mostly dictated by Gr. Furthermore, because no bandwidth limiting materials are used, ultrahigh working speeds can be reached.[2, 77]

Because of the short carrier lifetime, only photocarriers formed at the Gr near the metal contact can contribute to the $I_{phot}$, limiting the effectiveness of Gr-based PDs.[78] What's more, when the band structure of a Gr PD is the same on both the source and drain sides, the resulting $I_{phot}$ does not flow without a drain bias. This is because the photocarriers formed on the source and drain sides have opposite charges, which are balanced in the middle of the Gr channel.[2, 47] When the Gr PDs are biased by a drain bias, the $I_{dark}$ increases on the scale of μA. On the nanoampere scale, this is 1000 times larger than the normal $I_{dark}$ of Si-based PDs. To address this issue, a Gr PD with distinct metal contacts: palladium and titanium for the source and drain was suggested.[2] This device can yield nonzero net $I_{phot}$ because the difference in metal work function used in the source and drain causes a gradient in the Fermi level of the exfoliated Gr channel.

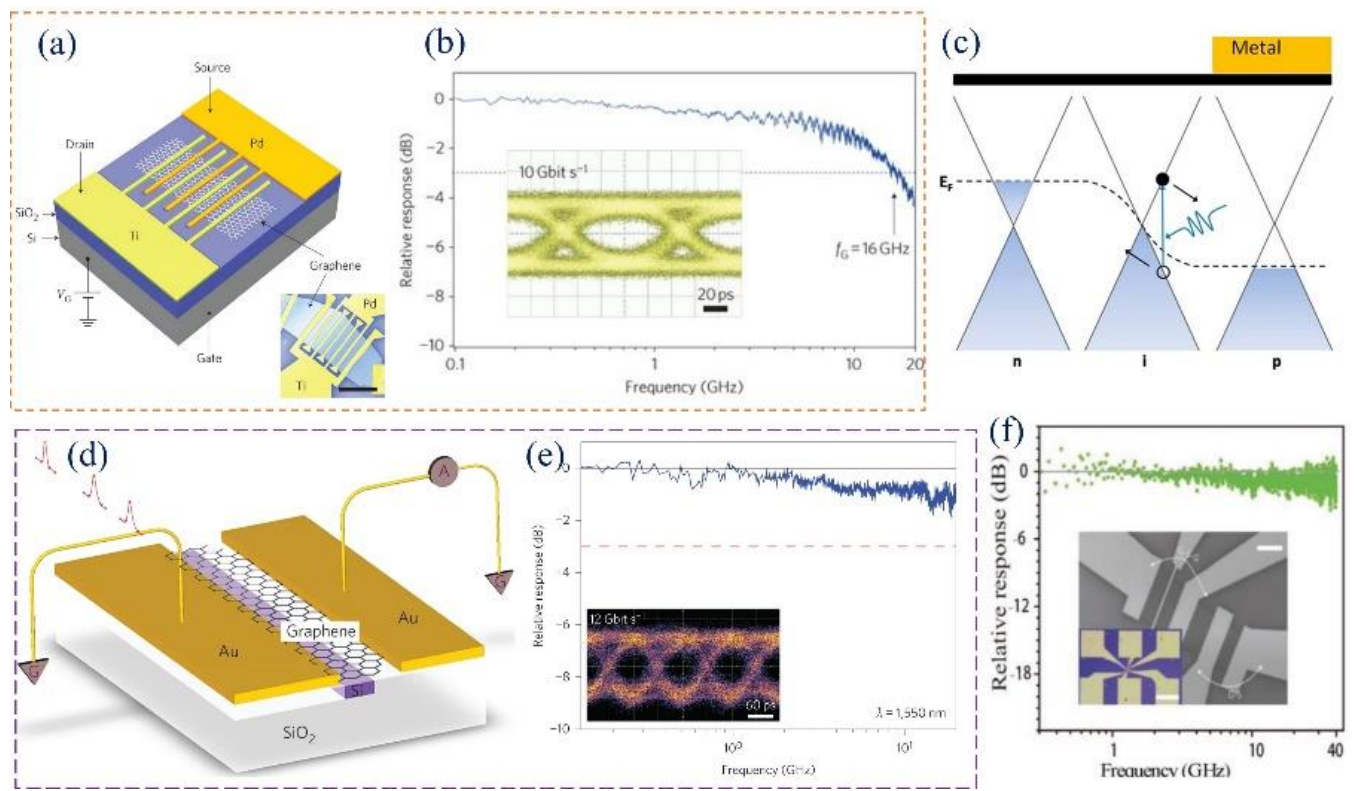

Fig.4: (a-b): MGrM PD and its operation in high-speed data communication. (a) 3D schematic of the MGrM PD. Bottom right: scanning electron micrograph of the MGrM PD. Scale bar, 5 μm. The spacing between the metal fingers is 1 μm and the finger width is 250 nm; (b) Relative photoresponse versus light intensity modulation frequency. The 3-dB bandwidth of this device

is ~16 GHz. Inset: receiver eye-diagram obtained using this MGrM PD, showing a completely open eye. Scale bar, 20 ps; Adapted with permission.[2] Copyright 2010, Springer Nature Limited. (c) energy-band profile in MGrM-PDs. Adapted with permission.[79] Copyright 2014, ACS; (d-e):Waveguide-integrated Gr PD. (d) Schematic of the device. Two metal electrodes contact the Gr and conduct the generated $I_{phot}$. To produce a potential difference in the Gr and couple with the waveguide's evanescent optical field, one of the electrodes is placed closer to the waveguide; (e) The relative a.c. photoresponse vs. light intensity modulation frequency shows ~1 dB attenuation of the signal at 20 GHz frequency. Inset: 12 Gbit/s optical data link test of the device, showing a clear eye opening. Adapted with permission.[80] Copyright © 2013, Springer Nature Ltd; (f) Ultrafast PD based on single-layer and few-layer Gr; the modulation frequency was ~40 GHz. Inset: SEM and optical images of the Gr device. Adapted with permission.[69] Copyright 2009, Springer Nature Ltd.

The photoresponse of transistor-based PDs composed of single-layer and few-layer Gr did not change for optical intensity modulations up to 40 GHz, and additional research revealed that the intrinsic bandwidth may surpass 500 GHz (**Fig.**4f).[69] In other study, the operation of the MGrM structure had been tested in high-speed optical communication using a $\lambda = 1.55$ μm.[2] The receiver's eye-diagram is shown in **Fig.**4b-inset with a data acquisition rate of 10 Gbit/s. **Fig.**4b depicts the relative photoresponse as a function of modulation frequency, where the 3 dB bandwidth is 16 GHz. The MGrM PD was developed on a Si wafer (1–10 kΩ cm) with a 300-nm thermal oxide layer. One finger-like electrode was made of palladium/gold (20/25 nm thick), and the other of titanium/gold (20/25 nm). The bi-layer Gr was used as an active layer. The active area of detector was $6 \times 6$ μm; the detector was connected to contact pads (each having an area of $80 \times 80$ μm), separated by 40 μm. Similarly, a waveguide-integrated Gr PD with high $R$, fast response, and broad spectral bandwidth has been demonstrated.[80] Using a metal-doped detector achieved a photoresponsivity of more than 0.1 A/W and an almost nonfluctuating response between 1450 and 1590 nm by using a metal-doped Gr junction coupled evanescently to the waveguide. This device achieved response rates surpassing 20 GHz and an instrumentation-limited 12 Gbit/s optical data link in zero-bias operation. In this structure, to provide a clean finish for Gr deposition, a Si waveguide was backfilled with $SiO_2$ and planarized. The Gr layer is electrically isolated from the underlying Si structures by a thin $SiO_2$ layer (~10 nm) put on the planarized chip. The optical waveguide mode connects to the Gr layer via the evanescent field, resulting in optical absorption and photocarrier generation. The $I_{phot}$ is collected by two metal electrodes situated on opposite ends of the waveguide. **Fig.**4e shows the device's a.c. photoresponse at zero bias, with a 1 dB attenuation of the signal at 20 GHz. To evaluate the performance of an optical data transmission at 12 Gbit/s, a pulsed pattern generator modulated a 1550 nm CW laser with a maximum 12 Gbit/s internal electrical bit stream (using an electrooptic modulator). Average optical power of ~10 dBm (10 mW) was launched into the waveguide Gr detector. The Gr detector's output electrical data stream was amplified and transferred to a digital communication analyzer to generate an eye diagram. At 12 Gbit/s, a clear and eye-opening graphic was obtained (**Fig.**4e-inset) [80]

Photoresponse measurements in epitaxial graphene (EG) devices with asymmetric metals (Au, Al) connected in a planar Au/EG/Al device configuration (**Fig.5**a-b) showed that at an excitation $\lambda$ of 632.8 nm, a biased Au/EG/Al PD (bias voltage of 0.7 V) reached an external photoresponsivity (or efficiency) of 31.3 mA/W.[81] The device exhibited the greatest photoresponse near the metal/EG contacts, and the $I_{phot}$ drops as one moves farther from the Au/EG or EG/Al interface (**Fig.5**c,). These findings lend credence to the notion that altering the work function of two metals contacting in planar geometry can significantly alter photoresponse by disrupting mirror symmetry in Gr-based planar devices. Near the metal/Gr contact, where the work-function difference between the two is greater, results in the formation of a *p-n* junction. The $\phi$s for EG, Au, and Al are reported to be 4.0 eV, 5.47 eV, and 4.06 eV respectively. Accordingly, the Au/EG/Al device is expected to have *p-n-n* junctions. The number of layers in EG affected $I_{phot}$ magnitude and $\tau_{res}$ regardless of incident photon energy or intensity. As shown in **Fig.5**d, as the number of EG layers increased, the $I_{phot}$ is improved in the four EG layer device due to the absorption of a large number of incident photons.[81]

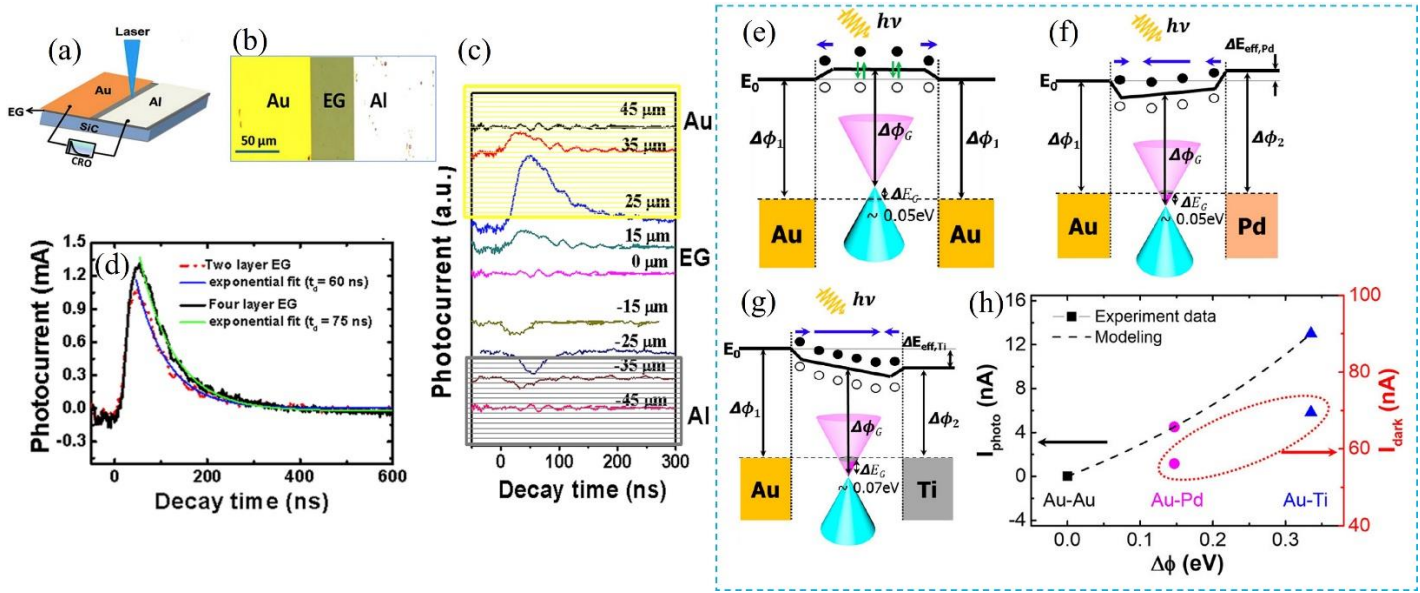

Fig.5: (a-d): Planar Au/Epitaxial Gr/Al device. (a) Schematic view of zero-bias Au/EG/Al device connected to an oscilloscope (CRO); (b) Optical image of the EG device contacted with Au and Al metals; (c) Photoresponse of the Au/EG/Al device at different positions excited with laser pulses (photo scan from Au to Al electrode); (d) Photoresponses (maximum $I_{phot}$) in devices with two and four layers of EG. Laser pulse duration: 7 ns, wavelength: 532 nm, and fluence: 0.94 mJ. Adapted with permission.[81]Copyright 2012, AIP Publishing. (e-h): CVD-Gr based PD with asymmetric metal contacts; (e-g) energy band models of Gr PDs with different source/drain metal combinations; (h) $I_{phot}$ vs. work function differential of the source and drain metal contacts under zero-bias operation with a halogen lamp at 0.25 μW/cm² power. The $I_{dark}$ in the 0 bias operation is denoted by the red dotted circle. Adapted with permission.[82] Copyright 2018, ACS.

Similarly, Au, Pd, and Ti were used in different combinations to construct CVD-Gr based PDs with an asymmetric metal contact arrangement.[82] With a difference in work function of 0.34 eV between the Ti and Au contacts, the maximum external photoresponsivity was measured to be 52 mA/W, which is ten times that of epitaxial Gr-based PDs. As demonstrated in Fig.5h, the $I_{phot}$ improved linearly in proportion to the effective work function of the source and drain metals. The $I_{dark}$ increased by 125% (70 nA) as the contact configuration changed from symmetric (Au-Au) to asymmetric (Au-Ti).

**Fig.5**e-g illustrates the proposed energy band model of Gr PDs with different source/drain metal combinations. Here $\phi_1$ and $\phi_2$ are the work function of Au and asymmetric metal respectively. $E_0$ is the vacuum level, and $E_f$ is the Fermi level of the metal and Gr. The doping condition of the Gr channel is described by $\Delta E_G$. The solid black and empty circles are the electrons and holes generated by the light absorption. $\Delta E_{eff}$ represents the difference in the effective work functions of the metals used for source and drain. Photocarriers can be collected more efficiently with asymmetric metal contacts due to the internal potential generated by the work function difference of both metal contacts. The $\Delta E_{eff}$ for the Au–Ti and Au–Pd cases are reported to be 0.34 eV and 0.15 eV respectively.

The spatial extent of the junction is 100-200 nm for both photovoltaic and photothermoelectric effects to ensue. Extending the junction region to capture more light is one approach. In a vertical (i.e., perpendicular to the surface of the device) $p$–$i$–$n$ semiconductor PD, this is done by using ion-implantation with energy (and dose) that are specifically chosen.[83]. However, in MGrM configuration, this is difficult. Another option is to improve Gr absorption in the ultrafast MGrM configuration. To address this issue, various solutions have been presented. For example, Gr can be integrated into an optical microcavity[40a] (> 20-fold improvement) or onto a planar photonic crystal cavity[84] (8-fold improvement) to take advantage of the trapped light's multiple passes through Gr, or it can be coplanarly integrated with a Si integrated photonic waveguide[80, 85] (>10-fold enhancement). In a planar optical microcavity that is integrated with a Gr transistor, the optical confinement caused by the microcavity efficiently controlled the efficiency and spectral selection of photocurrent generation, resulting in increased $I_{phot}$.[40b] Another possibility is to integrate plasmonic nanostructures on Gr[79, 86] to take advantage of the considerably amplified electromagnetic near-fields[87] related with LSPRs effect.[87] As was mentioned earlier, LSPRs are produced when the conduction electrons of a metal respond to incident radiation by oscillating in a resonantly coherent manner. The increased near-fields that form around the nanostructures boost light absorption in the materials that surround them. The best option would be to collect light in one part of the device and then deliver it to the junction region at the contact edge. This is accomplished by stimulating surface plasmon polaritons (SPPs) on the metal contacts. SPPs are surface-bound waves that propagate on a metal-dielectric interface and are caused by light interaction with the metal's free electrons. Their excitation is accomplished using an integrated diffraction grating, and their supply to the active region (junction at contact edge) increases overall absorption. In one such application, a plasmonic grating was coupled with Gr to utilize generated surface plasmon polaritons to deliver the captured photons to the junction region of a metal-Gr-metal PD.[88] This design resulted in a 400% increase in $R$ and a 1000% increment in photoactive length, as well as adjustable spectral selectivity.

Gr's photoresponsivity remains poor when compared to other classic semiconductor PDs due to its low absorption coefficiency, and Gr's photogeneration is frequently finite at metal/Gr interfaces.[89] A Gr/Si heterostructure photodiode applicable to the NIR-MIR working range was built by integrating Gr onto a Si optical waveguide on a Si-on-insulator (SOI) substrate. The waveguide absorbed evanescent light that propagated parallel to the Gr sheet, yielding a $R$ of 0.13 A/W at a 1.5 V bias for 2.75 μm light at RT.[90] Standard mechanical exfoliation was used to prepare monolayer Gr samples, which were then mechanistically deposited to the waveguide. It has been claimed that the width of the Gr ribbon can regulate the opening of the bandgap, offering nano-engineering to generate Gr nanoribbon with tailored absorption in the IR regime. For instance, sub-10 nm Gr nanoribbon array FET exhibited an $E_g$ of ~100 meV as an efficient NIR PD.[91]

Chemical contamination during the CVD process, as well as mechanical damage to Gr induced by transfer and exfoliation processes, can all substantially decrease Gr's photoelectric capabilities. In contrast, epitaxial Gr on a SiC single crystal wafer exhibits intrinsic characteristics and superior performance. In a study, a metal-Gr-metal PD was directly fabricated on epitaxial Gr synthesized by thermal breakdown of SiC (0001) wafer.[92] Under 780 nm laser illumination and 10 V bias voltage, the maximum $I_{phot}$ was ~$2.2 \times 10^{-7}$ A (~30 times larger than the $I_{dark}$). While under 0.05 mW/cm$^2$ illumination power, the highest $R$ and $EQE$ were $4.48 \times 10^{-2}$ A/W and 8.13%, respectively.

The introduction of a thin insulating oxide layer at the interface of a large area Gr/$n$-Si Schottky barrier solar cell has been shown to be successful in altering the Schottky junction and reducing $I_{dark}$.[93] Following the thermionic emission theory the electrical transport properties of a Schottky diode implanted with an interfacial oxide layer can be characterized as.[94]

$$I = I_o \left( e^{\frac{eV}{\eta k_B T}} - 1 \right) = AA^*T^2 e^{-\sqrt{\delta \chi}} e^{-\frac{\phi_B}{k_B T}} \left( e^{\frac{eV}{\eta k_B T}} - 1 \right) \quad \text{Equation 14}$$

Where $I_o$ is the reverse saturation current. $e, \eta, k_B, T, A, A^*$, and $\phi_B$ are the electronic charge ($1.60217663 \times 10^{-19}$ C), ideality factor, Boltzmann constant ($1.380649 \times 10^{-23}$ m$^2$ kg/s$^2$-K), temperature (K), photosensitive area, Richardson coefficient, and Schottky barrier height (SBH), respectively. $\chi$ and $\delta$ are the average barrier height and the thickness of the inserted film. The term $-\sqrt{\delta \chi}$ can represent the transmission factor across the interlayer, revealing that the interlayer contributes to lowering the reverse saturation current, which is the main component of the $I_{dark}$. The photovoltaic behavior of the Schottky junctions with and without the Gadolinium iron garnet, Gd$_3$Fe$_5$O$_{12}$(GdIG) interlayer was tested to see if the dark current suppression mechanism worked after the GdIG was inserted. GdIG is a transparent ferromagnetic insulator (garnet structure) with a dielectric constant of ~11.9 and shows excellent insulating properties even when it is very thin. Furthermore, the device's elevated temperature and chemical stability toward water vapor and oxygen allowing it to function for an extended period. Because of its intrinsic magnetooptical modulation capabilities, it may be able to detect polarized light at specified $\lambda$s.[95]

In order to manipulate the interface of a Gr/Si Schottky PD, a thin layer of GdIG film (1.97 nm thick) was introduced.[96] In this hybrid architecture (Fig.6a), the top layer is Gr, and the bottom layer is $n$-type Si ($n$-Si). The interface between Gr and Si forms a Schottky junction, and the GdIG thin film acts as an interlayer to facilitate carrier transit, as shown in **Fig.6**b. The Gr aided in the Schottky junctions formation and the collection of carriers. At a 2 V bias, the Gr/GdIG/Si structure significantly reduced dark current by 54 times. It also operated effectively in a self-powered mode, with an $I_{phot}/I_{dark}$ ratio of ~$8.2 \times 10^6$ and a $D^*$ of $1.35 \times 10^{13}$ Jones at 633 nm, since the interlayer improved the barrier height and passivated the contact surface, suppressing $I_{dark}$. It exhibited a broadband absorption capability covering UV-NIR range, a large $LDR$ and showed an operation speed of 0.15 ms with a steady response for 500 continuous ON/OFF cycles. As shown in Fig.6c, a Schottky junction is formed when semi-metallic Gr and $n$-type Si come in contact, causing the generation of the built-in electric field ($eV_{bi}$) and the rectifying Schottky barrier ($\Phi_B$) on the contact surface. In general, a Gr/Si Schottky junction often has a high density of surface states due to the intrinsic dangling bonds of Gr and the nonideal surface of Si, resulting in considerable carrier recombination. Furthermore, because the barrier height is lower, thermally excited electrons can penetrate the interface and create a reverse saturation current. As demonstrated in Fig.6d, after the insulating GdIG interlayer is inserted, Gr and Si are spatially separated, thereby improving $\Phi_B$ and $eV_{bi}$. At the barrier, thermally produced carriers are impeded, resulting in a suppression of the reverse saturation current. Additionally, a homogeneous interlayer with fewer structural defects helps to passivate the interface and reduce the density of surface states, lowering recombination current. As demonstrated in Fig.6e, holes are injected into the valence band of Gr while electrons flow into Si, resulting in a $I_{phot}$. Inserting a GdIG thin film permits photogenerated hole tunneling into Gr, and the impact ionization effect during the tunneling process acts as photogain, increasing the $I_{phot}$. When an external reverse bias voltage ($V_{bias}$) is given, Gr opens more available states for photoexcited holes, causing a higher $I_{phot}$, as seen in Fig.6f.[96]

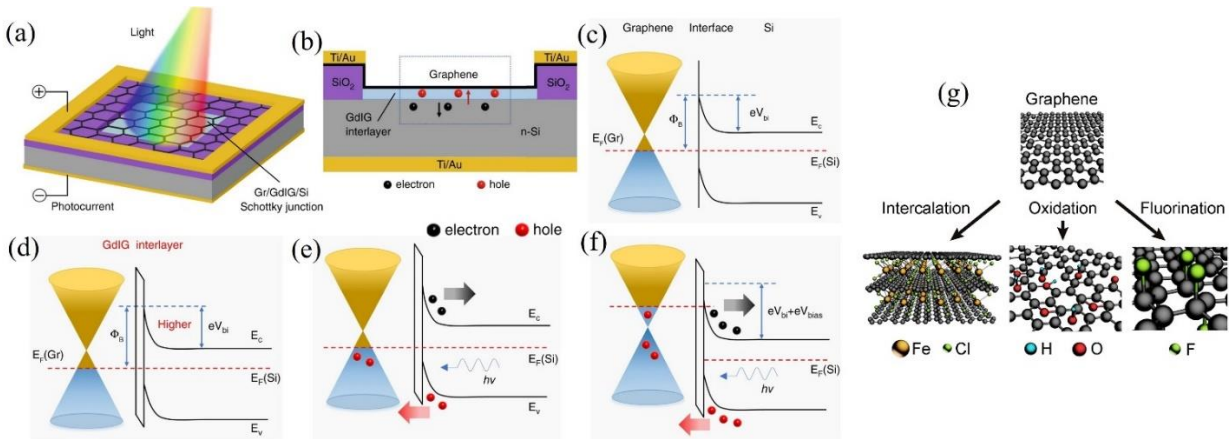

Fig.6: (a-f) Gr/GdIG/Si PD. (a) 3D schematic view; (b) Cross-section of the PD; (c) & (d) Gr/Si and Gr/GdIG/Si junctions under dark condition respectively; (e) Gr/GdIG/Si junction under illumination and (f) Gr/GdIG/Si junction under illumination with a reverse bias voltage. Here, $\Phi_B$, $V_{bi}$, $e$, $E_C$, $E_V$, $E_F(Gr)$, $E_F(Si)$, and $V_{bias}$ represent the SBH, built-in electric field, electronic charge, conduction band edge, valence band edge, Fermi level of Gr, Fermi level of Si, and the reverse bias voltage, respectively. The arrows indicate the direction of carrier movement. Adapted with permission.[96] Copyright 2022, Peirui Ji et al. (g) Examples of functionalized Gr materials. Adapted with permission.[97] Copyright 2018, De Sanctis et al.

Significant advances have recently been made in the synthesis and characterization of PDs based on chemically functionalized Gr. It is vital to distinguish between functionalization, which refers to changes in the structure or nature of the host material, and hybridization, which refers to the property combination of two or more materials (for example, 0D/2D, 2D/2D, 1D/2D, and so on). The functionalization can be employed to change the charge carrier dynamics in Gr, resulting in better photoresponse. Chemical functionalization is when chemical species are used to change the properties of a material. The chemical functionalization of Gr can be achieved using different paths (Fig.6f). Readers are encouraged to consult an article by Sanctis et al. for more information on such findings.[97]

## Black Phosphorus (BP)

BP, a layered semiconductor, is the most stable allotrope of phosphorus. Orthorhombic BP has potential application in NIR and MWIR photodetection.[98] The $E_g$ of bulk BP covers the range of 0.3 eV-1.2 eV and can increase to 1.8 eV by reducing its number of layer.[99] BP, like Gr, is made up of stacked phosphorene layers and may thus be mechanically exfoliated. The extremely anisotropic and intensely confined excitations in single layer BP were studied using polarization-resolved photoluminescence at RT. The results indicate that the light emitted by the single layer has a center located at 1.3 eV.[100] A vertical electric field was shown to successfully reduce BP's optical bandgap for MWIR (perhaps LWIR) photodetection by increasing the number of layers, with the photoresponse in a 5-nm-thick BP being stretched from 3.7 to and beyond 7.7 µm.[98] When excited by MIR light, the BP-based vertical vdW photodiode showed a high $R$ of ~1 A/W and an ultra-fast $\tau_{res}$ of 1.8 ns with a speed of ~200 MHz.[101]

BP has a strong intralayer anisotropy and could be used in high mobility FETs.[102] Each P atom in the atomic layer is connected to three neighbouring atoms, resulting in two unique $x$ and $y$ axis directions: armchair and zigzag. Anisotropic electric band dispersion occurs due to the highly anisotropic arrangement of P atoms, resulting in anisotropic electronic and optical properties, and thus optoelectronic features. A recent study shows that carbon doping results in high $\mu$ of 1995 cm$^2$/V-s.[103] The hole $\mu$ is high due to the significant in-plane anisotropy, with 1000 cm$^2$/V-s along the light effective mass direction and ~500 cm$^2$/V-s along the heavy effective mass direction. BP's samples also have a high conductivity, with thicknesses spanning from 2 to 5 µm. Other strategies for modulating the $E_g$ of BP include exerting strain,[104] employing an electric field, and alloying by composition.[105] BP, on the other hand, is very unstable since it degrades rapidly when exposed to air or humidity. Because of the limited light absorption and the lack of a junction to separate and convey the carriers, BP-based PDs feature poor photoresponse and response speed compared to other materials. Using a BP PD, a large ultraviolet $R$ of 105 A/W has been observed.[106] Despite its outstanding performance in the UV range, photoresponsivity fell by about five orders of magnitude once the excitation $\lambda$ extended beyond 500 nm. A tunable BP PD with $R$s of 518, 30, and 2.2 mA/W at 3.4, 5, and 7.7 µm, respectively, was recently demonstrated at 77 K.[98] A PD based on BP with a wideband $\lambda$ response from 400 nm to 3.8 m was reported.[107] It was discovered that BP is unstable and is rapidly affected by water and oxygen molecules in the air. To shield the BP channel from degradation, Al$_2$O$_3$ was utilized as a passivated layer. When the incidence intensity was low (~0.5 mW/cm$^2$ at 580 nm), the device delivered a $R$ of ~6 A/W and a $NEP$ of ~470 pW/Hz$^{1/2}$ at 2.5 µm (with $V_{DS}$ = 50 mV, $V_{GS}$ = -20 V).

A variety of techniques have been proposed and implemented to increase the amount of collected signal in BP PDs to enable high absorption. For example, tuning the gate voltage and photogating effects have been shown to make thin-film BP PDs much more sensitive to light in the MWIR.[99, 108] Furthermore, Si waveguides and plasmonic antennas have been shown to facilitate absorption in BP PDs operating in the SWIR (1-2 µm), and Fabry-Perot cavities have been employed in the MWIR.[109] Such integrated photonic structures have no direct effect on BP's intrinsic absorption. They do, however, enhance the amount of incident radiation gathered, which increases the signal-to-noise ratio of the sensor.

To address the weak absorption issue due to reduced light–matter interaction of BP, a method employing a slow light effect in photonic crystal waveguides (PhCWGs) was suggested and experimentally tested.[110] This technique could effectively compress the optical field and enhance light absorption. The $R$ of the BP PD on a 10 µm long PhCWG was increased by more than tenfold at around 3.8 µm when compared to the subwavelength grating waveguide (SWGWG) counterpart on a subwavelength grating waveguide. The BP PhCWG PD achieved a $R$ of ~11.31 A/W and a 0.012 nW/Hz$^{1/2}$ $NEP$ at a 0.5 V bias. Fig.7a-b depict a configuration of the proposed shared-BP photonic system that comprises of a SWGWG of identical length without the slow light effect for comparison and a PhCWG with the slow light effect for $R$ enhancement.

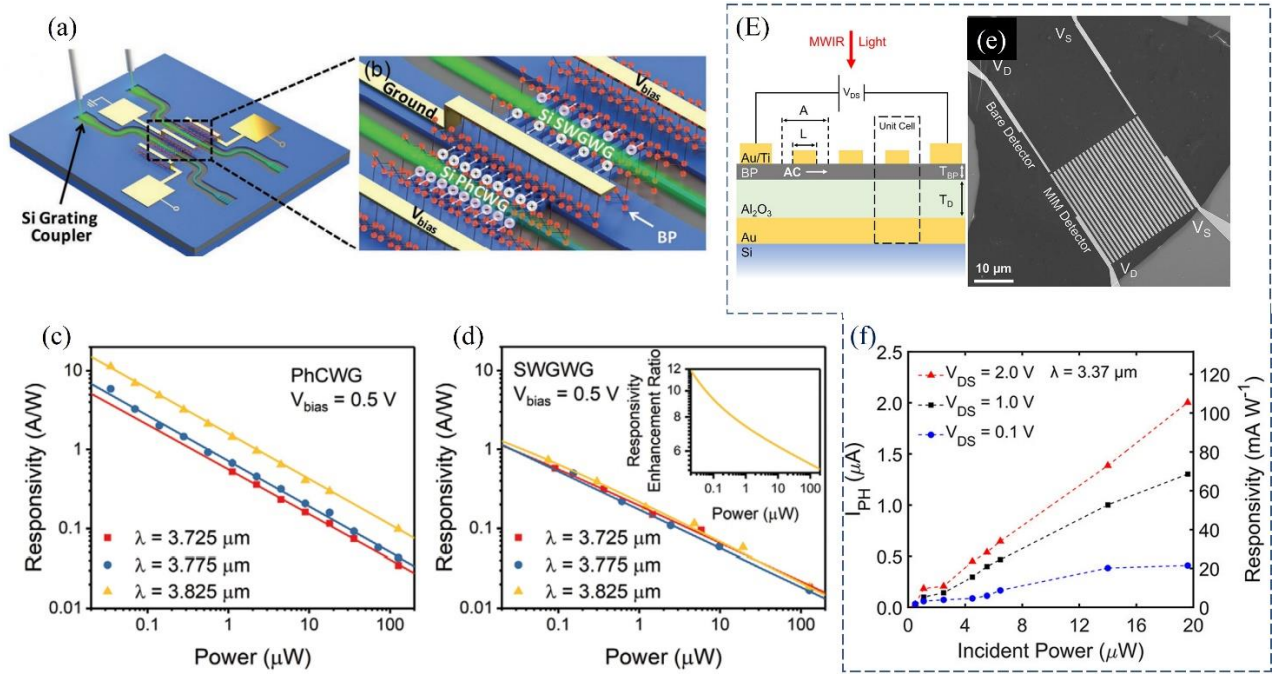

Fig.7: (a-d): Shared-BP photonic systems and their power-dependent photoresponse. (a) Shared-BP photonic system consisting of SWGWG and PhCWG systems; (b) Zoom-in view of the black square box in (a) showing the two BP PDs on the PhCWG and the SWGWG, respectively; (c) BP PhCWG PD, (d) BP SWGWG PD, at three λs. Solid lines: fitting curves by the Hornbeck–Haynes model. Inset of (d): variation of $R$ enhancement ratio with incident power at 3.825 μm. Adapted with permission.[110] Copyright 2020 WILEY-VCH Verlag GmbH & Co. KGaA, Weinheim. (E-e-f): MIM metasurface grating coupled with thin-film BP PD. (E) Device schematic: $A$ and $L$ are grating period and grating length respectively, $T_D$ and $T_{BP}$ are thicknesses of the dielectric spacer and the BP layer respectively. One grating unit cell has been labeled, and only three individual unit cells are presented. In this case, the PD bias was applied to the entire grating; (e) SEM image of two PDs (one with MIM grating and one without grating); (f) Variation of $I_{phot}$ and $R$ with respect to incident power at different device biases. Adapted with permission.[111] Copyright 2022, ACS.

Similarly, a resonant metal–insulator–metal (MIM) metasurface grating that is integrated with a thin-film BP PD has boosted the absorption in a BP PD.[111] The MIM's geometry was the factor in controlling the resonant absorption peak. When integrated with a thin-film BP PD, a MIM configuration increased RT $R$ from 12 to 77 mA/W at 3.37 μm. The proposed device architecture is displayed in Fig.7E. Each MIM was patterned on a Si substrate and comprised a 60 nm thick Au back reflector, Al₂O₃ dielectric layer, BP layer, and 50 nm thick Au/Ti gratings. At the maximum bias (2 V) and incident power (~20 mW), the PD's $R$ was 106 mA/W; $EQE = 4\%$ and $IQE = 6\%$ were obtained. The 90-10% $\tau_r$ and $\tau_d$ were 0.7 ms and 0.9 ms respectively.

Due to its inherent surface defects, the limited thickness of the absorption layer, and the instability of BP in the air, the metal-semiconductor-metal (M-S-M) PDs based on BP cannot simultaneously achieve high photoresponsivity and fast photoresponse, which restricts the application of BP M-S-M structured PDs. To address such problem, the IR PD with the multilayer BP was inserted between the top boron nitride protective layer and the bottom Au electrodes. This BP PD device exhibited a high photoresponsivity of ~1.55 A/W at a $\lambda$ of 1550 nm at RT, a fast photoresponse of ≈16 μs, a broadband photodetection of ~2 μm, and a high ON/OFF ratio of $10^3$.[112]

Similarly, to address the problems associated with the poor photoresponse and low response speed of BP, a NIR PD based on in-plane BP $p$-$n$ homojunction was fabricated on a single BP flake (~30 nm) by realizing $n$-doping using the field-induced effect from the $K^+$ center of the silicon nitride ($Si_xN_y$) capping layer.[113] The BP PD demonstrated broad detection that extended to the NIR range. When compared to a pristine BP PD, the photoresponsivity of the BP $p$-$n$ junction under 1550 nm improved 50-fold. The back-gated $p$-$n$ homojunction is depicted schematically in Fig.8a, where the exposed portion of the BP flake functions as a $p$-type channel and the $Si_xN_y$-covered BP functions as an $n$-channel due to the electrostatic doping from the $K^+$ center. Because of the built-in field, the BP $p$-$n$ junction has a higher $I_{phot}$ as compared to pristine BP (Fig.8b, red line). Under 1550 nm illumination, the photoresponsivity of the $p$-$n$ homojunction BP PD was 5 A/W. Fig.8c shows the $EQE$ and photoconductive gain measured at the 1550 nm. Under reverse source-drain bias $V_{DS}$ of 0.1 V, the BP $p$-$n$ junction yielded $\tau_r = 35$ μs and $\tau_d = 40$ μs (Fig.8d).

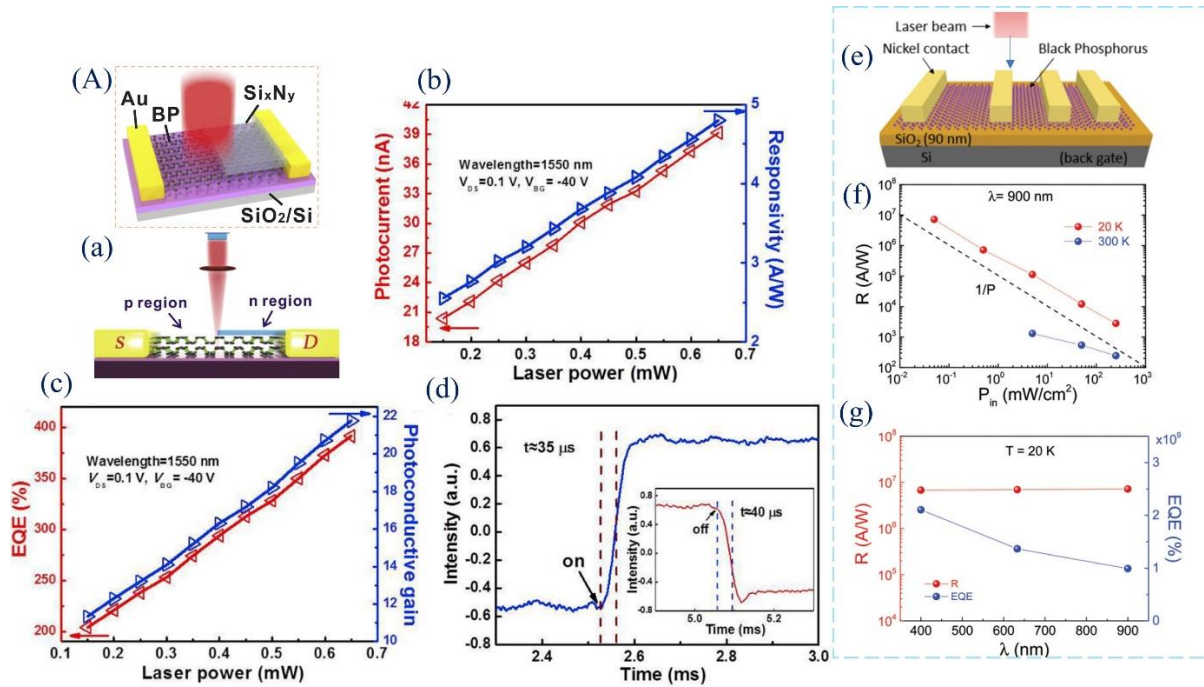

Fig.8: (a-d): BP flake $p$-$n$ homojunction. (a) Schematics of BP phototransistor based on field-induced homogeneous $p$-$n$ junction; lower image: cross section of in-plane BP $p$-$n$ homogeneous junction. The device was illuminated with $\lambda$ = 1550 nm and spot diameter of 2 μm ; (b) $J_{phot}$ and $R$ of the device vs. laser powers; (c) $EQE$ and Photoconductive gain of the BP $p$-$n$ junction device vs. laser powers; (d) 70–30 % $\tau_r$ and $\tau_d$ (insert) measured with a $\lambda$ = 1.5 μm illumination source. Adapted with permission.[113] Copyright 2020 Elsevier Ltd. (e-g): BP back-gate FET. (e) Side view of the fabricated device with Ni/Au contact electrode. The gate dielectric is 90 nm

Another method for improving the photoresponse of a BP-based device is to engineer a Schottky barrier between the metal and the BP. The use of a metal with an appropriate work function as a contact electrode not only facilitates hole injection towards the p-BP, but also increases photocarrier capture at the end of the source/drain contact. One such BP PD based on back-gate field-effect transistor (FET) configuration is shown in Fig.8e.[114] In this study, mechanical exfoliation was used to construct few-layer BP flakes on a 90 nm SiO$_2$ substrate; electron-beam lithography was used to create source/drain electrodes, and e-beam evaporation was employed to deposit Ni/Au metal contacts. Two $\lambda$s: 633 and 900 nm were used to characterize the $R$ values of the device at 20 K and 300 K. At 300 K, a high photoresponsivity of $6.7 \times 10^5$ A/W for 633 nm was observed. In the NIR region (900 nm), a high $R$ of ~$10^3$ A/W at 300 K and $7 \times 10^6$ A/W at 20 K were attained (Fig.8f). As shown in **Fig.8**g, the $EQE$ is increasing in 900 to 400 nm range due to the increased photon energy.[114]

Many Si photonic integrated devices have been successfully developed for optical fiber communications applications working in the NIR wavelength-band of 1.31/1.55 μm.[115] Recently, the MIR wavelength-band of 2-20 μm has become particularly appealing for several essential optical communication applications.[116] BP has been the subject of substantial research for the realization of PDs, and PV effect demonstrations of waveguide (WG)-integrated BP PDs have been performed. Metal-TMDC-metal (M-TMDC-M) and metal-BP-metal (M-BP-M) PDs exhibit smaller bandwidths, significantly lower $I_{dark}$s, and equivalent responsivities as compared to M-Gr-M PDs, unless the photogating effect dominates. **Fig.9** depicts more examples of metal-2D-metal PDs and waveguide-integrated Si-2D PDs with a variety of materials, including TMDC and BP, as well as various mechanisms. The following section provides more details on such studies.

Plasmonics can improve detector performance and reduce device length to a few micrometres, allowing for greater efficiency and a smaller device. **Fig.9**a depicts a SL Gr-plasmonic PD working around (1540 nm) telecom window.[117] This device is based on a plasmonic slot WG, where the plasmonic mode's subwavelength confinement resulted in improved light-Gr interactions and the narrow plasmonic slot of 120 nm provided short drift pathways for photogenerated carriers. The PD with a Gr coverage length of 2.7 μm showed no decline in response up to 110 GHz and an intrinsic $R$ of 25 mA/W. By increasing the device size to 19 μm, the intrinsic $R$ increased to 360 mA/W, corresponding to a high $EQE$ of 29%.

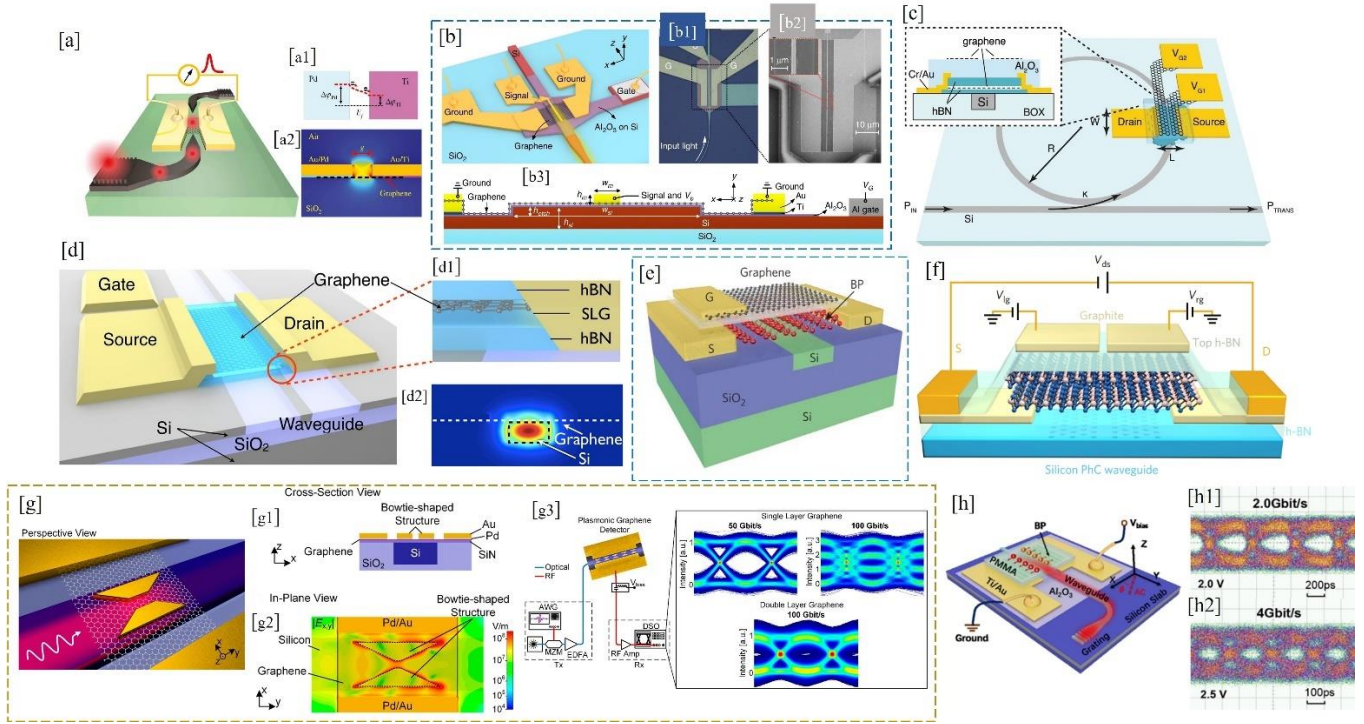

Fig.9: (a, a1, a2): SL CVD Gr-plasmonic integrated PD with bandwidth above 110 GHz.[117] (a) Gr-plasmonic hybrid PD schematics. (a1) Potential profile within the device displaying the drift of the photogenerated carrier. $\Delta\phi_{Pd}$ and $\Delta\phi_{Ti}$ respectively represent the difference between the Dirac point energy and the Fermi level in PD- and Ti-doped Gr; (a2) Cross-section of the device showing plasmonic effect at $\lambda$=1.55 μm. (b, b1, b2, b3): Si−Gr hybrid plasmonic WG PD. Adapted with permission.[118] Copyright 2020, Jingshu Guo et al. (b) Device configuration. (b1) Optical microscopy image; (b2) SEM images; (b3) Cross-section of Si−Gr hybrid plasmonic WG with the signal electrode in the middle and the ground electrodes on both sides (here, metal−Gr−metal sandwich structure was used). $V_b$ and $V_G$ are bias and gate voltages respectively. (c) Si ring resonator integrated GrPD. Adapted with permission.[119] Copyright 2021, Schuler et al. (d, d1, d2): hBN/SLGr/hBN PD on a buried Si WG. Adapted with permission.[120] Copyright 2015 ACS. (d) device schematics. Lower-right inset (d2): finite-element simulated electric field energy density of the fundamental TE WG mode. (e) BP PD integrated in a Si photonic circuit.[121](f) Schematic of the encapsulated bilayer MoTe$_2$ $p$–$n$ junction on top of a Si PhC WG. The carrier concentration in MoTe$_2$ is tuned via the split graphite gates; the distance between two gates is 400 nm; the dielectric layer is h-BN on top of the MoTe$_2$, and the thickness is 80 nm. The source (S) and drain (D) electrodes are thin graphite flakes connected to Cr/Au leads. Adapted with permission.[60] Copyright 2017, Springer Nature Ltd. (g, g1, g2, g3): Plasmonically enhanced WG-integrated Gr PD.[122] (h, h1, h2): Si/BP WG photodetector. (h) Configuration of the present hybrid Si/BP WG PD at 2 μm; (h1-h2) Observed eye-diagrams with different bit rates: (h1) 2.0 Gbit/s at 2 V); (h2) 4.0 Gbit/s at 2.5 V. Adapted with permission.[123] Copyright 2019 WILEY-VCH Verlag GmbH & Co. KGaA, Weinheim.

A hybrid plasmonic WG with an ultrathin, wide Si ridge (~100 nm thick Si) was used to fabricate a Si-Gr WG PD (**Fig.9**b, b1, b2, b3).[118] In this configuration, the light absorption in Gr was increased while metal absorption loss was minimized, which helped to improve $R$ to a greater extent. Additionally, a wide metal cap in the middle and the MGM sandwiched structures were inserted as the signal electrode and the ground electrode, respectively, to accomplish reduced Gr-metal contact resistances and a wider 3 dB bandwidth. Under zero bias, the photothermoelectric (PTE) effect dominates the photoresponse, however, when a bias voltage is supplied, the bolometric (BOL)/photoconductive (PC) effects predominate. At 2 μm, the presented Gr PD had a $R$ of ~70 mA/W and a 3 dB bandwidth of >20 GHz (setup-limited). The PDs also operated very well at 1.55 μm. For a bias voltage of −0.3 V (at 0.16 mW optical power), the $R$ was 0.4 A/W. The 3 dB bandwidth was greater than 40 GHz (experimental setup-limited).

By integrating a photothermoelectric GrPD with a Si micro-ring resonator, light absorption > 90% in a ~6 μm single layer Gr (SLG) channel along a Si WG was achieved, under critical coupling.[119] Cavity assisted light-matter interactions induced carriers in SLG to reach 400 K (carrier temperature) at an input power of 0.6 mW, resulting in a voltage $R$ 90 V/W, with a receiver sensitivity that enabled the GrPDs to operate at a $10^{-9}$ bit-error rate. On-chip ultrafast PD has been constructed and tested using a 2D heterostructure consisting of a Gr enclosed in a hexagonal BN. This 2D heterostructure-based PD demonstrated high-speed performance with a 3 dB cutoff at 42 GHz when coupled to the optical mode of a Si WG (**Fig.9**d, d1, d2).[120]

A WG-integrated and gate-tunable PD based on few-layer BP has been demonstrated for the telecom band (~1.55 μm) (**Fig.9**e).[121] At RT, this BP PD may function under bias with very low $I_{dark}$ and achieve intrinsic $R$ values of 135 mA/W and 657 mA/W in 11.5-nm and 100-nm-thick devices, respectively. The photovoltaic effect dominated the $I_{phot}$, and an operation at bit rates above 3 Gbit/s was achieved.

A Si photonic crystal (PhC) WG was integrated with MoTe$_2$-based lateral *p-n* junctions with an electrostatic split-gate arrangement allowed for different capabilities such as transistors, LEDs, and photodetectors (**Fig.9**f). To prevent natural oxidation of the MoTe$_2$, the entire device was enclosed by h-BN layers.[60]

A metal-Gr-metal photoconductive detector was constructed by co-integrating a high-speed plasmonically enhanced WG-integrated PD with a Gr layer.[122] Arrayed bowtie-shaped nanosized metallic structures were incorporated to excite Surface plasmon polaritons and strengthen the external $R$ of the PD. The 6 μm long device delivered a high external $R$ of 0.5 A/W under a −0.4 V bias with a frequency response surpassing 110 GHz. Furthermore, functionality over a broad spectral range covering S, C, and L bands of the optical communication windows was confirmed. The device's high $R$ and large bandwidth enabled 100 Gbit/s data reception of two-level OOK and four-level PAM-4 intensity encoded signals. The field enhancement effect induced by the plasmonic structure is shown in **Fig.9**g2, which shows an in-plane view of the extent of the $E_x$ and $E_y$ electric fields of the WG-integrated Gr PD, simulated by the 3D full-wave finite-element analysis. Substantial plasmonic field enhancements can be noticed at the borders and in the gaps of the bowtie-shaped structures. **Fig.9**g3 shows the measurement setup and observed eye diagrams of 50 Gbit/s on–off keying (OOK), 100 Gbit/s

PAM-4 intensity modulated optical signals with a SLGr device, and 100 Gbit/s OOK optical signals with a 2-layer Gr device (bottom figure).

BP thin films with optimized thicknesses (~40 nm) were introduced as the active material for light absorption, and Si/BP hybrid ridge WG PDs at 2 μm were demonstrated with a high $R$ of 306.7 mA/W (with a bias voltage of 0.4 V). The 3 dB-bandwidth was up to 1.33 GHz, and a 4.0 Gbit/s data receiving experiment is also demonstrated.[123] Similarly, in an integrated of Si-on-insulator (SOI) WG with BP PD, at a bias of 1 V, the BP PD achieved a $R$ of 23 A/W at 3.68 μm and 2 A/W at 4 μm and a $NEP < 1$ nW/Hz$^{1/2}$ at RT were observed.[124] The light-BP interaction was increased in this configuration by utilizing the optical confinement in the Si WG and grating structure to overcome the limitation of absorption length restricted by the BP thickness.

### 2D Tellurium (Tellurene)

Tellurium (Te), a metalloid element of group-VI, is well-known for its exceptional properties for numerous applications.[125] Under normal conditions, bulk Te is referred to as a 1$D$ vdW material ($\alpha$ phase). It has been predicted that the 2D structure of tellurium (tellurene)[126] has three configurations: the stable 1T-MoS$_2$-like ($\alpha$-Te) structure, the metastable tetragonal ($\beta$-Te) structure, and the 2H-MoS$_2$-like ($\gamma$-Te) structure. The covalent interactions in 2$D$ Te are very anisotropic, and the vdW force between the helical Te chains runs along the $c$-axis.[127] While decreasing the dimensionality to 2$D$, three more phases appear: $\beta$, $\gamma$, and 1T phases, which can be considered as reconstructions based on the $\alpha$ phase.[128] Te exhibits thickness dependent $E_g$s. For example, $\alpha$-phase Te has a wide $E_g$ tunability spanning from 0.3 eV for bulk to 1.2 eV for two layers (**Fig.10**d), but monolayer tellerene with the 1T phase exhibits metallic properties.[129] Because of its low bandgap and ability to shift optical transition from MWIR to NIR by changing the number of layers, 2D tellurium is regarded to be a suitable material for IR photodetection.[130] Strain has also been shown to have the capability to modulate Te's $E_g$.[131]

Large-area 2D Te fabrication is challenging because of its energy instability; nonetheless, large-area and stable 2D Te has recently been synthesized by lithium intercalation and solution processing.[132] Peng et al. have reported an unusual photoresponse in Te nanosheets that changes from negative to positive with thickness. The 2D Te structure was created from a layered $M$Te$_2$ ($M$ = Ti, Mo, and W) matrix using an excessive lithiation process. As a result, very thin Te layers with large sizes and clean surfaces were obtained. The photoresponse in Te nanosheets was prominently -ve when the thickness was < 5 nm and became +ve as the thickness increased.[133]

A combination of Au/Al$_2$O$_3$ optical cavity substrates and a solution-made, air-stable, quasi-2D tellurium (Te) nanoflakes were used for SWIR detection.[134] Via varying the thickness of the Al$_2$O$_3$ cavity, the peak $R$ of Te photoconductors could be changed from $\lambda$ = 1.4 μm (13 A/W) to $\lambda$ = 2.4 μm (8 A/W) with a cutoff $\lambda$ of 3.4 μm, completely covering the SWIR band. An optimal RT $D^*$ of $2 \times 10^9$ cm Hz$^{1/2}$/W was observed at a $\lambda$ of 1.7 μm. Fig.10b depicts the device's full-spectrum $R$. This detector has a maximum $R$ of 27 A/W at 78 K and 16 A/W at 297 K for $V_d$ = 5 V and = 1.7 µm. The photoresponse at 100 kHz was 5 times less than the photoresponse at 0.1 Hz due to a photoconductive gain present in the device (Fig.10c).[134]

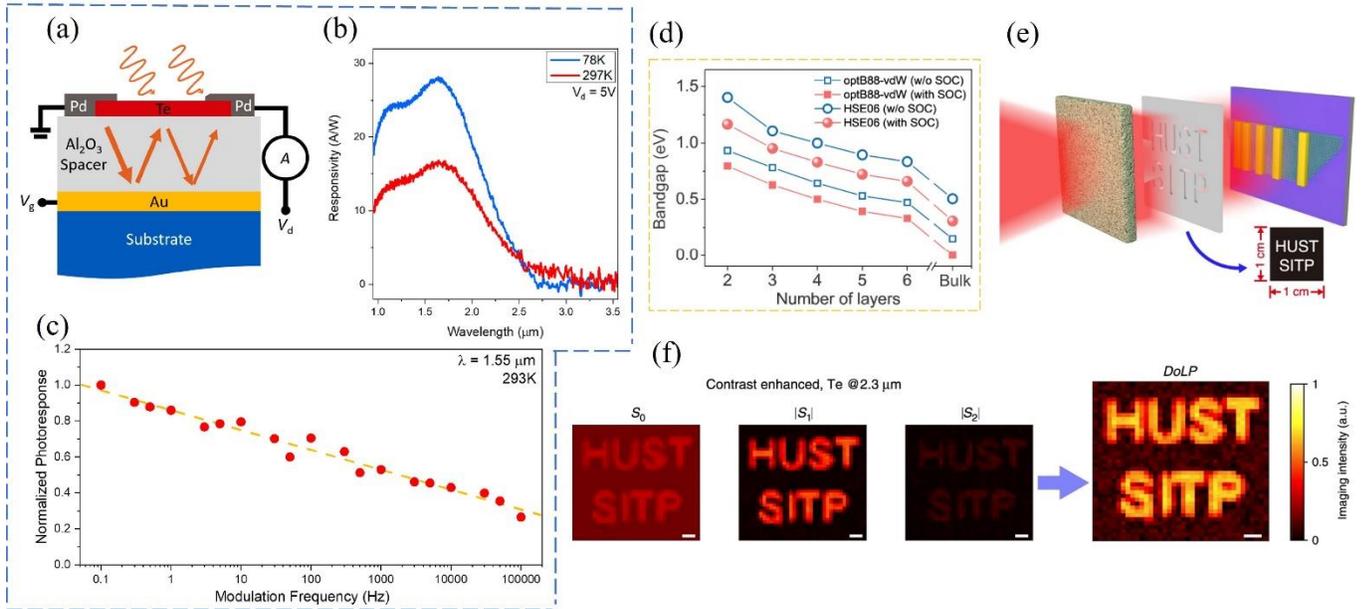

Fig.10: (a-c): Quasi-2D Te nanoflake SWIR photoconductor. (a) The optical cavity configuration adapted to fabricate SWIR photoconductors along with quasi-2D Te nanoflakes. Au film acts as the gate electrode and back-reflector and Al₂O₃ dielectric layer acts as a ∼ λ/4 spacer as well as the gate oxide; (b) Spectral responsivity/W of a Te photoconductor measured at 78 and 297 K under optimized gate bias and $V_d$ = 5 V; (c) Normalized photoresponse of a Te photoconductor versus modulation frequency; the device was excited using a sinusoidally modulated 1550 nm laser. Au/Al₂O₃ (150 nm) substrate was used for the measurements. Adapted with permission.[134] Copyright 2018 ACS; (d) Variation of $E_g$ with respect to number of Te layers; (e-f): Polarized infrared imaging. (e) Schematic of the imaging target, with a scattering media above the target. Inset: the schematic of the text pattern HUST and SITP; (f) Imaging degree of linear polarization (DoLP) results for the Te device under 2.3 µm illumination. The imaging contrast is strong for the target due to its intrinsic exceptional polarization detection ability. Scale bars in (f) is 1 mm. Adapted with permission.[135] Copyright 2020, Tong et al.

The 2D Te-based IR PD is known to exhibit polarization-sensitive SWIR photoresponse due to the anisotropic crystal structure of 2D Te. The broadband and ultrasensitive photodetection capabilities of hydrothermally produced and air-stable 2D tellurene nanoflakes were studied.[136] At RT, the tellurene nanoflakes demonstrated remarkable hole µs of ~458 cm²/V-s; the tellurene PD exhibited the maximal extrinsic responsivity of ~383 A/W, 19.2 mA/W, and 18.9 mA/W at λ = 520 nm, 1.55 µm, and 3.39 µm light, respectively. Because of the photogating effect, high gains of ~1.9 × 10³ and 3.15 × 10⁴ were observed at λs of 520 nm and 3.39 µm, respectively. The tellurene PD demonstrated extraordinarily high anisotropic behavior at the communication λ of 1.55 µm, yielding a wide bandwidth of 37 MHz.[136]

Although the hydrothermally synthesized approach is less expensive, the as-fabricated Te crystal quality was inferior to the more advanced CVD method. Very thin Te flakes (5 nm) were prepared via a hydrogen-assisted CVD method.[137] The Te flake-based transistor (11 nm thick) demonstrated a high ON/OFF ratio of $10^4$, a very low off-state current of $8 \times 10^{-13}$ A, and negligible hysteresis due to reduced thermally activated defects at 80 K. Furthermore, the Te-flake-based phototransistor displayed a large gate-dependent photoresponse: when the gate voltage varies from -70 to 70 V, $I_{on}/I_{off}$ increases by a factor of 40.

There have recently been some findings on polarized infrared imaging techniques that rely on a quasi-2D tellurium (Te) PD. The narrow $E_g$ of quasi-2D Te permits intense light-material coupling in the Vis to MIR (3.0 μm) spectral range, and the high $\mu$ and small effective carrier mass facilitate better photodetection performance.[127, 134, 138] In one of such studies, polarized IR imaging under scattering was made possible by using the photodetection properties of quasi-2D Te PDs.[135] The synthesized quasi-2D Te had a fairly high and anisotropic photoresponsivity of 1360 and 353 A/W, respectively, as well as a $D^*$ of $1.15 \times 10^{10}$ and $3.01 \times 10^9$ Jones in response to $\lambda = 1.06$ μm and 3 μm. Further, anisotropic photoresponse in scattering environments enabled polarized imaging.[135] Fig.10e depicts a schematic of the setup for the light image capture using the polarization imaging mechanism. Scattering media and a signal mask were set in series under light illumination to ensure the realization of a polarized light signal. When tested on a Te-based device under 2.3 μm illumination, a clear image of "HUST, SITP" was observed (Fig.10e-f). With respect to photoresponse performance, under 3.0 μm illumination, high $R$ of $\sim 3.54 \times 10^2$ A/W and $D^*$ of $\sim 3.01 \times 10^9$ Jones were attained due to excellent Te absorption.

The topological semimetals, platinum telluride ($PtTe_2$) has recently been reported to have a unique band structure (such as a type-II Dirac cone) and emerged as an intriguing contender for optoelectronics due to its unusual properties. Bulk $PtTe_2$ follows a $CdI_2$-type crystal structure.[139] However, large-scale $PtTe_2$ synthesis for future photoelectric and photodetection/imaging applications has received less attention. As a remedy, wafer scale 2D $PtTe_2$ materials have been created by injecting Te atoms into a pre-deposited Pt layer via a Te-vapor transformation approach.[140] The as-synthesized mosaic-like crystal structure of the 2D PtTe2 layers was constituted of domains of single crystals with highly preferential [001] orientation along the normal direction, limiting the effect of interface defects and enabling efficient out-of-plane carrier transport. This property, in tandem with $PtTe_2$'s broad absorption and the elegant vertical device architecture (Si CMOS compatible, Fig.11a), enabled the $PtTe_2$/Si Schottky junction PD to detect ultrabroadband (MIR spectral range: 3.04 -10.6 μm) with a high RT $D^*$ (0.93-6.92 × $10^9$ Jones) and a rapid $\tau_{res}$ of 2.4 μs. The proposed energy band model of the $PtTe_2$/Si vertical Schottky junction is shown in Fig.11c. Because the work function of multilayer $PtTe_2$ (4.80 eV) differs from that of $n$-Si (4.25 eV), when the mosaic-like 2D $PtTe_2$ layer meets the Si, a built-in potential is generated. On the Si side, a depletion layer will be formed, and the built-in potential will be directed from Si to $PtTe_2$. Both the mosaic-like 2D $PtTe_2$ layer and the Si substrate absorb light when exposed to UV-Vis-NIR light (200-1100 nm), and the photogenerated carriers are promptly separated by the built-in potential. Electrons would then transfer from the 2D $PtTe_2$ layer to the Si substrate, whereas holes would migrate into the 2D $PtTe_2$ layer and be acquired by the Gr electrode, resulting in a $I_{phot}$. Si is virtually blind to $\lambda$s > 1100 nm due to its bandgap constraint. The photoresponse in the longer $\lambda$ range is thus mostly attributable to the ultrawide light absorption of the mosaic-like 2D PtTe$_2$ layer. When illuminated with NIR-MIR light, the electrons in the 2D PtTe$_2$ layer can tunnel to the conduction band of the Si substrate due to the presence of an ultrathin depletion layer, resulting in a photoresponse of ~10.6 μm (Fig.11b).

Further, the high-resolution IR imaging capability of the PtTe$_2$/Si PD was demonstrated. Fig.11d depicts a schematics of the experimental setup and clear sensing imaging results based on a single device.

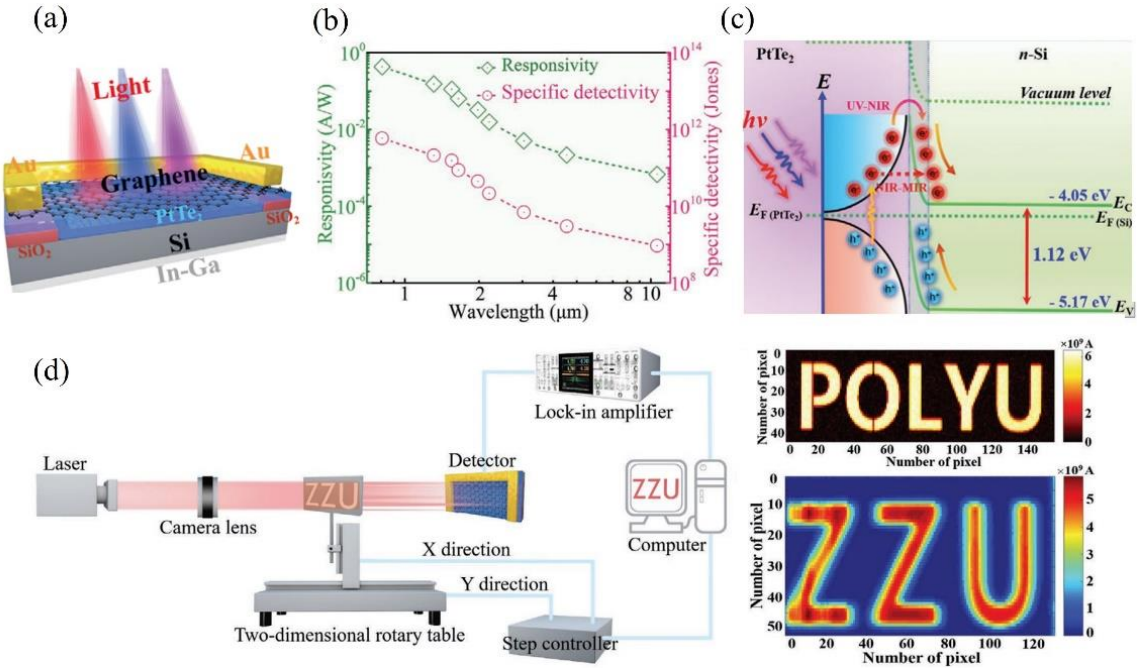

Fig.11: PtTe$_2$/Si Schottky junction PD, MIR photoresponse, working mechanism, and IR image sensing. (a) Schematics of the PtTe$_2$/Si vertical Schottky junction PD; (b) $\lambda$-dependent $R$ and $D^*$ of the device; (c) Energy band model of the PtTe$_2$/Si vertical Schottky junction PD; (d) Setup for a single-pixel imaging experiment. The observed images of "POLYU" and "ZZU" under 4.55 and 10.6 μm illumination, respectively.[140]

Similarly, there is a report about the direct tellurization of Pt nanofilms to fabricate large-area PtTe$_2$ films and the effect of growth conditions on the morphology of PtTe$_2$.[141] The electrical investigation of the as-grown PtTe$_2$ films revealed that they behaved like typical semimetals, which is consistent with the results of the first-principles density functional theory (DFT) simulation. Furthermore, the PtTe$_2$-based PD demonstrated a broad photoresponse to incoming radiation in the $\lambda$ = 200-1650 nm range, with the strongest photoresponse at 980 nm. The PtTe$_2$-based PD exhibited $R$ and $D^*$ of 0.406 A/W and 3.62 × 10$^{12}$ Jone, respectively. In addition, the $EQE$ was as high as 32.1%. The $\tau_r$ and $\tau_d$ were reported to be 7.51 and 36.7 μs, respectively. The $I$-$V$ characteristic (nonlinear behavior) of the PtTe$_2$/Si heterojunction without light illumination reveals a pronounced rectifying tendency, with the rectification ratio estimated to be ~10$^4$ at ±1 V. By employing a numerical power law, $I_{phot} = AP_\lambda^\theta$, where $\theta$ is an exponent, $A$ is a constant related with 980 nm NIR light, and $P_\lambda$ is 980 nm laser power density, an exponent value of 0.851 for the PtTe$_2$-based NIRPD was obtained, revealing a weak rate of recombination activity due to the relatively low density of defect states (Fig.12c). As shown in Fig.12d, $R$ and $EQE$ decrease with increasing light intensity due to increased carrier recombination, which impedes $I_{phot}$ generation. Finally, an image sensor comprised of an 8 × 8 PtTe$_2$-based PD array capable of recording five NIR images with a better resolution under 980 nm was fabricated (Fig.12a).

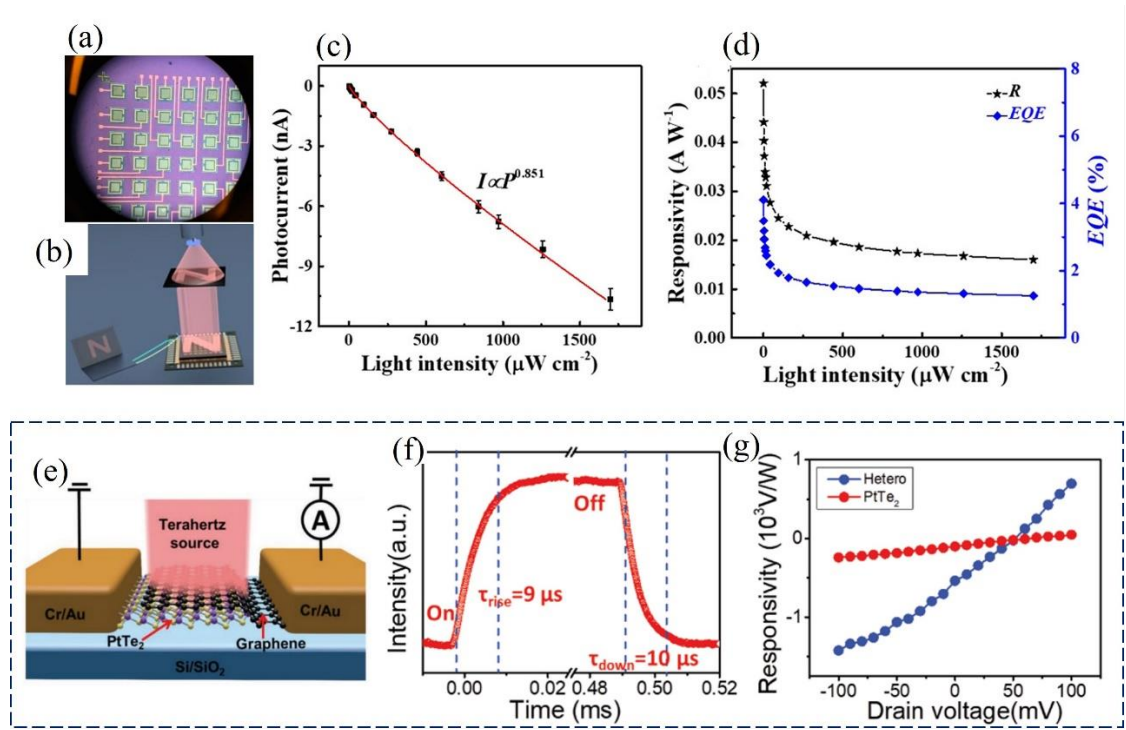

Fig.12: (a-d): PtTe$_2$-based PD. (a) Digital photograph of the PtTe$_2$-based NIRPD array; (b) PtTe$_2$-based image sensor in operation; (c) The quantitative relationship between the $I_{phot}$ of the PtTe$_2$-based NIRPD and light power density; (d) The variation of $R$ and $EQE$ of the PtTe$_2$-based NIRPD with respect to varying intensities. Adapted with permission.[141] Copyright 2020 ACS. (e-g): PtTe$_2$-Gr heterostructure; (e) Illustration of the electrical configuration for PtTe$_2$-Gr vdW heterostructure based THz detector; (f) The response/recovery times for the heterostructure THz device under zero bias; (g) The voltage responsivity of the two devices under different bias at 0.12 THz. Adapted with permission.[139] Copyright 2019 WILEY-VCH Verlag GmbH & Co. KGaA, Weinheim.

A planar metal/PtTe$_2$/metal structure based THz PD with bow-tie-type planar contacts demonstrated a high photoresponsivity of ~1.6 A/W without a bias voltage and a $\tau_{res} < 20$ µs at RT.[139] To improve millimeter-scale wavelength absorption, a bow-tie antenna with a subwavelength gap was connected to the detecting material to make a channel. In the same work, a PtTe$_2$-Gr heterostructure-based detector was designed via electron beam lithography and tested (Fig.12e). In self-powered mode, this device produced a $R$ of 537 V/W at RT and was capable of surpassing 1.4 kV/W with a $\tau_{res} < 9$ µs (Fig.12f). As shown in Fig.12g, the heterostructure (PtTe₂-Gr) device has a substantially higher responsivity than the pristine PtTe₂ device, owing to the high light absorption and real-space asymmetrical contacting between these two different Dirac materials.

Due to their competitiveness as light-absorbing materials for photovoltaic applications, perovskite materials (formula $ABX_3$, where $A$, $B$, and $X$ are organic cation, metal cation, and halide anion, respectively) have recently piqued the attention of researchers. The highest reported power conversion efficiency of solar cells based on this material is 23%,[142] which is attributed primarily to the better intrinsic optoelectronic properties of organic-inorganic trihalide perovskites (OITPs), including strong optical absorption and a low nonradiative recombination rate.[143] Ascribed to the well-suited interface nature between organic semiconductors and carbon electrodes, carbon electrode-molecule junctions (CEMJs)-based transistors exhibited outstanding field-effect performance with sensitive photoresponsive ability. A nanoscale hybrid PD was fabricated by combining OITP active material and CEMJ nanoarchitecture into a single device This hybrid PD showed high performance simultaneously in $R$ ($56 \times 10^7$ A/W), photo gain ($53 \times 10^8$), $D^*$ ($\sim 28 \times 10^{15}$ Jones), $LDR$ ($\sim 92$ dB), and a broad spectral response ($\sim 300$–$800$ nm).[144]

In a recent study, an ionic liquid, 1-butyl-3-methylimidazolium tetrafluoroborate (BMIMBF₄), was added as an additive into methylammonium lead triiodide (MAPbI₃) NWs to passivate defects and generate nanochannels, allowing for fast charge transfer.[145] Consequently, the prolonged stability and performance of MAPbI₃ NWs were enhanced significantly. When tested on Au/MAPbI₃ NW/Au structured device (with 0.6 mmol BMIMBF₄), the $D^*$, $LDR$, and $NEP$ of the MAPbI₃ NW PD reached $2.06 \times 10^{13}$ Jones, 160 dB, and $1.38 \times 10^{-15}$ W/Hz$^{1/2}$, respectively at irradiance power of $1.45 \times 10^{-6}$ mW/cm². After nearly 5000 hours in an open environment, the unencapsulated PD retained 100% of its initial performance.

A halide perovskite, CH₃NH₃PbI₃ granular wires (PGWs) have been tested in the design of flexible high-performance photosensor arrays.[146] The $D^*$ of these PGWs was $3.17 \times 10^{15}$ Jones. Furthermore, $8 \times 8$ integrated flexible PGW photosensor arrays capable of detecting incident photonic signals with high resolution were produced using selective surface treatment procedures. The distinctive band-edge modulation along the PGW axial direction is expected to result in significant inhibition of $I_{dark}$ generation in the device, resulting in high performance.

Irrespective of their salient opto-electronic features, the use of hybrid perovskites is restricted due to their intrinsic instability towards oxygen and moisture. It is known that the hydrophobic ligands at the ends of Ruddlesden-Popper (RP) perovskites make them much more stable against humidity and air.[147]

Table 1: A comparison of the performance of several photodetectors based on 2D materials

| 2D materials | $E_g$ [eV] | $\lambda$ [µm] | Detection range | $R$ (A/W) | Detectivity [cm Hz$^{1/2}$/W] | Response time [s] | Ref |
|---|---|---|---|---|---|---|---|
| Gr | 0 | 12.2 | Vis-LWIR | $160 \times 10^{-4}$ | | $10^{-7}$ | [148] |
| Gr | 0 | 1.55 | SWIR | 83 | $10^8$ | $6 \times 10^{-7}$ | [149] |
| Gr | 0 | 1.47 | SWIR | 0.2 | | | [150] |
| | | 10 | LWIR | 0.4 | | | |

| Material | Bandgap | Wavelength | Range | Responsivity | Detectivity | Response time | Ref |
|---|---|---|---|---|---|---|---|
| Two-layer graphene | 0 | 1.3 | SWIR | 4 | | | [151] |
| | | 2.1 | SWIR | 1.9 | | | |
| | | 3.2 | MWIR | 1.1 | | | |
| Graphene nanoribbons | 0.1 eV | 1.47 | NIR | 1.5 | | | [152] |
| | | 10 | LWIR | 0.18 | | | |
| BP | 0.3 | 3.39 | MWIR | 82 | | $10^{-4}$ | [153] |
| BP | 0.3 | 3.68 | MWIR | 23 | | | [124] |
| BP | 0.3 | 1.56–3.75 | SWIR–MWIR | | $10^6$ | $65 \times 10^{-15}$ | [154] |
| BP | 0.3 | 1.55 | SWIR | 1.55 | $5.5 \times 10^9$ | $16 \times 10^{-6}$ | [112] |
| BP | 0.3 | 2 | SWIR | 8.5 | $1.7 \times 10^9$ | | [155] |
| PCWG-BP | 0.3 | 3.825 | MWIR | 11.31 | | | [156] |
| Metal dichalcogenides | | | | | | | |
| $MoTe_2$ | | 0.915 | NIR | 1479 | $8.2 \times 10^{11}$ | $18 \times 10^{-3}$ | [157] |
| $MoS_{1.89}$ | 0.26 | 2.7 | SWIR | $28 \times 10^{-3}$ | $0.88 \times 10^9$ | | [158] |
| $MoS_{2.15}$ | 0.13 | 7.8 | MWIR | $220 \times 10^{-4}$ | | | [159] |
| Td-MoTe | | 10.6 | LWIR | $41 \times 10^{-6}$ | $91 \times 10^5$ | $3.1 \times 10^{-7}$ | [160] |
| $Td-WTe_2$ | | 3.8 | MWIR | 250 | | | [161] |
| $ReSe_2$ | 0.9 | 0.98 | NIR | $4.96 \times 10^3$ | | $1.4 \times 10^{-3}$ | [162] |
| $ReSe_2$ | 0.9 | 0.785 | NIR | $4.2 \times 10^4$ | | $1.9 \times 10^{-3}$ | [163] |
| $ReSe_2$ | 1 | 1.064 | NIR | $\approx 4 \times 10^4$ | | $30 \times 10^{-3}$ | [164] |
| $n\text{-}WSe_2$ | | 0.852 | NIR | $23 \times 10^{-3}$ | $2.3 \times 10^{10}$ | $68.3 \times 10^{-3}$ | [165] |
| $p\text{-}WSe_2$ | | 0.852 | NIR | $37.7 \times 10^{-3}$ | $10^{10}$ | $61.2 \times 10^{-3}$ | [165] |
| $WSe_2$ | | | NIR | 0.6 | | | [166] |
| $WSe_2$ | | | visible−NIR | 224 | | | [167] |
| $Bi_2Te_3$ nanosheets | 0.165 | 1.55 | SWIR | 74 | $3.8 \times 10^9$ | 0.42 | [168] |
| $Bi_2Te_3$ nanowires | 0.165 | 1.55 | SWIR | 778 | $1.2 \times 10^9$ | 0.5 | [168] |
| $Bi_2Te_3$ nanoplate | 0.21 | 0.85 | NIR | $55.06 \times 10^3$ | $5.92 \times 10^7$ | 1.043 | [90] |
| $Bi_2S_3$ | 1.3 | 0.78 | NIR | 4.4 | $10^{11}$ | $10^{-5}$ | [91] |
| $Bi_2S_3$ nanobelts | 1.3 | 0.8 | NIR | 201 | $10^{10}$ | $5 \times 10^{-5}$ | [169] |
| $Sb_2Te_3$ | 0.28 | 0.98 | NIR | 21.7 | $1.22 \times 10^{11}$ | 238.7 | [170] |
| $PtSe_2$ | 0.3 (BL) | 10 | LWIR | 4.5 | $\sim 7 \times 10^8$ | $10^{-3}$ | [171] |
| $PdSe_2$ | 0.1 | 10.6 | LWIR | 42.1 | $1.10 \times 10^9$ | $74.3 \times 10^{-3}$ | [172] |
| $PdSe_2$ | 0.1 | 1.06 | NIR | 708 | $1.31 \times 10^9$ | | [173] |
| Miscellaneous | | | | | | | |
| $Bi_2O_2Se$ | | 0.85–1.55 | SWIR | 101 | $1.9 \times 10^{10}$ | <30 ms | [174] |
| Black-AsP | 0.15 | 8.2 | MWIR | $15 \times 10^{-3}$ | $4.9 \times 10^9$ | $0.54 \times 10^{-3}$ | [175] |
| Te | 0.31 | 3.4 | MWIR | 13 | $2 \times 10^9$ | 0.57 | [176] |
| Tellurene | 0.27 | 1.55 | SWIR | $19.2 \times 10^{-3}$ | | | [54] |
| GeAs | 0.6 | 1.6 | SWIR | 6 | | $10^{-3}$ | [177] |
| GeSe | 1 | 0.85 | NIR | $28 \times 10^{-2}$ | $41 \times 10^8$ | $74 \times 10^{-3}$ | [178] |
| PbS | 0.4 | 0.8 | NIR | 1621 | $172 \times 10^9$ | 0.3 | [179] |
| GaSe | | | UV−visible | 149 | | | [180] |
| SnS | 1.1 | 0.8 | NIR | 300 | $6 \times 10^9$ | $36 \times 10^{-3}$ | [181] |
| SnS | 1.1 | 0.85 | NIR | 1604 | $3.42 \times 10^{11}$ | $7.6 \times 10^{-3}$ | [182] |
| SnTe | 0.197 | 0.98 | NIR | $698 \times 10^{-3}$ | $3.89 \times 10^8$ | 1.45 | [183] |
| InSe | | | visible-NIR | 157 | | | [56] |

## 4.2 2D Heterostructures for IR Detection

Blending distinct *2D* materials in vertical vdW heterostructures or coupling 2D materials with other nanomaterials allows for diverse amalgamation of the individual components' exclusive optoelectronic capabilities in a single device with several qualities.[184] In general, a *2D* heterojunction is an interface between two layers of different 2D materials. The alignment of the energy bands at the interface is critical to the photoelectrical behavior of a heterojunction. In accordance with the band alignment there are three types of heterojunctions in semiconductor interfaces: straddling gap (type-I), staggered gap (type-II), and broken gap (type-III).[185]

The following section will examine various heterostructures and their corresponding infrared (IR) detection techniques. These include 2D/0D systems: graphene/quantum dot (Gr/QD) and transition metal dichalcogenide/quantum dot (TMDC/QD), as well as 2D/2D systems like graphene-based and transition metal dichalcogenide-based heterostructures.

### 4.2.1. 2D/0D Heterostructure

Gr is known to possess extremely high $\mu \sim 200000$ cm$^2$/V-s[66, 72] and broad $\lambda$ absorption, prompting the development and deployment of Gr transistors as ultrafast IR PDs in high-speed optical communications (Section 4.1).[2] However, due to the poor light absorbance of Gr ($\sim$2% for single layer Gr), the $R$ of the IR detectors is very low ($\leq 6.1$ mA/W). As a result, they can only detect extremely bright IR light.[2] Furthermore, the hot carrier effect, rather than the photovoltaic effect, was discovered to be one of the causes of the photoresponse of Gr. Gr's lack of a bandgap and low inherent light absorption pose certain difficulties for its utilization in practical applications.[186] But, pairing Gr with the right IR light absorbers is a feasible and straightforward strategy to get around Gr's weak light absorption and improve the photoresponsivity of a Gr based heterostructure.[40b, 77, 187] For example, by altering the Gr layer with IR light absorbing QDs, the $R$ of a Gr-based IR detector can be significantly increased. If the carriers triggered by IR light can migrate to the Gr film, their $\mu$ will be greatly increased, thereby increasing the $R$ of the QD-based IR detector.

Further, the size of QDs has a notable influence on the $E_g$ of QDs in Gr/QD heterostructured PDs. This factor can change the light absorption band, the photoconductive gain, and the overall merits of PDs.[63b]

$$E_g^{QD} = E_g^{bulk} + \frac{2h^2}{D^2}\left(\frac{1}{m_e} + \frac{1}{m_h}\right) - \frac{3.6e^2}{4\pi\varepsilon\varepsilon_o D}$$
<div align="right">Equation 15</div>

Where $E_g^{bulk}$ is the bulk material bandgap; $D$ is the diameter of QDs; $e$ is the electronic charge; the second and third terms come from the confinement effect and Coulomb interaction, respectively. Smaller QDs have larger $E_g$, require larger energy from light to trigger the electron transfer. Therefore, QDs of appropriate size can be employed to detect distinct light waves. In PbS QDs, the $E_g$ may be lowered from 3 to 2.5 eV by increasing the size of QD.[188]

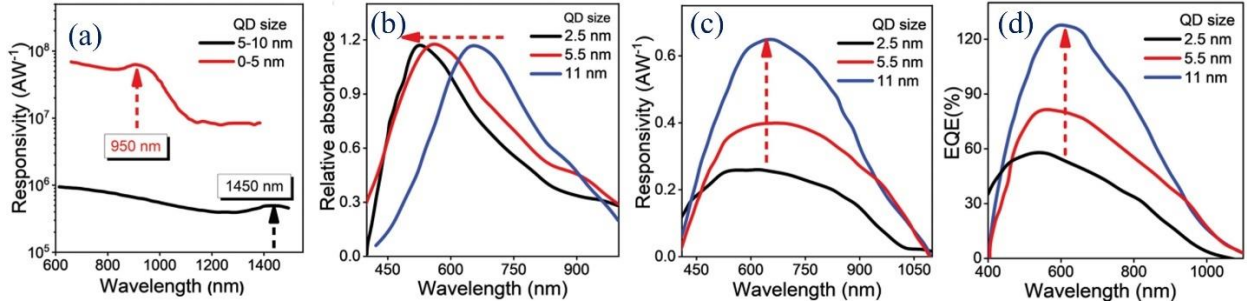

Fig.13: PbS and Ge QDs size effect on photoresponse properties. (a) Variation of $R$ with $\lambda$s when the PbS QD size is 0 to 5 nm and 5 to 10 nm; (b) Relation between absorption and $\lambda$s when the Ge QD size are 2.5, 5.5, and 11 nm; (c) Variation of $R$ with $\lambda$s when Ge QDs sizes are 2.5, 5.5, and 11 nm; (d) Variation of $EQE$ with $\lambda$s when Ge QD size are 2.5, 5.5, and 11 nm. Adapted with permission.[189] Copyright 2016 Elsevier B.V.

The variation of $R$ with respect to PbS QD size in a Gr/PbS QD heterostructured PD is displayed in Fig.13a. With an increase in QD size and $E_g$, the $R$ can approach as high as $\approx 10^7$ A/W. The photodetection range is also heavily influenced by the size of the QDs, with the 0-5 nm PbS QD heterostructured PDs having a maximum $R$ at the $\lambda = 950$ nm ($E_g = 1.31$ eV) and the 5-10 nm size QDs having a maximum $R$ at the $\lambda = 1450$ nm ($E_g = 0.86$ eV) (Fig.13a). This is explained by the fact that the electron transition in the smaller QDs requires more light energy to be activated. Fig.13b illustrates how the size of the Ge QDs decides the performance of Gr/Ge QD heterostructured PD. Ge QDs' light absorption shifts into higher optical bands as their size decreases, allowing for light absorption in higher energy ranges. The PDs displayed a $R$ of 0.65 A/W and an $EQE$ of 130% (Fig.13c-d).[195]

Bulk PbS is a semiconductor with a $E_g$ of 0.41 eV and an ionization potential of ~4.95 eV.[190] Yet, because of the quantum confinement effect, the ionization potential of PbS QDs can be higher. The $E_g$ for PbS QDs is predicted to be ~1.2 eV (with absorption peak at 990 nm). Because the effective masses of holes and electrons in PbS are approximately identical, the conduction and valence band levels of PbS QDs will differ from those of PbS bulk. As shown in the inset of **Fig.14**b, the energy levels for the conduction and valence bands of PbS QDs are calculated to be 4.15 eV and 5.35 eV, respectively. The PbS QD ($p$-type) has a Fermi energy near its valence band.[191] Conversely, at the Dirac point, the Fermi level of intrinsic Gr is ~4.6 eV.[192] As a result, holes in PbS will migrate to the Gr sheet, causing a $p$-type doping effect in Gr at the PbS QD/Gr heterojunction. However, as illustrated in **Fig.14**b, a drop in the Fermi levels in Gr may result in higher contact resistances for electron injection, thus decreasing the effective electron $\mu$ of the transistor.[193]

A gain of $10^8$ electrons/photon and a $R$ of $10^7$ A/W were observed in a hybrid PD comprising single layer or bilayer Gr bounded with a thin film of PbS CQD. When illuminated, the photogenerated carriers isolated at the interface of Gr/PbS CQD, trapping one type of carrier in the QD layer and circulating the other type of carrier in the Gr channel via the photogating effect, delivering very high optical gain and photoresponsivity. With a $D^*$ of $7 \times 10^{13}$ Jones, the device benefited from gate-tunable sensitivity and speed, as well as spectral selectivity from Vis-SWIR range. In this heterostructure, PbS QD served as a tunable light absorber in the NIR.[77] Reports say that the way the sensor works is that NIR light causes charges to form in PbS QDs, which can regulate the Fermi level and, in turn, the conductivity of the underlying Gr film.[194] In one such studies, NIR photoconductors made of CVD-grown monolayer Gr coated with PbS QDs (synthesized via solution processing) are reported to have better photoresponsive properties.[194] The devices were solution-processed on different substrates, including flexible ones, and demonstrated high $R$ of ~$10^7$ A/W (Fig.*14*c), which are significantly higher than the $R$s of the Vis light detector based on CVD-grown Gr and PbS thin films obtained by electron beam deposition.[195] Furthermore, the ligand layer covering the surface of the QDs was shown to be critical to the device's photoresponsivity, since the ligand layer dominated charge transfer from the QDs to the Gr film. The devices fabricated on flexible plastic showed exceptional bending stability.

Integrated circuits (ICs) based on Si-based complementary metal oxide semiconductor (CMOS) technology have been at the center of the technological revolution in image sensors over the past few decades. The difficulty in integrating nonsilicon electrooptical materials with Si ICs, on the other hand, has been an obstacle to realizing their vast potential for detection beyond the Vis range.[196] The photodetection performance of a hybrid decorated (N, S codecorated) Gr-PbS CQD phototransistor (Si CMOS-compatiable) is reported.[197] The resulting device has an ambipolar feature that can be tuned with the gate and has a low gate bias of less than 3.3 V at RT. With a gain of $10^5$ and a $\tau_{res}$ of 3 ms, broadband spectra from Vis to NIR and SWIR light can be detected. With a low driving voltage of 1 V, the phototransistor displayed a high $R$ of $10^4$ A/W and a $D^*$ of $10^{12}$ Jones when illuminated by SWIR light (1550 nm).

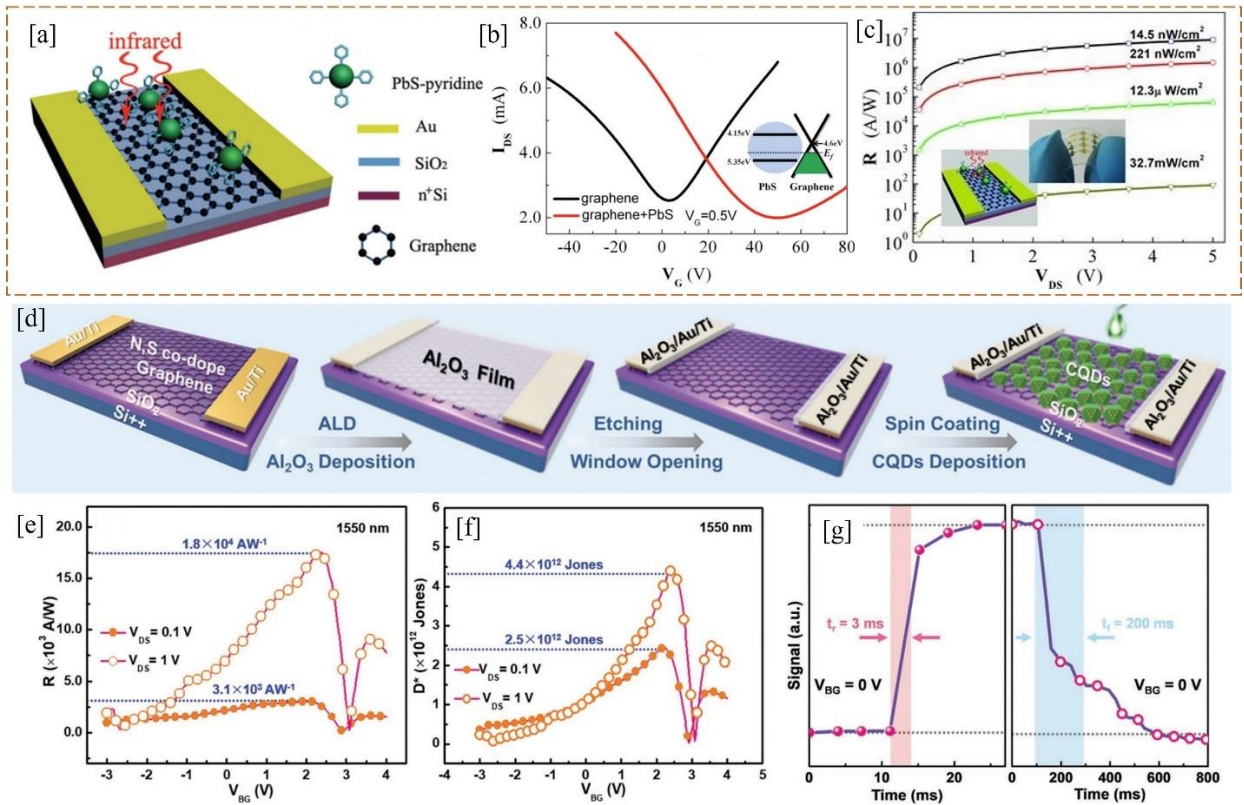

Fig.14: (a-c): CVD-Gr/PbS QDs PD. (a) Schematics of a Gr transistor modified with PbS QDs under light illumination; (b) Transfer characteristics ($I_{DS} \sim V_G$, $V_{DS} = 0.5$ V) of Gr transistors with or without the addition of PbS QDs on the Gr film. where $I_{DS}$ is channel current; $V_{DS}$ voltage applied between the source and drain electrodes; $V_G$ is gate voltage. Inset: Energy diagram of the heterojunction of PbS QD and Gr; (c) $R$ of a PbS QDs/Gr photoconductor as functions of applied voltage characterized under different light irradiance. $\lambda$: 895 nm. Adapted with permission.[194] Copyright 2012 WILEY-VCH Verlag GmbH & Co. KGaA, Weinheim; (d-g): Gr-PbS CQD PDs. (d) The fabrication steps of hybrid Gr-PbS CQD phototransistor; (e) The $R$ of the hybrid PD as a function of the gate voltages for fixed driving voltage of 1 and 0.1 V under $\lambda = 1550$ nm illumination; (f) The $D^*$ vs. gate voltage for fixed driving voltages of 1 and 0.1 V under $\lambda = 1550$ nm illumination; (g) The rising and falling edges observed at $\lambda = 1550$ nm. Adapted with permission.[197] Copyright 2018 WILEY-VCH Verlag GmbH & Co. KGaA, Weinheim.

The development and characterization of a Gr/PbS QD hybrid structure applicable to NIR PDs has been reported.[198] A spin casting machine was used to deposit blends of Gr flakes and PbS QDs on a glass substrate. The Gr-PbS QD composite's higher absorbance and quenched photoluminescence intensity supported greater photoinduced charge transfer between Gr and the PbS QDs. A Gr/PbS QD-based PD structure was devised for device application. When compared to PbS QDs alone, the produced PD with Gr/PbS QDs had a five-fold larger $I_{phot}$, a 22% faster $\tau_r$, and a 47% faster $\tau_d$.

Fullerene ($C_{60}$), a 0D carbon allotrope, has a diverse set of chemical and physical properties and tested in optoelectronic applications.[199] The strong and variable UV light absorption of $C_{60}$ molecules allows for the formation of a UV functional composite with Gr.

Theoretical and experimental investigation shows the existence of very efficient charge transfer at the Gr/C$_{60}$ interface.[200]

A sensitive UV PD (sensitive to 200−400 nm) based on the Gr/C$_{60}$ heterointerface yielded UV $R$ of ∼10$^7$ A/W (**Fig.15**a-c).[201] Extremely high photoconductive gain was achieved by efficient exciton seperation at the heterointerface and increased optical absorption. The gate voltage can alter the response time at the heterointerface because of the $e^-$-$h^+$ recombination process. The use of all-carbon hybrid materials ensured stable operation and allowed for the validation of a model 5 × 5 (2-dimensional) PD array. **Fig.15**b depicts the $R$ and $I_{phot}$ measured at $V_{DS}$ = 250 mV at different gate biases and optical illumination power density. The $R$ of the device reached ∼10$^6$ A/W under 405 nm and a power density ∼5 μW/cm$^2$. The relationship between $I_{phot}$ and C$_{60}$ film thickness (**Fig.15**c) shows that with an increase in C$_{60}$ film thickness, the $I_{phot}$ first grow and then gradually get saturated, showing that the photoresponse is mostly dominated by C$_{60}$ molecules near the heterointerface. The *IQE* begins to decrease as the thickness of the C$_{60}$ layer increases (inset of *Fig.15*c). However, it stays ∼2.0 × 10$^4$ for all devices, even under a ∼5 mW/cm$^2$ laser power density.

Because carbon is ubiquitous in nature, inexpensive to extract, and nontoxic, the carbon quantum dots (C-QDs) coupled Gr/Si Schottky junction structure has the potential to be used to produce competent, ecologically benign, and cost-efficient PDs. A C-QDs (5 nm in size)–coupled Gr/Si Schottky junction PD exhibited a $R$ of 290 mA/W.[202] Carbon evaporation was used to create C-QDs with a high light absorption rate for UV light. This allowed for easy control over the distribution and pattern of the QDs, which can be easily maneuvered by a mask. Vis-a-vis Gr PDs, in Gr/ C-QDs heterostructure PDs, photoresponsivity was augmented by 4.3% when the C-QDs were coupled with the Gr/Si Schottky-junction. The visible spectrum was covered completely by the wideband spectral response. What's more, the device's $\tau_r$ was 0.93 μs and its $\tau_d$ was 2.24 μs, which are considerably faster than the $\tau_{res}$ of Gr PD. *Fig.15*i depicts the schematics of the C-QDs -integrated Gr/Si Schottky-junction PD and *Fig.15*j shows the band model, where $E_f$ is the Fermi level. Gr is $p$-type doped and the $E_f$ is decreased during the Gr transfer and device construction processes. When the $n$-doped Si is in contact with Gr, thus forming a Schottky junction, the $E_f$s of Si and Gr are leveled, and the energy band of the $n$-Si is bent. $W_g$, and W$_{Si}$ are the work functions of Gr and Si, respectively. $\phi_{bn}$ is SBH between Gr and the $n$-Si (∼the difference between their work functions). Incident photons elicit electrons from the valence band to the conduction band, initiating to the generation of $e^-$-$h^+$ pairs. When the energy of the photon > $\phi_{bn}$, the photogenerated carriers create a $I_{phot}$. When C-QDs are paired with Gr, electrons from the Gr are transported to the QDs. The $E_f$ of Gr is reduced, the SBH (the width of the depletion layer) increases, and electrons and holes are efficiently separated.

Si colloidal QDs have special advantage because of the ample supply and nontoxicity of Si, the stability of Si QDs, and their compatibility with Si-based technologies.[203] Doping Si QDs with boron (B) and phosphorus (P) causes Si QDs to induce LSPR in the MIR region.[204] In terms of dopant activation efficiency, oxidation resistance, and dispersibility, B-doped Si QDs are currently preferable to P-doped Si QDs for device applications.[204a] This makes it likely that B-doped Si QDs could be used as plasmonic antennas to elevate Gr optical absorption more in the MIR region.[39a, 205] Given that heavy-B-doping-induced band-tail states have already expanded the optical absorption of Si QDs from a typical UV-Vis window into the NIR regime,[206] heterostructured phototransistors based on Gr and Si QDs should offer UV-to-MIR ultrabroadband photodetection capability. The photodetection range of Si-QD/Gr hybrid phototransistors has been tremendously expanded by using doping-induced tunability of Si-QD optical absorption (**Fig.15**d).[207] The electron-transition-based sub-bandgap optical absorption of B-doped Si QDs extended the photodetection range of QD/Gr hybrid phototransistors from the usual UV−Vis region into the NIR region (leading to photogating for Gr), while B-doped Si QDs' LSPR stretched the photodetection range to the MIR region; B-doped Si QDs improved Gr's MIR absorption. This UV-to-MIR ultrabroadband photodetection of QD/Gr hybrid phototransistors achieved peak $R$ ($\sim10^9$ A/W), gain ($\sim10^{12}$), and $D^*$($\sim10^{13}$ Jones). In the MIR and UV-to-NIR regions, the reported values of $NEP$ were $\sim10^{-10}$ and $10^{-18}$ W/Hz$^{1/2}$ respectively.

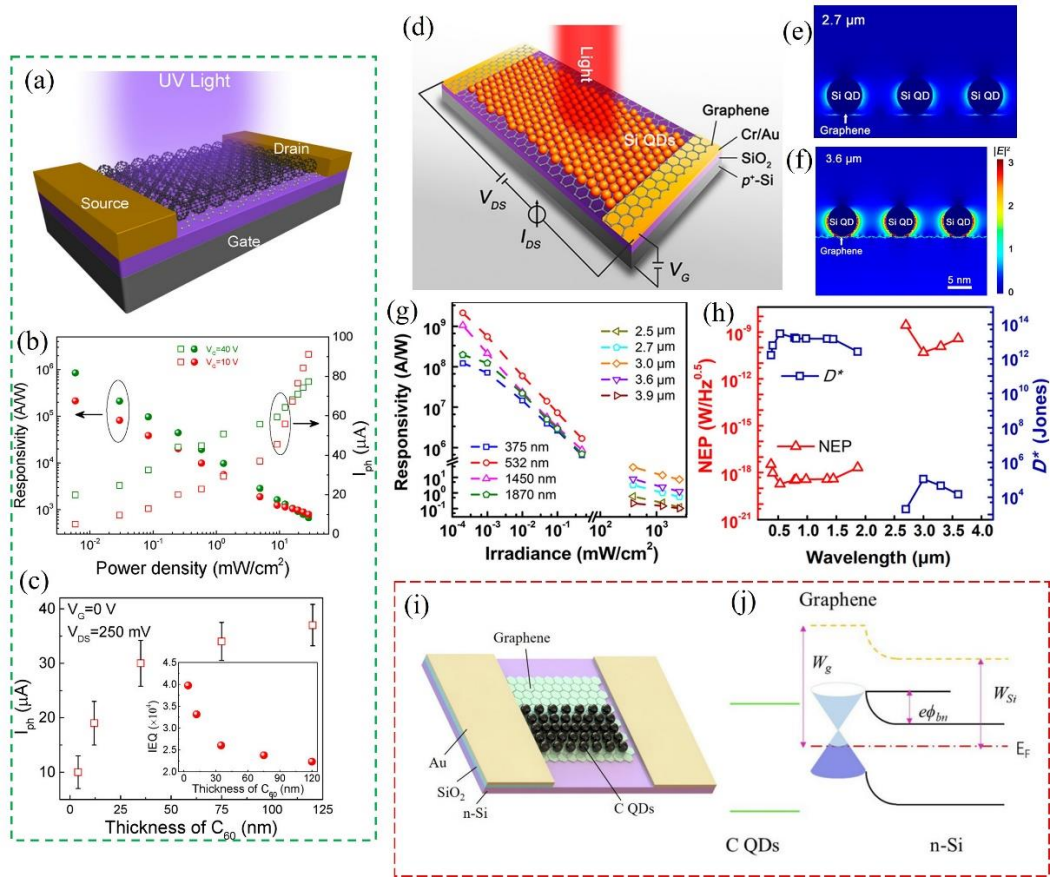

Fig.15: (a-c): Gr/C$_{60}$ phototransistor. (a) Schematic of the Gr/C$_{60}$ phototransistor on SiO$_2$/Si; (b) $R$ and $I_{phot}$ of the hybrid Gr/C$_{60}$ device vs. optical illumination power density at different gate biases; (c) $I_{phot}$s measured in Gr/C$_{60}$ phototransistors with different thicknesses of C$_{60}$ at $V_G = 0$ V, $V_{DS} = 250$ mV, and the illumination power of~5 mW/cm$^2$. (c inset) $IQE$ of different Gr/C$_{60}$ hybrid devices. Adapted with permission.[201] Copyright 2018 ACS. (d-h): LSPR based Si QD/Gr PD. (d) The structure of B-doped Si-QDs/Gr hybrid phototransistor; (e-f) Cross-sectional distribution of the $|E|^2$ at B-doped Si QDs and Gr; (g) Variation of $R$ with laser irradiance at different laser $\lambda$s at $V_G = 0$ V and $V_{DS} = 1$ V; (h) Spectral dependence of the $NEP$ and $D^*$ of the device. The UV-to-NIR and MIR measurements were conducted at RT and 77 K, respectively.

The variation of the $R$ with illumination irradiance of the phototransistor at $V_{DS} = 1$ and $V_G = 0$ V is illustrated in Fig.15g. The reduction in $R$ as irradiance increases is typical for PDs. $R$ varies between 0.22-44.9 A/W for MIR illumination at a low irradiance of 375 mW/cm², and $R$ varies between $1.2 \times 10^8$ and $2.2 \times 10^9$ A/W for UV-to-NIR illumination at a low irradiance of 0.2 µW/cm². A finite-difference time-domain (FDTD) simulation of the distribution of the square of electric field ($|E|^2$) at Si QDs and the underlying Gr was performed to understand the photoresponse of Si-QD/Gr phototransistors in the MIR region using $\lambda = 2.7$ and 3.6 µm (Fig.15e-f). The LSPR of Si QDs clearly introduces strong electromagnetic fields in their vicinity. When the illumination is at $\lambda = 3.0$ µm, the electromagnetic field beneath Si QDs is significantly enhanced. The increased electromagnetic fields would considerably enhance Gr's optical absorption, resulting in a strong photoresponse.

Wu et al. reported plasmon-enhanced PTE conversion in CVD Gr $p$-$n$ junction-based PDs integrated with Au NP plasmonic nanostructures,[208] which are intended to function as optical antennas, converting incident light into plasmonic oscillations and significantly increasing the electromagnetic field close to the Gr $p$-$n$ junction (GPNJ) area (Fig.16a). The increased generation of $e^-$-$h^+$ pairs in Gr triggered by the strong near-field light would result in increased $I_{phot}$ generation after the $e^-$-$h^+$ pairs are segregated due to the temperature gradient. To juxtapose the plasmon enhancement impact of $I_{phot}$ generation at CVD Gr $p$-$n$ junctions with that at Gr-metal junctions, Au NPs (average diameter of 115 nm) were dispersed (28 particles/µm²) onto the entire device, including the Gr $p$-$n$ junction area and electrodes. PTE conversion plasmonic enhancement factors in Gr $p$-$n$ junctions ranged from 2.5 to 4.0, while those in Gr-electrode junctions ranged from 1.8 to 3.2. Similarly, plasmon resonance increased the $I_{phot}$ responsivity of the Gr $p$-$n$ junction from 0.075 to 0.3 mA/W while increasing the photovoltage responsivity from 1 to 4 V/W. When the device was subjected to global illumination (632.8 nm laser, power, 0.2 mW, Fig.16c), the net $I_{phot}$ produced by three zones—the Gr $p$-$n$ junction and two Gr-metal junctions—was nonzero. A current shift of about 7 nA was observed (Fig.16d). Au NPs with avg. diameters of 21, 115, and 220 nm were employed to induce LSPR effect at 632.8 nm excitation, and it was discovered that 115 nm NPs excited with a 632.8 nm laser generated the maximum $I_{phot}$.

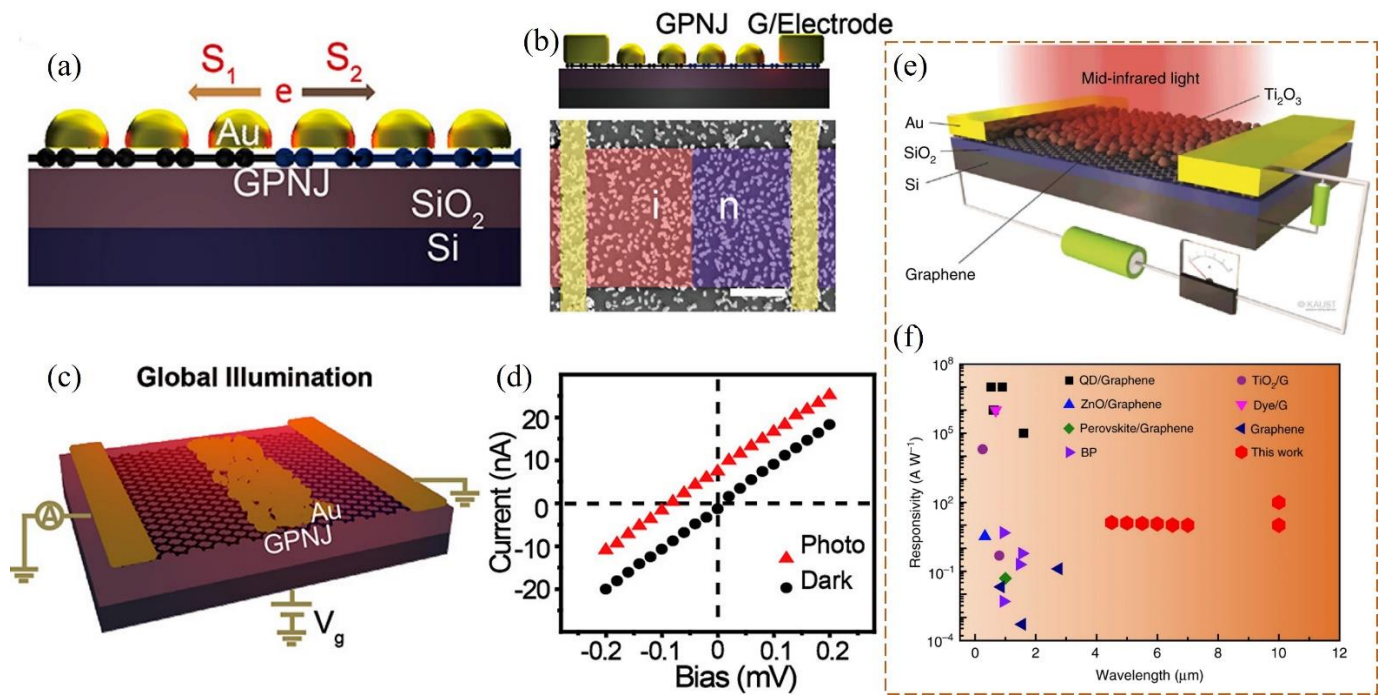

Fig.16: (a-d): CVD Gr *p–n* junction (GPNJ) with Au NPs. (a) PTE at the GPNJ with Au NPs. Hot electrons diffuse with different Seebeck coefficients, $S_1$ and $S_2$; (b) Top: side view of the schematics of the device geometry. Bottom: false color SEM image of a GPNJ PD fully covered with transferred Au NPs. Scale bar: 2 μm; (c) Schematic drawing of the detection of global light with the GPNJ PD combined with plasmonic nanostructures; (d) Current versus source–drain bias in the dark (black circle) and with defocused 633 nm laser. Adapted with permission.[208] Copyright 2013 ACS. (e-f): Gr/Ti$_2$O$_3$ PD. (e) A schematic phototransistor PD based on 0D-Ti$_2$O$_3$/Gr heterostructure; (f) Anology of the performance of the hybrid Gr/Ti$_2$O$_3$ PD with the state-of-the-art. Adapted with permission.[209] Copyright 2018, Xuechao Yu et al.

Fig.16e shows a MIR hybrid PD made by combining Gr with a small $E_g$ semiconducting Ti$_2$O$_3$ ($E_g$ = 0.09 eV) NP, demonstrating a peak $R$ of 300 A/W at RT over a broad $\lambda$ range of ~10 μm (LWIR).[209] The resulting $R$ was ~larger than that of any commercial MIR PD by a factor of two orders of magnitude. This work used Ti$_2$O$_3$ (a stronger light absorber) and Gr with fast carrier mobility to fabricate a high-performance PD. When the photogenerated carriers emerged in Ti$_2$O$_3$, the numerous holes moved to the Gr channel, trapping the electrons in Ti$_2$O$_3$. Further, the device displayed unchanged broadband photoresponse with high $R$ in the 4.5 to 10 μm range.

Cuprous oxide (Cu$_2$O), a naturally *p*-type material with a hole concentration of $10^{15}$/cm$^3$, has emerged as an effective absorbing layer for low-cost photovoltaic applications.[210] Its direct bandgap ranges from 2.0 to 2.2 eV.[211] Copper can be thermally oxidized to Cu$_2$O under favorable conditions, and copper is also a preferred substrate for Gr CVD growth.[212] Cu$_2$O QDs can thus be sandwiched between the Cu and the Gr, while oxygen penetrates through the Gr grain boundaries or defects by adjusting the oxidation duration and temperature.[213] Following metal deposition, a transfer-free Gr/Cu$_2$O QD PD structure can be obtained. In this direction, a new mechanism for carriers to multiply was proposed and tested in a Gr/Cu₂O QD system by measuring the quantum capacitance of Gr.[214] In a manner comparable to doping, a single electron or hole is trapped inside a QD when a single photon is absorbed by the QD, moving the fermi level up or down (Fig.*17*b-c). The volume of the QDs determines the measure of Fermi level tuning and given that Gr is in contact with Cu₂O QDs, the Fermi level in Gr will shift concurrently with the Fermi level alignment, raising the carrier concentration $n_g$ in Gr to $10^{11}$-$10^{12}/cm^3$, where the volume of the QDs determines the carrier density. The source of the internal current gain in this mechanism, which is identical to field-effect gating, is the photoactivated carrier concentration in Gr is regulated by light irradiation. The Gr/Cu₂O QD hybrid PD with the benefits of both transfer free Gr and Cu₂O QDs achieves a high $R$ of $10^{10}$ A/W and fW-scale light detectivity at RT. Furthermore, the device demonstrated outstanding flexibility (maintaining a $R$ of over $4 \times 10^6$ A/W). The $R$ rises as incident power decreases (Fig.*17*d). When the incident light power is lowered to 20.37 fW, the RT $R$ touches $12 \times 10^9$ A/W at 0.19 V with a congruent gain of $3.3 \times 10^{10}$ (Fig.*17*e). Different Cu₂O QDs will absorb the incident photons. However, the amplification effect (which generates extra charge carrier concentration in Gr) of a few QDs due to an electrochemical potential change of 0.3 eV is almost identical to that of a single QD, explaining why a lower incident power correlates with a larger observed $R$ (Fig.*17*d). When the illumination power increases, more and more Cu₂O QDs are photoexcited, but the number of extra carriers in Gr remains ~$10^{10}$, triggering a decrease in $R$.

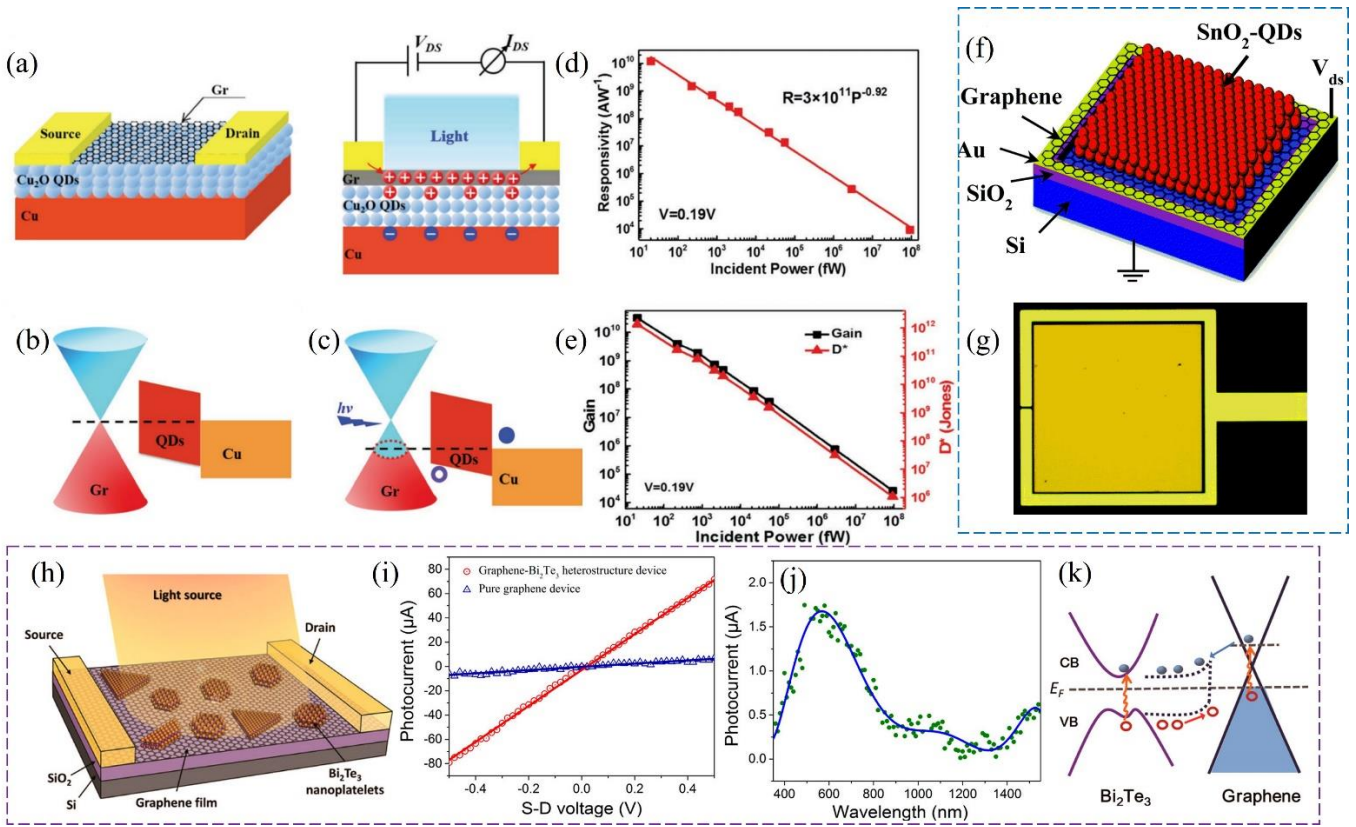

Fig.17: (a-e): Gr/Cu₂O QD hybrid PDs. Photoconductive results of the hybrid PD under 450 nm illumination. (a) Structure and operational mechanism of the PD. Energy band diagram of the

arXiv copy

Because of their narrow, tunable bandgap over the whole IR spectrum,[217] and excellent optical characteristics,[218] HgSe (mercury selenide) and (mercury telluride) HgTe CQDs have been investigated for IR detection. On integrated electrode topologies, photoconductive detectors based on these QDs have also been demonstrated utilizing spin-coating, spray-casting, or inkjet printing techniques.[217a, 218b, 219]

HgTe CQDs have demonstrated the maximum spectrum modulation for colloidal nanomaterials from MIR[219] to terahertz[220] and their films have shown a very high $\mu$ of >1 cm²/V-s.[221] There have been numerous reports demonstrating single-band short-wave imaging,[222] mid-wave thermal imaging,[223], and dual-band IR imaging.[224] The spectral detection range of Gr-based vertical junctions can therefore be considerably expanded by integrating Gr with HgTe CQDs. A spectral sensing range of 2 μm has been demonstrated in HgTe CQDs/MoS₂ hybrid PDs.[225] The sensing range has been further extended to 9 μ using HgSe CQDs/Gr nanocomposites.[226] Introducing an IR-transparent ionic glass LaF₃ (as a substrate to allow back-side illumination) gate structure to the Gr electrode in a 2D/0D $p$-$n$ heterojunction IR PD (having Gr and HgTe nanocrystal) enabled modulating the Fermi level of Gr between electron-doped and hole-doped states, permitting the design of a reconfigurable $p$-$n$ junction between Gr and HgTe nanocrystals.[227] It also diffused the $p$-$n$ doping all through nanocrystal thin film due to its ambipolar characteristic. The consequential built-in electric field, generated within the QD thin film, enabled very low $I_{dark}$ and allowed the entire QD thin film to interact with light with a fast photoresponse time in addition to preventing charge recombination, thereby improving phototransport. When illuminated with 1550 nm, the IR PD operated as a self-powered PV device with a high $D^*$ of $10^9$ Jones and an increased $SNR$ and temporal response.

Because of its novel properties in the NP dimension, CdS (chalcogenide binary semiconductor compound) is an interesting material for the synthesis of CdS QDs ($E_g$ of ~1.7–2.4 eV), which find applications in energy harvesting and optoelectronic devices: solar cells,[228] [229] LEDs,[230] photocatalysis,[231] and gas sensors.[232] CdS QDs were deposited on a Gr film via chemical bath deposition (CBD)[233] and a flexible hybrid Gr/CdS QDs PD was fabricated, which measured a photoelectric response of ~40 A/W (at 450 nm) under weak illumination.[233]

Tin dioxide (SnO₂) QDs have particular advantages in optoelectronics due to their environmental friendliness, Sn abundance, and great stability.[234] Furthermore, the Gr/B-Si Schottky PD can be built based on mature Si semiconductor technology. SnO₂ QDs have a good compatibility with established Si technology. Fabricating high performing $SnO_2$ QDs/Gr/Si Schottky PDs by combining $SnO_2$ QDs with Gr is therefore desirable. $SnO_2$ QD film (23 nm thick) coupled with CVD grown Gr to form a Schottky junction with bulk silicon (B-Si) have been proved to significantly enhance the parameters of the $SnO_2$ QDs/Gr/B-Si Schottky PD (Fig.17f-g).[215] This hybrid PD has showed sensitivity to a broad range of $\lambda$s in the UV-Vis-NIR range, with a high $R$ of 967 A/W (~4 orders greater than standard B-Si Schottky PDs) and $D^*$ of $1.8 \times 10^{13}$ Jones, as well as fast $\tau_r$ and $\tau_d$ of 0.1 and 0.23 ms, respectively. The hybrid structure's high light absorption and the elevated built-in potential of the Gr/B-Si Schottky junction that facilitated effective segregation of photoexcited $e^-$-$h^+$ pairs, were the reasons for the improved performance.

WS$_2$ CQDs were used as the photo absorbing medium and Gr as the conducting channel in a phototransistor.[235] At a low bias voltage ($V_{ds}$ = 1 V) and an optical power as low as 0.8 mW/cm$^2$, a $R$ of ~$3.1 \times 10^2$ A/W, fairly high $D^*$ of ~$8.9 \times 10^8$ Jones, and a low $NEP$ of ~$9.7 \times 10^{-11}$ W/Hz$^{0.5}$ were obtained, can be further altered by varying the gate bias.

The optical gap in TMDCs and PbS QDs (1-2 eV) significantly limits their application in NIR range photodetection, especially in telecom bands at 1550 nm (0.8 eV).[236] Other prominent class of 2D topological insulators: $Bi_2Te_3$,[237] $Bi_2Se_3$, $Sb_2Te_3$ (with $E_g$s ~0.15-0.3 eV in the MIR range), would be a better alternative.[238] $Bi_2Te_3$ is a typical layered material via vdW interactions. Its hexagonal unit cell is similar to that of Gr. Because of the relatively small lattice misfit, $Bi_2Te_3$ nanocrystals can grow epitaxially on a Gr template[239] over a large area.[239b] This results in a novel vdW heterostructure with an atomically gapless interface that allows photoexcited carriers to travel and separate effectively. In that direction, Gr/$Bi_2Te_3$ heterostructure was built.[216] Taking into consideration of the aforementioned advantages, better $Bi_2Te$-light interaction and reduced recombination of photoexcited carriers, this Gr/$Bi_2Te_3$ device displayed a maximum $R$ (35 A/W at $\lambda$ = 532 nm) and a high photoconductive gain (maximum ~83), much better in comparison to single layer Gr-based devices. Furthermore, the small $E_g$ of $Bi_2Te_3$ (0.15–0.3 eV) augmented the device's detecting $\lambda$ range to NIR (980 nm) and telecom band (1550 nm).

Fig.17i depicts the $I_{phot}$ outcomes of two distinct devices based on single layer Gr and a Gr/$Bi_2Te_3$ heterostructure in absence of gate bias. The $I_{phot}$ of the Gr/$Bi_2Te_3$ heterostructure device is ~ten times greater than the $I_{phot}$ of the Gr device. The heterostructured device shows photoresponse over a wide range of $\lambda$s: UV, Vis, and NIR as shown in **Fig.17**j. A measurable current of ~0.5 μA was observed at 1550 nm (telecom C band). To ascertain the enhanced $I_{phot}$ an energy diagram was proposed (**Fig.17**k). Because the fermi level of as-grown Gr samples is below the Dirac point and the fermi level of as-grown $Bi_2Te_3$ nanocrystals (n-type) is above the bottom of the bulk conduction band, a schottky-like junction occurs at the Gr/$Bi_2Te_3$ interface as electrons from $Bi_2Te_3$ are infused into Gr. Therefore, the direction of band bending comes from $Bi_2Te_3$ to Gr (akin to a semiconductor/metal contact). When illuminated by light, driven by the built-in electrical field, photoexcited electrons in Gr can flow into the conduction band of $Bi_2Te_3$, whereas photogenerated holes remain in Gr's valence band. $Bi_2Te_3$, meanwhile, can absorb light and generate $e^-$-$h^+$ pairs. Because of the energy barrier, electrons will be trapped inside $Bi_2Te_3$, whereas holes can be injected into Gr's valence band. As a result, photogenerated carrier recombination can be controlled, and the number of majority carriers (holes) in Gr can be increased, resulting in an increased $I_{phot}$.

Bulk tin di-selenide ($SnSe_2$) is a semiconductor; belongs to the TMDC family and has an indirect $E_g$ of 1.0 eV.[240] When the dimension of $SnSe_2$ particle size diminished, the onset of quantum confinement effects causes the discretization of energy levels. The $E_g$ in $SnSe_2$ QDs therefore is larger.[241] Due to their configurable $E_g$ and excellent quantum efficiency, $SnSe_2$ QDs can be utilized in fast and highly responsive phototransistors. An UV detector, monolayer Gr/$SnSe_2$ QDs had demonstrated photoresponsivity of ~$7.5 \times 10^3$ A/W and a $\tau_{res}$ of ~0.31 s (corresponding to UV light, $\lambda$ = 405 nm).[242] Sonication and a laser ablation techniques were followed to prepare $SnSe_2$ QDs. The $n$-$n$ heterostructures formed by single layer Gr and $SnSe_2$ QDs increased light absorption and photocarrier migration, significantly enhancing the device photoresponsivity.

Because of its direct $E_g$ of ~3.4 eV, strong electron saturation velocity, high exciton binding energy (higher than bulk value of 60 meV), environmentally benign ZnO NPs are very desirable UV sensitizers.[243] Additionally, when the ZnO NP radius equals the Debye length (19 nm), the maximum surface/volume ratio gives a photoconductive gain, allowing for orders of magnitude better UV photoresponsivity and $I_{phot}/I_{dark}$ ratios than the theoretical limit in bulk ZnO optoelectronics.[244] However, the $\mu$ of ZnO single crystals is typically ~100-200 $cm^2$/V-s, which can result in dismal photoresponsivity or photoconductive gain. This constraint could be circumvented in Gr/ZnO hybrids through the transfer of photoelectrons that were originated in ZnO-NP sensitizers to Gr, which possesses a charge transport channel that is far more efficient. The atomic layer deposition (ALD) technique was used to create a ZnO QDs/Gr composite.[245] The PD comprising ZnO QDs/Gr composite and polymers displayed very high UV light sensitivity, with a maximum $R$ of 247 A/W at $\lambda$ = 325 nm. The high active surface/volume ratio of the ZnO QDs/Gr composite is thought to be responsible for this device's high photoresponsivity. The Gr's excellent carrier transport and collection efficiency resulted in a fast transient response with a $\tau_{res}$ of ~tenths of milliseconds.[245]

Liu et al.[246] demonstrated Gr/ZnO-NP nanohybrid UV detectors, exhibiting a high $R$ of ~$5.0 \times 10^3$ A/W (at 5 V bias, $\lambda$ = 340 nm, and 3 mm W/$cm^2$ illumination power density), which was several orders of magnitude greater than the ZnO-nanostructure and Gr/ZnO nanohybrid PDs that had been reported previously.[245, 247]

Due to the semi-metallic nature of Gr, PDs based on Gr/QDs typically endure high dark conductivity of the Gr channel. A hybrid TMDC-QD phototransistor was constructed and tested as a solution to this issue.[248] It consisted of $p$-type CPbS QDs and a few very thin layers of $n$-type $MoS_2$ ($\geq$2 layers). This hybrid detector benefited from strong and size-tunable light absorption in the QDs, high $\mu$ in the $MoS_2$ layer, and operation in the channel's depletion mode, resulting in a low $I_{dark}$. The NIR $R$ of the detector was $10^3$ A/W, with $\tau_d$ of ~0.3-0.4 s.

2D TMD molybdenum disulfide ($MoS_2$) is among the most researched and commonly employed materials. It possesses a high absorption coefficient, excellent stability, a high in-plane carrier mobility (200 cm$^2$/V-s), and a layer-dependent $E_g$.[249] Because of its unique optical features not seen in bulk materials, monolayer MoS$_2$ is widely used in the fabrication of PDs.[250] Because of quantum confinement, monolayer MoS$_2$ has a direct $E_g$ of 1.8 eV, but bulk MoS$_2$ has an indirect $E_g$ of 1.3 eV.[17, 251] For a few layers of MoS$_2$, the absorption coefficient is typically ~ 0.6 × 10$^6$/cm, corresponding to an absorption length = 1/ absorption coefficient = 17 nm.[252]

At RT, the $D^*$ of 10$^{12}$ Jones was measured experimentally in a MoS$_2$/HgTe QDs hybrid PD at $\lambda = 2$ μm, two orders of magnitude greater than previous results from HgTe-based PDs (Fig.18a-b).[225] In this design, a TiO$_2$ buffer layer between the MoS$_2$ channel and the HgTe QD layer shielded the MoS$_2$ channel while also acting as an *n*-type electron acceptor to generate an effective *p-n* junction with the HgTe QDs at the interface. This made it easier for charges to move to the MoS$_2$ channel, which led to gate-modulated low $I_{dark}$ current. The reported detector outperforms currently used extended-InGaAs, InAs, or HgCdTe detector technologies that require thermoelectric cooling during RT operation. In this HgTe QD based PDs, for the first time, gain of ~10$^6$ and $R$ of ~10$^6$ A/W were achieved. A carrier recirculation mechanism is thought to be responsible for the high $R$. Fig.18a depicts the device architecture. The device fabrication process began with the exfoliation of a few layers of MoS$_2$ on the SiO$_2$/Si (285 nm) substrate. Using photolithography and electron beam deposition, the Ti/Au source-drain electrodes were then created. Subsequently, a thin TiO$_2$ buffer layer was deposited using ALD. Finally, layer-by-layer HgTe CQDs were spin coated on top of the devices, yielding a thickness of 80-90 nm. The energy band of HgTe QDs aligns with TiO$_2$/MoS$_2$ to form a type-II band alignment (Fig.18b). As light strikes the HgTe QDs, the built-in field transports the photogenerated electrons into the MoS$_2$ channel via the thin TiO$_2$ layer, while the holes are trapped in the QD layer. The transferred electrons are then made to drift to the drain by applying the source-drain bias. The carrier transit time in QDs is orders of magnitude smaller than the trapping lifetime due to the high $\mu$ of MoS$_2$ (18 cm$^2$/V-s) and the small channel length (5 μm). As a result of a single $e^-$-$h^+$ photogeneration, numerous electrons are recirculated in the MoS$_2$ channel, resulting in a photoconductive gain of 10$^6$. The trap-state passivation process can be used to further improve the carrier lifetime, which is supposedly dictated by trap-assisted (indirect) recombination via the defects or traps in the TiO$_2$ or HgTe layer.

Bi$_2$O$_2$Se is a new 2D semiconductor with a high $\mu$(450 cm$^2$/V-s) and good stability at RT.[253] Even though progress has been made in obtaining a very high $R$ (>10$^5$ A/W) and a $D^*$ of ~10$^{15}$ Jones in Bi$_2$O$_2$Se for Vis spectrum, on account of its intrinsic indirect optical bandgap and weak light absorption,[254] the photoresponse performance of Bi$_2$O$_2$Se near its band gap diminishes rapidly to < 0.1 A/W at $\lambda = 1550$ nm in PDs. However, when integrated with CQDs, 2D Bi$_2$O$_2$Se semiconductor channel, might be able to provide a high-speed charge transport pathway. In 2D Bi$_2$O$_2$Se/PbSe CQDs heterostructure, field-effect tuned broadband IR photodetection was explored.[255] On the basis of the type II interface constructed with PbSe CQDs, which allowed charge injection from PbSe CQDs to Bi$_2$O$_2$Se upon illumination, an IR response beyond 2 μ was observed. Due to the fast charge transfer dynamics at the interface, the photoresponse speed (< 4 ms) in hybrid PD was greatly enhanced from the intrinsic sluggish response in either Bi$_2$O$_2$Se or PbSe CQDs, which was restricted by minority carrier trapping. It was discovered that the degree of photogating could be easily controlled by modifying the interfacial band alignment via external field-effect modulation, imparting an extremely sensitive $R$ of $>10^3$ A/W at 2 μm. UV photoelectron spectra (UPS) measurements to gauge the energy band alignment when PbSe CQDs and $Bi_2O_2Se$ were come into contact, yielded the work functions of PbSe and $Bi_2O_2Se$ as 4.07 and 4.37 eV respectively. The valence band maximum (VBM) of bare PbSe and $Bi_2O_2Se$, on the other hand, were determined to be 0.28 and 0.58 eV, respectively, below their Fermi level ($E_f$). A type II band diagram was proposed based on the $E_g$s of PbSe (0.6 eV) and $Bi_2O_2Se$ (0.8 eV), as well as the estimated work function and valence band position (Fig.18d). Due of the lower work function of PbSe CQD, electrons will be transported to $Bi_2O_2Se$ when they come into contact. As shown in Fig.18d, the charge transfer formed interfacial dipoles, which hampered further electron injection into $Bi_2O_2Se$. Despite the large conduction band offset of 0.4 eV, electron injection into $Bi_2O_2Se$ was still beneficial during excitation. The photoresponse spectra of the compared devices are shown in Fig.18e. Fig.18f compares the power dependence of device $R$ to that of bare PbSe CQDs under 1456 and 2000 nm IR illumination.

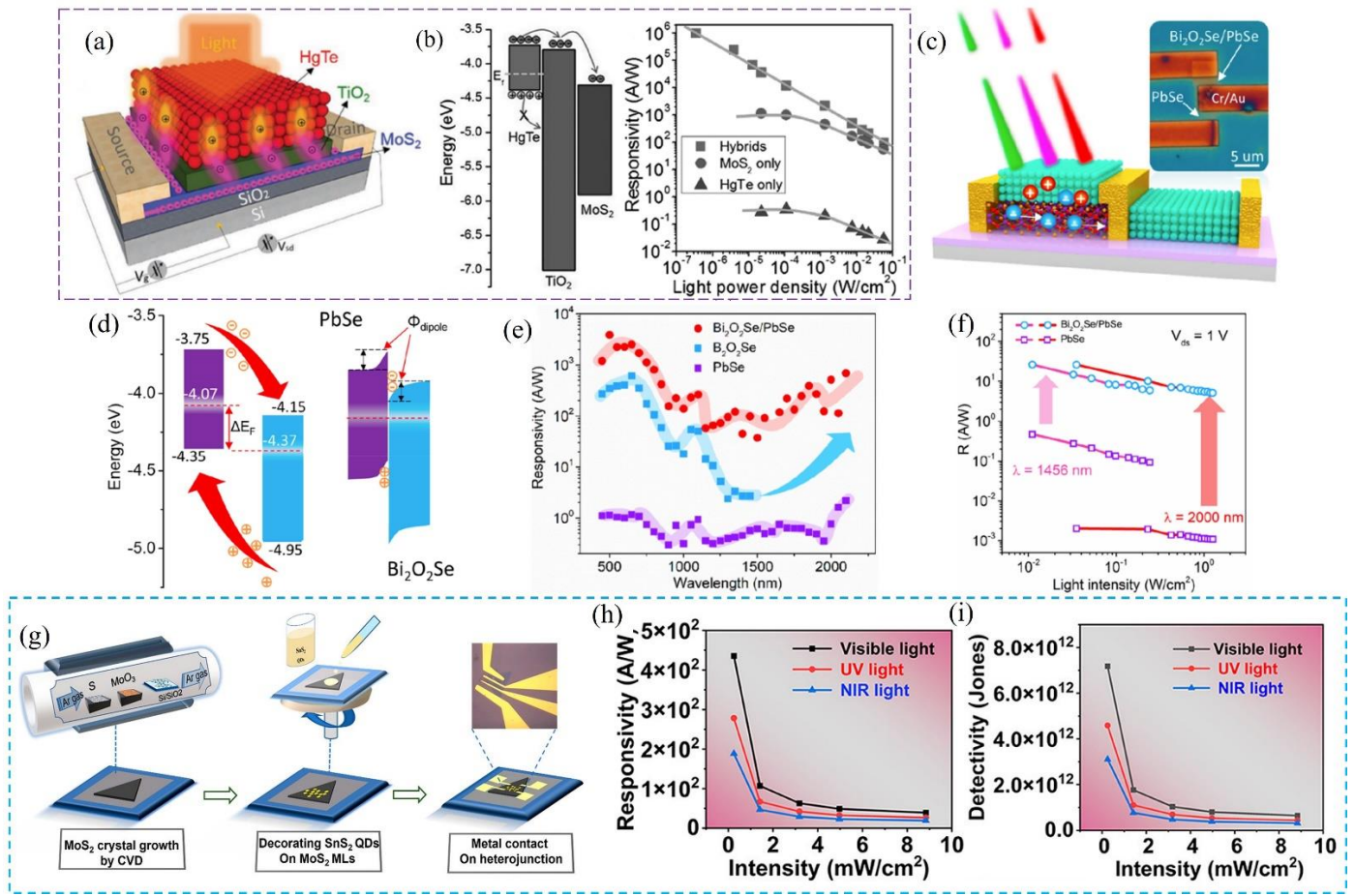

Fig.18: (a-b): $MoS_2$/HgTe QDs (a) Schematics of hybrids device with light illumination and photogenerated carrier separation and transport in the hybrid PDs; (b) Band alignment of $MoS_2$, $TiO_2$, and HgTe QDs, type-II band alignment at the $HgTe/TiO_2$ interface; b right image: $R$ vs. light power density of $MoS_2$, HgTe and hybrid PDs, the $\lambda$ is 635 nm, the applied bias and gate are 1 and 0 V, respectively.[225]; (c-f): PbSe CQD/$Bi_2O_2Se$ hybrid PD. (c) Illustration of the hybrid PD under light excitation. Inset: an optical microscope image of the as-fabricated device;

A Vis-NIR light detector based on a monolayer MoS$_2$ nanosheet (~3.6 nm)/Si heterostructure demonstrated $10^{13}$ Jones detectivity and an ultrafast $\tau_r$ (~3 μs).[257] A Vis light PD based on a vdW heterostructure of SnSe$_2$/MoS$_2$ demonstrated a high $R$ of $9.1 \times 10^3$ A/W.[258] Nevertheless, because MoS$_2$ absorbs more light in the Vis and NIR ranges than in the UV range, the spectral range of most reported MoS$_2$-based PDs is confined to the Vis and NIR regions. One method for producing a broad (UV-NIR) spectral response is the fabrication of a heterojunction with a high $E_g$ and UV-sensitive QDs, which not only extend the spectral region detection to UV but also assist in injecting photogenerated carriers to boost the device's photoresponse.[259] In that regard, $n$-type SnS$_2$-QDs have been used to construct high $R$ Monolayer MoS$_2$/SnS$_2$-QD-based mixed-dimensional PD (Fig.$18$g).[256] In this study, MoS$_2$ crystals were grown on SiO$_2$/Si using a CVD method, and SnS$_2$-QDs were synthesized using a low-temperature solution processing method. By spin coating SnS$_2$-QDs on MoS$_2$ monolayers, a 0D/2D SnS$_2$-QDs/MoS$_2$ hybrid device was fabricated. The thicknesses of the SnS$_2$-QDs and MoS$_2$ crystals are reported to be 1.5 nm and 0.7 nm (monolayer) respectively. Due to the difference in energy level, the photogenerated carriers in SnS$_2$-QDs were moved to MoS$_2$. That is, a single layer of MoS$_2$ served as a transport conduit for photogenerated carriers delivered from SnS$_2$-QDs. Several heterojunctions were formed by the unconnectedly distributed SnS$_2$-QDs on monolayer MoS$_2$, resulting in an electric field that aided in the effective isolation and transportation of photoexcited charge carriers. Furthermore, the NIR photoresponse in the constructed device was caused by electron doping of the MoS$_2$, which was caused by charge transfer from the SnS$_2$-QD to the conduction band of the MoS$_2$ monolayer. Under UV, Vis, and NIR light illumination, the photoresponsivities of the fabricated SnS$_2$-QDs/MoS$_2$ device were 278, 435, and 189, A/W, respectively (Fig.$18$h). Because of the higher absorption of monolayer MoS$_2$ in the Vis range, the device demonstrated the maximum photoresponsivity in the Vis regime. Further, the device showed $D^*$ of $4.589 \times 10^{12}$, $7.19 \times 10^{12}$, and $3.11 \times 10^{12}$ Jones, respectively, under UV, Vis, and NIR lights (Fig.$18$i). The $\tau_r$ of the fabricated device was 100 ms.

### 4.2.2. High performance PDs with 2D/2D Heterostructures

Aside from employing single films as IR absorbers, much research has focused on stacks of 2D materials, where the adjacent layers interact with one another via vdW forces. This is owing to the fact that the amalgamation of various 2D crystals with multifarious electrical, optical, and mechanical and physical properties can result in robust optoelectronic performance and device functionality. 2D van der Waals heterostructures are a viable venue for utilizing several physical phenomena in a wide range of new spintronic device applications.[260]Because of the vdW bonding characteristic of these 2D materials, direct heterostructure fabrication is possible without taking into account elements that are strictly necessary in conventional semiconductor heterostructures: epitaxial conditions, surface properties, and lattice mismatches.[261] Further, a plethora of novel phenomena arising from variations in electron distribution and transport processes at the interface of the vdW junction can be examined in heterostructures.[262] It has been reported that a photogenerated exciton that could be directly transformed into interlayer excitons in a $WS_2/WSe_2$ heterostructure.[263] The lack of undesirable surface states and dangling chemical bonds in 2D materials is vital for IR optoelectronics, as a clean surface minimizes carrier generation-recombination processes.[264],[265]

2D/2D heterostructure based PDs are generally photodiodes and phototransistors. PDs that operate based on the photovoltaic mechanism are frequently referred to as photodiodes. Photodiodes often consist of *p-n* junctions formed by two semiconductors (typically doped) with distinct major carrier types. Schottky barrier (or junction) contacts can be formed by metals and semiconductors having different work functions. A considerable built-in electric field generally exists at the interface of a *p-n* junction (or Schottky barrier), which plays a critical role in the photogenerated carrier dissociation process. Due to the presence of a built-in electric field, a photodiode has a rectifying property, and it can operate in two modes. The photodiode is generally reversely biased in PV mode (the external electric field direction is the same as the built-in electric field), which increases carrier extraction and results in a lower $I_{dark}$, improved $D^*$, and a faster response speed. In the photoconductive mode, the photodiode is forward biased (the external electric field is directed in the opposite direction as the built-in one), resulting in a high gain. Since the presence of built-in potential, a photodiode may also operate without any external voltage bias, which has been dubbed as "self-driven mode." In this mode, the device is more energy-efficient. This property is especially appealing for photodiodes based on 2D layered materials because of their ultrathin nature, which provides the device with a complete depletion zone.

The current flowing through a Schottky junction can be ascertained via the following relations:[257, 266]

$$I = I_s \left[ \exp\left( \frac{eV}{nkT} \right) - 1 \right]$$ 
<div align="right">Equation 16</div>

$$I_s = A_1 A^* T^2 \exp\left( -\frac{e\varphi_b}{kT} \right)$$
<div align="right">Equation 17</div>

Where $I_S$ is the reverse saturation current; $V$ is the bias voltage; $e$ is the electronic charge; $k$ is the Boltzmann constant; $n$ is the ideality factor; $T$ is the absolute temperature; $A_1$ is the area of the device; $A^*$ is the effective Richardson constant (112 A/ $cm^2$-$K^2$ for *n*-type Si), and $\phi_b$ is the Schottky barrier height. Provided *I–V* characteristic curves are obtained, $\varphi_b$ can be obtained by fitting the *I-V* data with the aforesaid equations. In general, the Schottky diode operates in two modes: (a) band to band excitation mode (where $hv > E_g$) and (b) internal photoemission mode ($\phi_{SBH} < hv < E_g$, where SBH is Schottky barrier height).

A phototransistor is usually a three-terminal device that can be controlled. It works by having a photoconductive and photogating effect, with the carrier channel sandwiched between a top gate dielectric layer and a bottom gate dielectric layer and the source and drain electrodes bridging it horizontally. The nature of channel materials have a significant impact on phototransistor performance: crystal quality, crystal symmetry, doping level, and thickness. Aside from the specifics of device assembly, applying externally appropriate gate voltage and bias voltage to a PD with a transistor structure can significantly alter performance. Additionally, it is modifiable by combining it with different chemical species to create hybrid structures that effectively absorb light while allowing photogenerated $e^-/h^+$ to transfer into the conducting channel. Such devices that use hybrid channel materials are referred to as hybrid phototransistors.

Gr provides an appealing framework for broadband Vis/IR photodetection. Previous attempts to improve its sensitivity by including light-absorbing colloids, waveguides, or antenna design resulted in positive results, albeit at the expense of reduced photon detection bandwidth. However, in 2D-2D heterostructures, as one of the components, Gr has exhibited some useful features. A Schottky diode (Schottky junction with metal/semiconductor contact) generally functions in two different modes: internal photoemission mode ($\Phi_{SBH} < h\nu < E_g$) and band-to-band excitation mode (when $h\nu > E_g$), demonstrating that, depending on the formed Schottky barrier height, Gr-based Schottky diodes can be suitable IR detectors.[267] The Gr/2D semiconductor heterostructures are attractive for IR photodetection due to interlayer interactions that manifests in combined features of high carrier mobility and broadband $I_{phot}$ production. Furthermore, the Gr/TMDC heterostructures shows ultrahigh photoconductive gain due to the photogating effect generated by the trapping electrons (or holes) in localized states of $n$-type TMDC ($p$-type TMDC) as in the case of hybrid Gr/MoTe$_2$ PDs, where detection of broadband NIR wavelength with photoresponsivity of 970.8 A/W, photoconductive gain of $4.69 \times 10^8$, and detectivity of $1.55 \times 10^{11}$ Jones at 1064 nm were demonstrated.[268] An investigation of the magnetic proximity effect between monolayer Gr and 2D ferromagnet CrBr$_3$ showed a significant Zeeman splitting field in Gr and allowed for the design of 2D spin logic and memory devices.[269] Gr is often used as a rapid carrier-drifting layer in 2D heterostructures, where out-of-plane carrier drift is substantially more competent than in 2D/3D and 2D/0D heterostructures.[270] Because of the significant absorption of WSe$_2$, a Gr/WSe$_2$/Gr heterostructure has been proved to be an ultrafast IR PD with a picosecond $\tau_{res}$ and excellent detection efficiency. [271]

There is a dearth of reports on the fabrication of 2D materials and their vertical/lateral heterostructures based on CVD technique. Surface adsorbates from the environment are likely to have a major effect on device performance if not encased.[272] Environmental factors are influencing photoresponse in 2D single layer PDs; their effect on heterostructure devices, however, is probably going to be far more complicated.[273] The literature data for the widely used detector design based on Gr/MoS$_2$ heterostructures show a perplexing divergence: varying signs of $I_{phot}$ from positive to negative.[274] Furthermore, a large range of $R$s spanning many orders of magnitude (from $R = 3.2 \times 10^{-2}$-5 $\times 10^8$ A/W) have been found.[275] According to a recent study, adsorbates can modify the magnitude and sign of the $I_{phot}$.[276] In this study, the effect of surface adsorbates and charge carrier extraction on the photoresponse of Gr/MoS$_2$ heterostructure devices was examined.[276] As a substrate, crystalline sapphire was employed, and Gr and MoS$_2$ were grown via CVD and metal-organic chemical vapor deposition (MOCVD), respectively. Under both ambient and vacuum conditions, lasers with specific $\lambda$s ranging from UV-Vis spectral range were used to generate charge carriers. It was found that the type of doping in Gr can be governed by the number of species that adhere to the surface of MoS$_2$, leading to a $I_{phot}$ that changes in size and sign depending on the adsorbate density.

An in-plane lateral Gr/MoS$_2$ heterostructure was fabricated using one-step CVD technique.[275b] With the edge contact of Gr and MoS$_2$ in the channel, a Schottky junction was formed. It exhibited good rectification properties, ON/OFF ratio of ~10$^6$. As a PD, it exhibited a $D^*$ of ~$1.4 \times 10^{14}$ Jones and a $R$ of $1.1 \times 10^5$ A/W, attributed to strong absorption, adequate photoexcited carrier separation, and rapid charge migration at the Schottky junction.

A self-powered Gr/MoS$_2$ hybrid phototransistor with a $\tau_{res}$(0.13 ms), high ON-OFF ratio (1428), and $R$ (3.0 A/W) under zero bias was demonstrated.[277] A few-layer MoS$_2$ (obtained by mechanical exfoliation of bulk MoS$_2$) and high-mobility Gr (prepared via CVD) were used as photo sensitizers to absorb light and expressways for carrier movement, respectively. An asymmetric metal contact (T$_i$/Au (10 nm/50 nm) and Pd/Au (10 nm/50 nm)) design was adopted to sever the mirror symmetry of the internal (built-in) electric-field profile (which is the case with conventional phototransistor channels), thus providing a strong internal electric-field. A photovoltage manifests at the two asymmetric Gr/electrode interfaces with a photodiode mechanism when illuminated with 632.8 nm light. This photovoltage worked as a bias voltage to drive the large number of photocarriers that were infused from the MoS$_2$ layer through the Gr channel. Because of the photovoltage, the phototransistor could operate as a self-powered device with a very low $I_{dark}$ (1 nA).

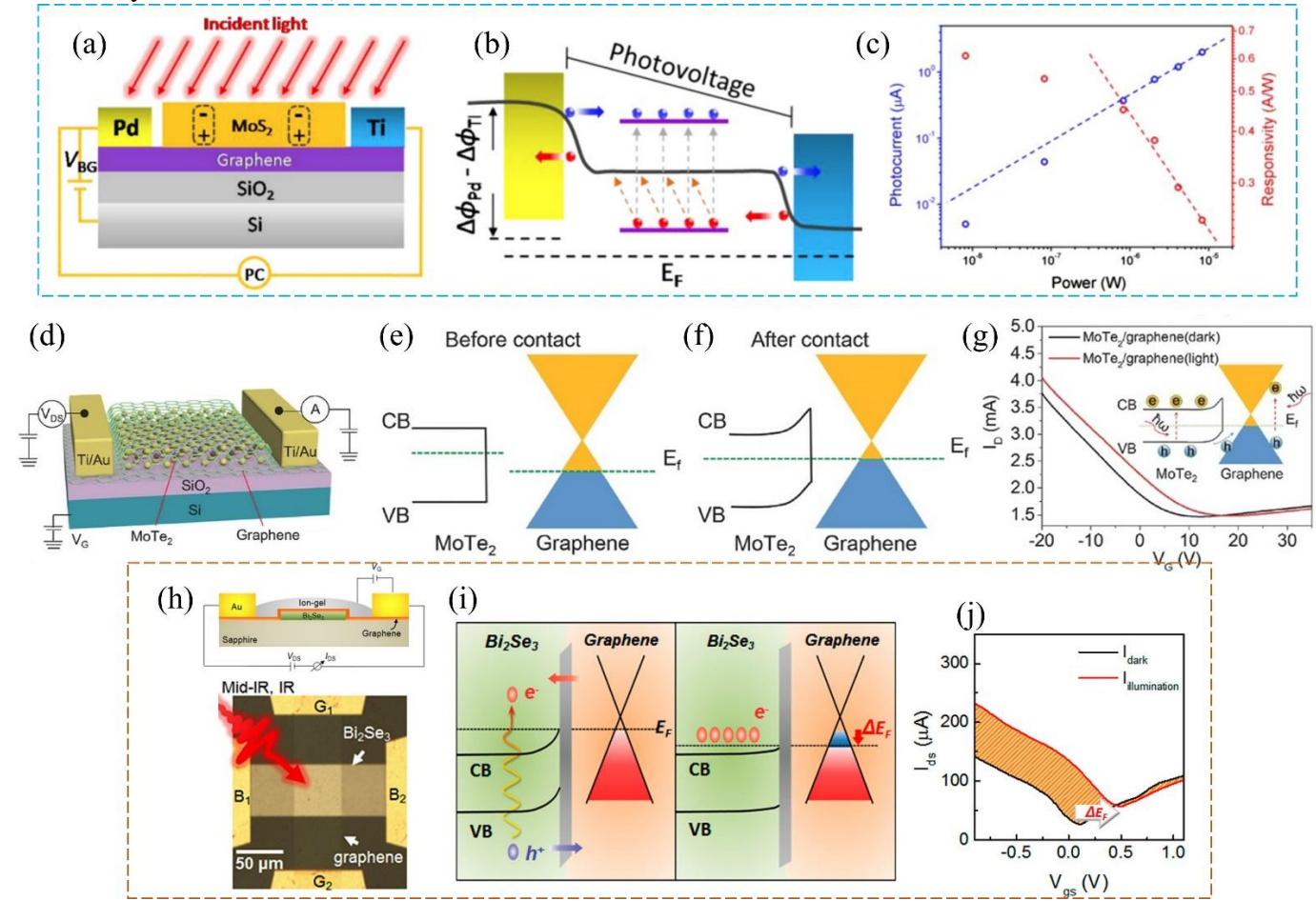

Fig.19: <u>(a-c)</u>: Gr/MoS$_2$ hybrid phototransistor. (a) charge generation and separation process; (b) energy diagram of the Gr/MoS$_2$ hybrid phototransistor with asymmetry metal contact under 632.8 nm irradiation. $\Delta\phi_{Pd}$ and $\Delta\phi_{Ti}$ are the difference between the Dirac point energy and the Fermi level in pd- and Ti-doped Gr, respectively; (c) $I_{phot}$ and the $R$ of the phototransistor vs. optical power.[277] <u>(d-g)</u>: MoTe$_2$/Gr phototransistor. (d) Schematic diagram of the MoTe$_2$/Gr phototransistor. V$_{DS}$: source–drain voltage; $V_G$: gate voltage. Monolayer Gr is on top of MoTe$_2$. The Mo and Te atoms are shown in yellow and gray balls, respectively; (e) Band diagram of MoTe$_2$ and Gr before they are in contact. CB: conduction band, VB: valence band. (f) Band diagram of MoTe$_2$ and Gr after contact; (g) Transfer curves with ($\lambda$ = 980 nm) and without light illumination. Inset: Illustration of the photoexcited holes transfer process from MoTe$_2$ to Gr under light incident.[268] <u>(h-j)</u>: Gr/Bi$_2$Se$_3$ heterostructure. (h) *Top*: Experimental set up of $I$–$V_{gs}$ measurement circuit using the ion-gel gate bias voltage ($V_{gs}$) and the Gr bias voltage. The potential of the bottom Bi$_2$Se$_3$ layer was allowed to float, while the current of the top Gr transistor was measured with $V_{ds}$ = 0.5 V. *Bottom*: Image of the Gr/Bi$_2$Se$_3$ heterostructure PD having MBE-grown Bi$_2$Se$_3$ and wet-transferred CVD Gr; (i) Band diagram and photoexcited hot-carrier transfer mechanism under light illumination. The red and blue dots are e$^-$ and h respectively. Vertical arrows: photoexcitation, and lateral arrows represent tunneling of hot electron (red arrow) and hole (blue arrow); (j) $I$–$V_{gs}$ characteristics of the measured Gr PD with Photocarrier separation occurs at the Gr/electrode interface due to the strong internal electrical field induced by asymmetric band bending, resulting in an $I_{\text{phot}}$ that generates a photovoltage in the phototransistor channel. As depicted in Fig.19a-b, this photovoltage is anticipated to supplant the external bias voltage required for a normal phototransistor to function with the photoconduction mechanism. The $I_{\text{phot}}$ exhibits a non-linear decrease with the optical power due to the onset of poor excitation efficiency especially in the lower optical power range (Fig.19c). The $R$ also increases nonlinearly as the optical power diminishes (Fig.19c), in contrast to the standard phototransistor, which has a linear relationship.

A MoTe$_2$/Gr heterostructure PD for an efficient NIR light detection[268] has shown a high $R$ of 970.82 A/W (at 1064 nm), a broad photodetection range (Vis-1064 nm), an exceedingly high photoconductive gain of $47 \times 10^7$, and $D^*$ of $15.5 \times 10^{10}$ cm Hz$^{1/2}$/W due to the efficacious photogating effect induced by electrons trapped in the localized states of MoTe$_2$.[268] Furthermore, despite hundreds of cycles of bending (1.2% tensile strain), devices on a flexible substrate retain a good photodetection ability, with a high $R$ of 60 A/W at 1064 nm at $V_{\text{DS}} = 1$ V. As seen in **Fig.19**f, because the work functions of these two materials differ, the energy bands after contact slope downward toward MoTe$_2$ and form an internal electric field that directs from MoTe$_2$ to Gr. When illuminated, the VCNP (voltage of charge neutral point) rises to a greater voltage (15 V) than when exposed to darkness (10 V) as shown in **Fig.19**g, implying the presence of a photogating effect. Due to the built-in electric field, the $e^-$-$h^+$ pairs generated by light photons in MoTe$_2$ (more photoactive than Gr), will be separated at the MoTe$_2$/Gr interface (**Fig.19**g inset). Specifically, holes are transported to the Gr layer, and electrons are trapped in the localized states in MoTe$_2$. This mechanism will result in a photogating effect, in which electrons trapped in MoTe$_2$ function as a local gate, attracting additional holes in Gr, lowering resistance, and increasing $I_{\text{phot}}$ under incident light. Because of Gr's ultrahigh $\mu$, holes injected into it can be extracted quickly, resulting in a significant increase in $I_{\text{phot}}$.

A monolayer Gr/few-layer-InSe heterostructure PD was developed by depositing anodic-bonded Gr on top of mechanically exfoliated few-layer-InSe..[279] The covering Gr layer protected the ultra-thin InSe layers. This detector, under illumination condition (λ = 532 nm with a power density of 12.6 W/m$^2$), showed $R = 9.4 \times 10^2$ A/W and $EQE = 21.8 \times 10^4$ % with $V_{\text{ds}} = -50$ mV and $V_{\text{g}} = 0$ V. These values were four orders of magnitude greater than the values observed in the few-layer InSe device ($R = 0.101$ A/W and $EQE = 23.5$% with $V_{\text{ds}} = 5$ V and $V_{\text{g}} = 0$ V). Because InSe layer is only a few nanometers thick, photoelectrons can potentially be extracted vertically into the Gr layer from the entire volume.[279]

2D Bi$_2$Se$_3$ belongs to a group of compounds called Tetradymites ($M_2X_3$, where $M$ = Bi or Sb, and $X$ = chalcogens (S, Se, or Te). Theoretical and experimental studies indicate that Bi$_2$Se$_3$ has a direct $E_{\text{g}}$ of ~0.35 eV and known to exhibit thermoelectric properties.[280] A MIR PD based on a Gr/Bi$_2$Se$_3$ heterostructure demonstrated RT performance with broadband detectio n and high $R$ (1.97 and 8.18 A/W at MIR and NIR, respectively).[278] Utilizing broadband absorption of MIR (3.5 μm) and IR photons in a small-bandgap $Bi_2Se_3$ (~300 meV)[281] topological insulator, and an effective hot carrier dissociation and effective photogating across the $Bi_2Se_3$/Gr interface, improvements in spectral range and $R$ were achieved simultaneously. Further, $D*$, and response speed under MIR illumination were $1.7 \times 10^9$ cm·Hz$^{1/2}$/W ($V_{gs} = 0.08$ V at RT) and 4 ms, respectively. Photoexcitation occurs in both Gr and $Bi_2Se_3$, and hot carriers are efficiently split into Gr and $Bi_2Se_3$ via tunneling. Because of the asymmetric potential, holes easily tunnel from $Bi_2Se_3$ to Gr (Fig.19i, left panel). As a result, negative charges build up in the lower $Bi_2Se_3$ (Fig.19i, right panel), causing $\Delta n$ (photoinduced adjustment of the carrier concentration) on the top Gr transistor. The Gr channel has a high $\mu \sim 4000$ cm$^2$/V-s in CVD-Gr, making its conductance particularly sensitive to variable carrier density. The transfer curve under light illumination ($I_{illum}$) with a laser power of 50 μW changes considerably toward positive $V_{gs}$ when compared to the dark transfer curve $I_{dark}$, as illustrated in Fig.19j, and a horizontal shift of the Dirac point ($V_{shift}$) of 0.68 V, implying the onset of photogating effect. This means, the trapped electrons in the $Bi_2Se_3$ layer generate an increase in hole concentration in Gr since they operate as a gate bias. Photogenerated holes tunnel from $Bi_2Se_3$ to Gr, accumulating electrons at the conduction band's bottom (Fig.19i) inducing photogating effect (Fig.19j) clearly culminating in a positive shift of the neutral point.

In a Si waveguide integrated design, a Gr/h-BN heterostructure (single layer Gr encapsulated by h-BN) offered strong $R$ as well as ultrafast pulse response at 1550 nm.[120] Similarly, strong $D^*$ up to $2.6 \times 10^{13}$ Jones (greater than that of $MoS_2$-based devices by a factor of 100–1000) was experimentally measured in a $MoS_2$/h-BN (7-15 nm)/Gr heterostructure due to the huge electron barrier at the Gr/h-BN junction, while photoexcited carriers were successfully tunneled via the low barrier at the $MoS_2$/h-BN junction. $R$ of 180 A/W and high $I_{phot}/I_{dark}$ of $10^5$ were concomitantly achieved.[282]

Si is a conventional photoactive material in deep-UV (DUV, 190-350 nm) PDs. Due to its electronic band structure, Si-based detectors absorb both UV and Vis light. For DUV sensing, spectral selectivity, especially blindness to visible light, is very important. Optical filters are often used to limit the transparent band to the DUV spectral region. These filters, however, diminish the intensity of the incident DUV light as well as the responsiveness of detectors. High spectral selectivity and $R$ are thus required at DUV wavelengths.

The indirect band gap of multilayer hexagonal boron nitride (h-BN) is ~5.95 eV, which corresponds to a DUV wavelength of 208 nm. The h-BN has been commonly used as an encapsulation layer in 2D heterostructures to tune the dielectric and reflex index environments, as well as to improve light management.[283] CVD of h-BN was used to produce photosensitizers and realized a photogating effect in the DUV spectral region.[284] The device with h-BN photosensitizers produced distinct photoresponses and an $I_{phot}$ of $57.1 \pm 7.7$ nA, whereas the device without h-BN photosensitizers produced an $I_{phot}$ of $10.6 \pm 2.1$ nA. The photogating effect was ascribed to the gate voltage modulation of the Gr FETs. The photogating effect manifesting in the device is shown in Fig.20c. Photoelectrons or holes generated in h-BN close to the proximity of the Gr channel induce modulation of the back-gate voltage, $V_{bg}$. Fig.20b shows the $I_{phot}$ and $R$ dependence of the devices with and without h-BN photosensitizers vs. $V_{sd}$. The device's (with h-BN photosensitizers) photoresponse grew linearly with source-drain voltage ($V_{sd}$), and the DUV $R$ reached a maximum of ~$19.6 \times 10^3$ A/W, correlating to a quantum efficiency of $9.3 \times 10^6$ %. The minimum $NEP$ reached $3.09 \times 10^{-13}$ W/Hz$^{1/2}$ and the $\tau_{res}$ was 200-300 ms. The device's $\tau_{res}$ was influenced by direct carrier transfer from the h-BN photosensitizer to the Gr channel, as well as the buildup of photocarriers in the h-BN photosensitizers.[284]

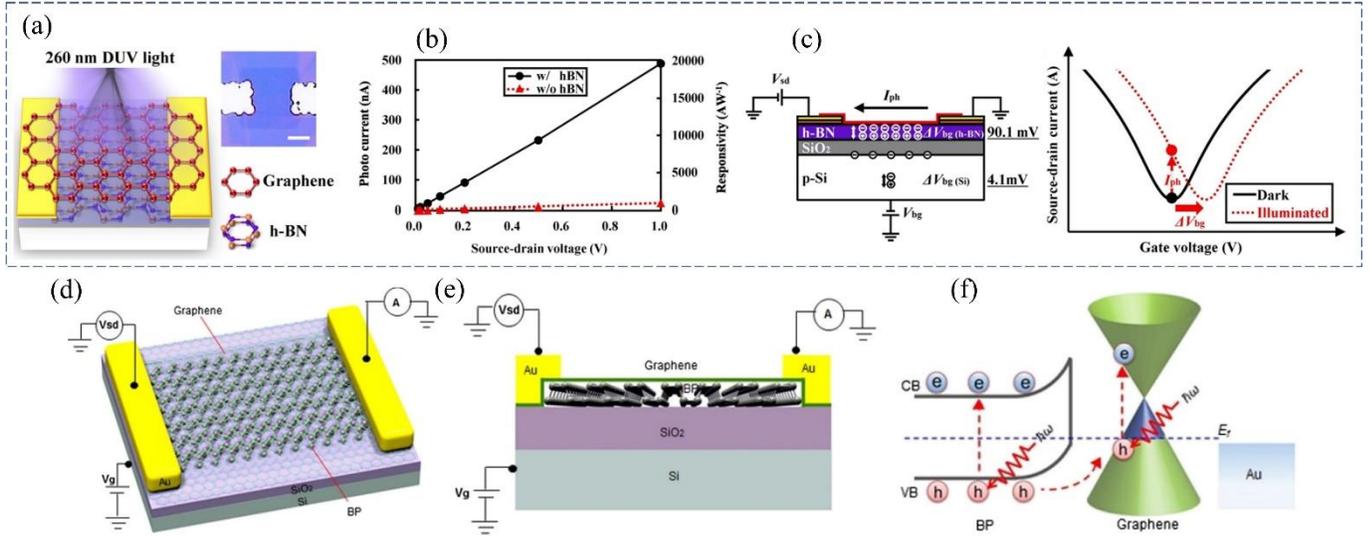

Fig.20: (a-c): Gr/h-BN heterostructured DUV PD. (a) Gr/h-BN heterostructure. CVD-grown multilayer h-BN and monolayer CVD Gr were transferred onto a SiO$_2$/Si substrate. The Cr/Au source/drain electrodes are connected to the Gr channel. Inset: Image of the Gr/h-BN heterostructure. Scale bar: 5 µm; (b) $I_{phot}$ and $R$ of devices with and without h-BN layers; (c) Mechanism of $I_{phot}$ generation via the photogating effect under DUV light irradiation. Photocarrier behavior and $\Delta V_{bg}$ in the device.[284] (d-f): Gr/BP heterostructured PD. (d) 3D schematics of the Gr/BP heterostructured PD, in which monolayer Gr is coated on top of the BP flake to enhance the $I_{phot}$ response at 1550 nm. ($V_{sd}$: source–drain voltage, $I_{sd}$: source–drain current, and $V_g$: gate voltage); (e) Schematic of the Gr/BP heterostructure-based PD and the circuit used to perform the two-terminal electrical measurements; (f) Band diagram of the Gr/BP heterostructure and the photoexcited hot carrier transport process under illumination related to zero gate voltage ($V_g$); dotted line indicate the Fermi levels ($E_F$). Adapted with permission.[285] Copyright 2017 ACS.

The layer-by-layer etching process, which exposes BP to oxygen and water molecules in the air, is responsible for the rapid degradation of the BP, which has already been acknowledged as a significant drawback associated with the usage of BP.[286] Because of the substantial Schottky barriers between BP and metal electrodes, efficient photocarrier movement is severely hampered, resulting in relatively low $R$ in the bulk of reported BP-based PDs.[11b] A Gr/BP (35 nm thick) vertical heterostructured PD measured a very high $R$ ($3.3 \times 10^3$ A/W), a photoconductive gain (~$11.3 \times 10^8$), and a $\tau_r$ of 4 ms at NIR $\lambda$ of 1550 nm.[285] The encapsulation of monolayer Gr in this device architecture (**Fig.**20d-e) effectively prevented deleterious effects on BP, ensuring the device's long-term stability under ambient conditions. Upon light illumination, the photoexcited $e^-$-$h^+$ pairs created in BP are isolated and injected into Gr. Since the energy barrier at Gr/Au is much lower than that at BP/Au, direct contact between Gr and the source-drain electrodes provided a high-speed pathway for dynamic charge carrier separation (**Fig.20**f).

Using the rapidly evolving CVD technology, large-area, fine quality Gr sheets have recently been fabricated.[287] Direct deposition of 2D materials on Gr films has emerged as a viable method for producing high uniformity and controllable quality Gr-based heterostructures.[288] However, large-scale fabrication of such high-performance heterostructures is still a challenge. Lead iodide ($PbI_2$), a semiconductor ($E_g$ of ~2.4 eV) has shown promising optical properties. $PbI_2$ (a layered hexagonal crystal structure, P-3m1 symmetry group, $a = b = 4.56$, $c = 6.98$, $\alpha = \beta = 90°$, $\gamma = 120°$), in which $PbI_2$ layers are linked by relatively weak vdW interactions. By vapor deposition of $PbI_2$ on a Gr/polyethylene terephthalate (PET) substrate (with < 200 °C), a submeter-sized, vertically stacked heterojunction of $PbI_2$/Gr was created on a flexible PET film. This film was then utilized to create a bendable Gr/$PbI_2$/Gr sandwiched PD with high sensitivity (45 A/W-cm$^2$), response time ($\tau_r$ = 35 ms, $\tau_d$ = 20 ms), and high-resolution imaging capabilities (1 μ).[289]

Van der Waals heterostructures (vdWHs) are made up of 2D materials with a dangling-bond-free surface that weakly bonds to neighboring materials via vdW interactions.The weak vdW interactions, on the other hand, cause an intrinsic vdW gap within the vdW interfaces, resulting in an additional tunneling barrier. In vdWH-based phototransistor devices, the interface tunnelling barrier checks the injection of photogenerated carriers (**Fig.21**a). These intrinsic vacancies at the edge serve a purpose; the dangling bond of nonlayered chalcogenide semiconductors bridges the vdW gap via orbital hybridization between the vacancies and the dangling bond. A general approach to bridging the vdW gap has been proposed.[290] A large number of chalcogen (S, Se, and Te) vacancies can be created via annealing 2D TMDCs. The vacancies of active chalcogens (S, Se, and Te) are linked to the dangling bond of nonlayered chalcogenide semiconductors, giving rise in a covalently bonded interface. This approach significantly improves overall performance when compared to a phototransistor based on a mixed-dimensional heterostructure.

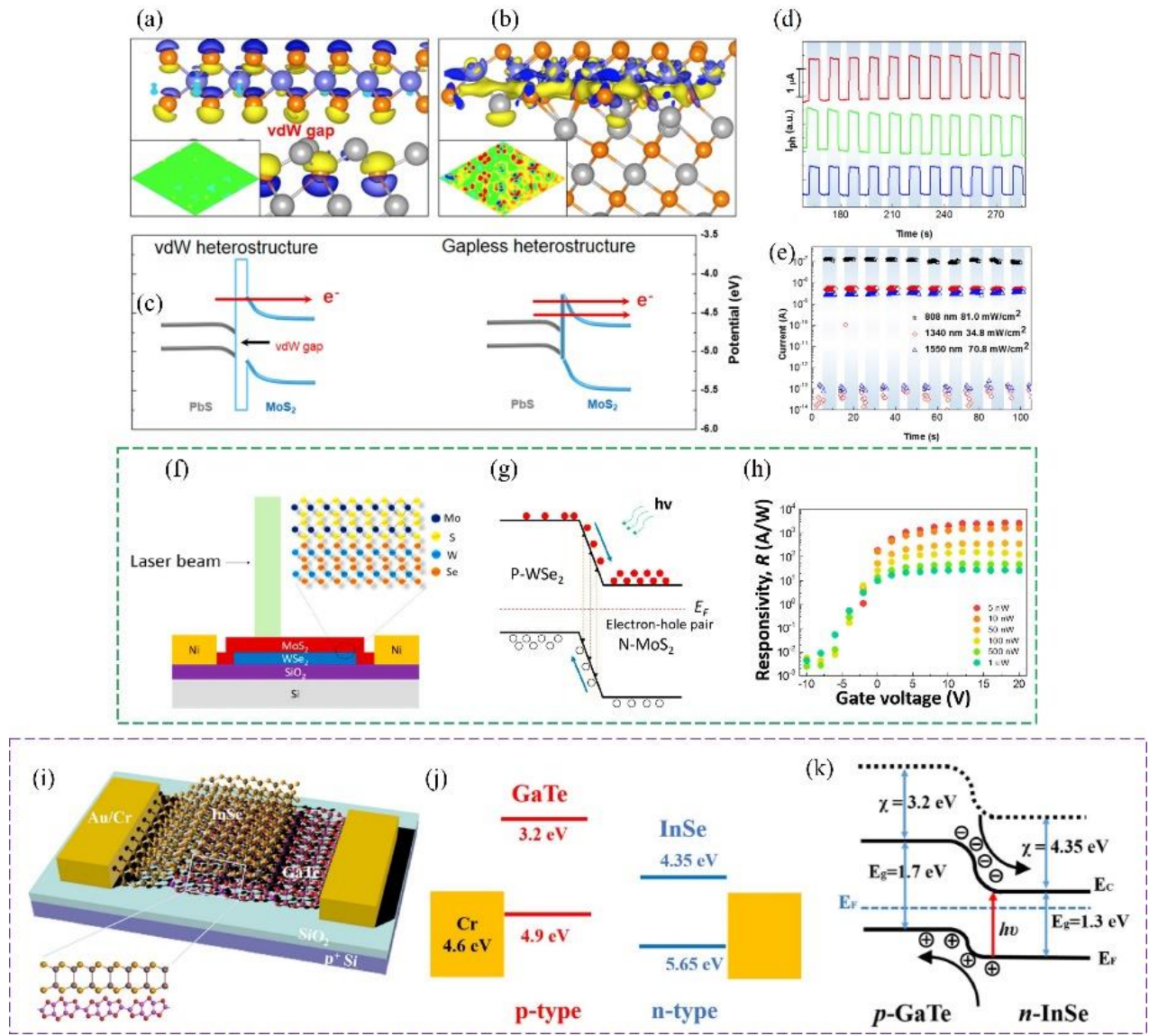

Fig.21: <u>(a-e)</u>: MoS$_2$/PbS heterostructures. Interface of MoS$_2$/PbS heterostructures schematic and time-resolved response. The charge density difference of MoS$_2$/PbS heterostructures with vdW interface (a) and gapless interface (b); (c) The band diagram of MoS$_2$/PbS heterostructures. Left: Typical vdW heterostructures with vdW gap. Right: Gapless heterostructures without vdW gap; (d) Time evolution of $I_{phot}$ with $V_{gs}$ = 80 V at different $\lambda$s (red line 808 nm with 12.5 mW/cm$^2$, blue line 1550 nm with 4.3 mW/cm$^2$, and green line 1940 nm with 70.8 mW/ cm$^2$). (e) Time-resolved response at depletion region ($V_{gs}$ = -20 V) with different $\lambda$ (808, 1340, and 1550 nm). Adapted with permission.[291] Copyright 2019 ACS. <u>(f-h)</u>: WSe$_2$/MoS$_2$ vdW PN heterojunction. (f) Schematic diagram of the device under laser irradiation at the junction; (g) Energy band diagram of the WSe$_2$/MoS$_2$ vdW PN heterojunction with photogenerated carriers under focused laser beam irradiation; (h) Photoresponsivity of the transistor based on the WSe$_2$/MoS$_2$ vdW heterojunction, indicating a high $R$ of 2700 A/W at an illumination power of 5 nW and gate voltage of 20 V, for $V_{DS}$ = 5 V, $V_{GS}$ = −10 to 20 V, and $\lambda$ of 532 nm. Adapted with permission.[292] Copyright 2020 ACS. <u>(i-k)</u>: GaTe/InSe heterojuncton and its characterization. (i) Illustration of the GaTe/InSe vdW $p$–$n$ junction. (j) band diagrams for GaTe, InSe, and Cr

A gapless heterostructure has been created in which S dangling bonds of nonlayered PbS (104.8 nm thick) are connected to the bonding sites of $MoS_2$ (with factitious S vacancies via annealing saturated $MoS_2$) via strong orbital hybridization.[291] The S dangling bonds of nonlayered PbS are connected to the bonding sites via chemical bonds. The strong interface coupling allows gapless heterostructures to attain $R$ and photogain ($G$) values in excess of $10^5$, and detectivity ($D^*$) >$10^{14}$ Jones. Besides, the gapless heterostructure displayed periodic transient photoresponse with fast $\tau_r$ and $\tau_d$s of ~47 and 49 μs, respectively. In this study, charge density difference (CDD) technique was employed to discriminate the net charge between vdW interface (SPb−SMoS, **Fig.21**a) and gapless interface (PbS−MoS, **Fig.21**b). The charge redistribution was found to be nonlocal in the CDD projected on the (0001) plane of the gapless heterostructure interface (**Fig.21**b inset), which suggests that there are active electronic states in the interface. For a vdW interface, however, charge redistribution occurred exclusively on the surfaces of $MoS_2$ and PbS separately. Furthermore, the projected CDD (Fig.21a inset) showed that no chemical bonds were formed at the vdW interface. Despite the presence of numerous dangling Pb atom bonds on the surface, the S atoms of $MoS_2$ do not form bonds with the Pb atoms. The band alignment model of $PbS/MoS_2$ heterostructure is shown in **Fig.21**c. Vis-a-vis vdW heterostructures, the vdW gap of a gapless heterostructure disappears. As a result, the absence of additional tunnel barriers ushers in improved charge injection efficiency. The periodic transient photoresponse was very stable and reliable at $\lambda$s of 808, 1550, and 1940 nm (**Fig.21**d). Further, at $\lambda$s of 808, 1340, and 1550 nm (and a gate voltage of 20 V), the heterostructure exhibited very high photon-triggered ON/OFF ratios of $1.6 \times 10^6$, $1.8 \times 10^5$, and $4.5 \times 10^4$ respectively (**Fig.21**e).

A PD based on the vdW heterostructure of $WSe_2$ and $MoS_2$ was fabricated.[292] The $MoS_2$ channel served as a phototransistor channel, while the $WSe_2/MoS_2$ p-n junction in the out-of-plane configuration operated as a charge transfer layer. This culminated in a photoconductive gain of $10^6$. The phototransistor exhibited a photoresponsivity of ~2700 A/W, a $D^*$ of $50 \times 10^{10}$ Jones, and a $\tau_{res}$ of 17 ms. A schematic depiction of a device configuration with back-gate control is shown in Fig.21f. In this study, 2D $WSe_2$ was employed to create a vertical built-in electric field for seggregating photogenerated $e^-$-$h^+$ pairs. Fig.21g illustrates the energy band profile between p- $WSe_2$ and n-$MoS_2$ when illuminated by a laser beam. Laser irradiation generated $e^-$-$h^+$ pairs at the p-n junction interface; electrons moved to the $MoS_2$ channel, while holes were moved to $WSe_2$ with the internal electric field driving. Fig.21h depicts the PD's $R$ as the gate voltage varies from -10 to 20 V under varied laser intensities. Therefore, when the gate voltage scans from - to + under a source-drain voltage of 5 volts, $D^*$ and $\tau_{res}$ can be significantly improved.[292]

SnS$_2$ is an IVA-VIA group compound with strong chemical stability, economically viable, and environmental friendliness, rendering it a suitable photodetection material.[294] It has a appropriate $E_g$ of ~2.2 eV, a high optical absorption coefficient, and thus strong photoconductive properties to UV and Vis light.[295] An integration of Gr with very high $\mu$ and SnS$_2$ with a large yield of photocarriers can make a vertical Gr/SnS$_2$ vdW heterostructure PD with very high performance. Taking this fact as a cue, a Gr/SnS$_2$ PD[296] was fabricated (SiO$_2$/Si substrate) that showed an ultrahigh $R$ of $6.35 \times 10^5$ A/W (at 365 nm light illumination), which is substantially higher than that of single Gr or SnS$_2$. What's more, the PD demonstrated a high $EQE$ of $2.15 \times 10^8$ % and a $D^*$ of $3.68 \times 10^{11}$.

InSe, GaSe, and GaTe, which are indium and Ga-based MMCs, have also shown a great deal of potential for use in high-performance electrical and optoelectronic applications.[297] Gallium telluride (GaTe) is a prominent III-VI layered semiconducting material with a monoclinic crystal structure (space group, C2h3) and a large number of individual layers linked together by weak vdW interactions. In both bulk and single layer, GaTe shows a direct $E_g$ of ~1.7 eV at RT, thus making it an excellent optoelectronic material in its multilayer state.[298] In InSe, each layer contains a honeycomb lattice that is made up of four covalently connected Se-In-In-Se atomic planes. At ambient and liquid-helium temperatures, $\mu$s approach $10^3$ cm$^2$/V-s and $10^4$ cm$^2$/V-s, respectively. The $E_g$ rises by more than 0.5 eV as thickness decreases from bulk to bilayer InSe.[299] Furthermore, the field effect-induced 2$D$ electron gas (2DEG) at the surface of multilayered (20 nm) InSe crystals demonstrated low-temperature mobilities of up to 2000 cm$^2$/V-s.[300]

A vertical GaTe/InSe vdW $p$-$n$ heterojunction demonstrated gate-tunable current-rectifying property with a rectification factor of 1000.[293] It also had a high power conversion efficiency and rapid photodetection under zero bias. When excited by a 405 nm laser, the zero-biased PD had a high $EQE$ of 4.2% and a high $R$ of 13.8 mA/W. A $\tau_{res}$ of 20 µs was achieved because of the stable and swift carrier transmission through the built-in electric field in the depletion region. Charge separation in the vertical 2D $p$-$n$ junction was fast and efficient, allowing for zero $I_{dark}$, high speed, and little power utilization.[293] Both InSe and GaTe are III-VI materials. However, their crystal structures are distinct (**Fig.21**i inset). Based on their band structure, the ideal GaTe/InSe heterojunction is expected to have a type II band alignment (**Fig.**21j. As shown in the energy-band model (Fig.21k), for the Fermi levels to pin together, the energy bands of the GaTe bend downward while the energy bands of the InSe bend upward, thus forming a type II heterojunction creating a path for an unimpeded charge transfer process due to the absence of energy barriers. Endowed with an built-in field, during light irradiation, excited minority carriers adjacent to the depletion area in both GaTe and InSe can easily flow into each other, allowing for efficient $I_{phot}$ generation under zero-bias operation. The fast dynamic response time for rising and falling signals (20 µs) illustrates a wide bandwidth capable of tracking transient optical signals.

A GaTe (4.5 nm)/MoS$_2$ (3.5 nm) $p$-$n$ heterojunction transistor with self-driven photoelectric properties demonstrated forward biased rectifying operation with ambipolar transport that included electron and hole carriers.[301] Photoexcited $e^-$-$h^+$ pairs were easily segregated due to the type II band alignment and the presence of a built-in potential generated at the GaTe/MoS$_2$ interface under illumination ($\lambda$ = 633 nm laser), effectively producing self-driven $I_{phot}$ in less than 10 ms. The considerable built-in potential at the abrupt interfaces of fundamentally different material systems with distinct crystal symmetries nonetheless permits effective charge transfer mechanisms at the interface. The heterojunction's photoresponse and $EQE$ were 1.36 A/W and 266%, respectively and ON–OFF ratio was ~340.[301]

Similarly, An ultrathin and tunable $p$-GaTe (14.1 nm)/$n$-MoS$_2$ (5.5 nm) vdW heterostructure has shown a rectification ratio, $EQE$, and photoresponsivity of $4 \times 10^5$, 61.68%, and 21.83 A/W, respectively.[302] Surpassing the $D^*$s of commercial Si and InGaAs PDs,[8] the $D^*$ of this structure reached $8.4 \times 10^{13}$ Jones.

GaSe ($p$-type) has an indirect $E_g$ of 2.05 eV at the Brillouin zone's center, just 25 meV above the conduction band minimum.[303] Due to the fact that the gap between the direct and indirect band edges is so small, electrons can be transferred with little energy between these two minima. The photoresponse of GaSe PDs is reported to extend into the UV region.[304] The WS$_2$ is a typical TMDC with high $\mu$ and $n$-type conductivity.[305] When bulk WS$_2$ is exfoliated into a monolayer, it transforms from an indirect $E_g$ (1.4 eV) semiconductor to a direct $E_g$ (2.1 eV) material.[306] A very thin layer of WS$_2$ has a very strong ability to absorb light.[307] A high photoresponsivity UV-Vis PD based on $p$-GaSe and $n$-WS$_2$ can therefore be envisaged to deliver better results. A $p$-GaSe/$n$-WS$_2$ heterojunction with top and bottom Gr electrodes has yielded a photoresponsivity of ~149 A/W at 410 nm with a $\tau_{res}$ of ~37 $\mu$s.[180] A robust photoresponse at zero bias was furthered by the heterojunction's built-in potential.

A gate-tunable ultrathin $p$–$n$ diode built on a $p$-type BP (11 nm)/$n$-type monolayer MoS$_2$ (0.9 nm) vdW $p$–$n$ heterojunction has shown a highest photodetection $R$ of 418 mA/W at the $\lambda$ of 633 nm and photovoltaic energy conversion with an $EQE$ of 0.3%.[308] In this case, $R$ reached 1.27 mA/W and 11 mA/W at forward and reverse bias, respectively. At -2 V reverse bias, the maximum $R$ was 100 times greater than that of a pristine BP phototransistor.[11b]

A heterojunction PD for Vis - NIR detection range was constructed by vdW assembly between a few-layer BP (22 nm thick) and a few-layer MoS$_2$ (12 nm thick).[309] This heterojunction with electrical characteristics that are electrically tunable by a gate voltage achieved a broad range of current-rectifying behavior with a forward-to-reverse bias current ratio surpassing $10^3$. The $R$ of this PD was ~22.2 A/W (at $\lambda$ = 532 nm) and 153.3 mA/W (at $\lambda$ = 1.55 $\mu$m) with a $\tau_{res}$ of ~15 $\mu$s. At RT, the maximum $D^*$ was ~$3.1 \times 10^{11}$ Jones (at $\lambda$ = 532 nm) and $2.13 \times 10^9$ Jones at $\lambda$ = 1550 nm.

Rhenium disulfide (ReS$_2$) is a novel and fascinating 2D semiconductor with a direct $E_g$

of ~1.5 eV between monolayers and multilayers and an anisotropic crystal structure with a distorted 1T phase.[310] A ReS$_2$ PD fabricated using CVD delivered a high $R$ of 16.14 A/W.[311] The anisotropic crystal structure of a polarized light-sensitive PD based on pure ReS$_2$ has shown an even greater $R$ of $10^3$ A/W.[312]

The stacking process was employed to fabricate a photodevice based on a hybrid vdW heterostructure of ReS$_2$ (8 nm) and Gr (0.7 nm).[313] The device had a $R$ of ~7 × $10^5$ A/W, $D^*$ of $1.9 × 10^{13}$ Jones, and a fast $\tau_{res}$ < 30 ms. Due to the direct bandgap, high quantum efficiency, and intense light absorption of the multilayer ReS$_2$ and the high carrier mobility of Gr, very large $I_{phot}$s are formed in the heterostructure. Under the applied gate voltages, the ReS$_2$/Gr heterostructured device delivered a high $I_{phot}$ due to the photogating effect triggered by the Gr/ReS$_2$ junction. Mechanical exfoliation was used to produce Gr and ReS$_2$.

2D covalent organic frameworks (2D COFs) are new generation layered materials with high thermal stabilities, permanent porosity, a large surface area, diverse structure, and low mass densities.[314] These are a type of organic crystalline material that has predesigned π-electronic skeletons and highly ordered topological structures that are covalently built from 2D building blocks. These building blocks combine to form periodic planar networks, which then stack to form layered structures that are atomically precise in the vertical direction.[315] 2D COF-ordered electronic systems can transport charge carriers due to their high crystallinity and stacking alignment, implying their potential optoelectronic applications.[316] As a result, the optoelectronic features of 2D COFs hold promise for programming by selecting appropriate monomer combinations from a vast pool of 2D π-electronic building blocks. COF$_{ETBC–TAPT}$/Gr heterostructure demonstrated good capability, with a photoresponsivity of $3.2 × 10^7$ A/W at 473 nm and a $\tau_{res}$ of 1.14 ms.[317]

### 4.2.3. PDs with Pyroelectric and LSPR effect

As was already said, pyro-phototronic effects have been observed in ZnO, CdS with a wurtzite structure, perovskites, and some organic materials like crystalline rubrene. The noncentrally symmetric crystal structures in these materials are stated to be the origin of the pyroelectric effect.[176, 318] With a temporal change in the temperature across the material, a short-lived current generates because of the pyro-phototronic effect that aids in improving $\tau_{res}$ and photodetection capabilities of PDs.

Because of their propensity to support surface plasmons, metallic nanostructures are well known to be exceptional tools for manipulating light at the nanoscale. In $n$-ZnO nanowires (NWs), the transient rate of temperature change is a major limiting factor in its performance improvement. In Au NP coated $n$-ZnO nanowires (NWs)/$p$-Si heterojunction, the LSPR effect originated at the nearby junctions of Au NPs and ZnO NWs, substantially increased d$T$/d$t$ of the ZnO material, thus boosted the photoresponse properties of the NIR PD.[319] In comparison to the pristine ZnO-based PD, the $\tau_{res}$ of the Au-coated NIR PD was reduced from 113 to 50 μs at the rising edge and from ~200 to 70 μs at the falling edge. The optical detectivity and $R$ of the Au-coated ZnO-based PD were augmented by 266% and 212, respectively. **Fig.**22a depicts the schematic preparation procedure of Au NPs/$n$-ZnO NWs/$p$-Si PD. **Fig.22**b shows a section of a single period of the pyroelectric device's $I-t$ response curve. When the light source is turned ON and OFF, two photoresponse current peaks appear, which are triggered by the transient pyroelectric current generated by the device's pyroelectric effect. As stated previously, $I_{pyro} \propto$ d$T$/d$t$. The temperature of ZnO rises instantly (d$T$/d$t > 0$) due to laser-induced heating, and pyroelectric polarization is generated in ZnO.[320] Under a zero bias voltage, the built-in electric field separates charge carriers at the interface of ZnO and Si, and electrons and holes are independently transported to the ZnO and Si layers, forming photocurrent. The pyroelectric polarization electric field ($E_{py}$) generated by the pyroelectric effect is also aligned along the same direction as the built-in electric field from the ZnO side, resulting in an increase in the conduction band ($E_c$) and valence band ($E_v$) of the ZnO side at the $p$-$n$ junction contact. When the device is continuously lighted (i.e. d$T$/d$t = 0$), the pyroelectric current vanishes, the photovoltaic current takes over, and the curve declines and stays at a low value. In absence of the light, the brief temperature drop (d$T$/d$t < 0$) of the ZnO material causes the formation of $E_{py}$ in the opposite direction to built-in electric field. Carriers flow to the interface because of $E_{py}$ action, and some carriers can reverse and cross the potential barrier, resulting in a reverse current peak. Thus, the pyroelectric current gain was produced by thermal energy transfer assisted by the photothermal effect of a plasmonic Au nanostructure and the subsequent injection of hot electrons from the LSPR effect of Au NPs into the ZnO.

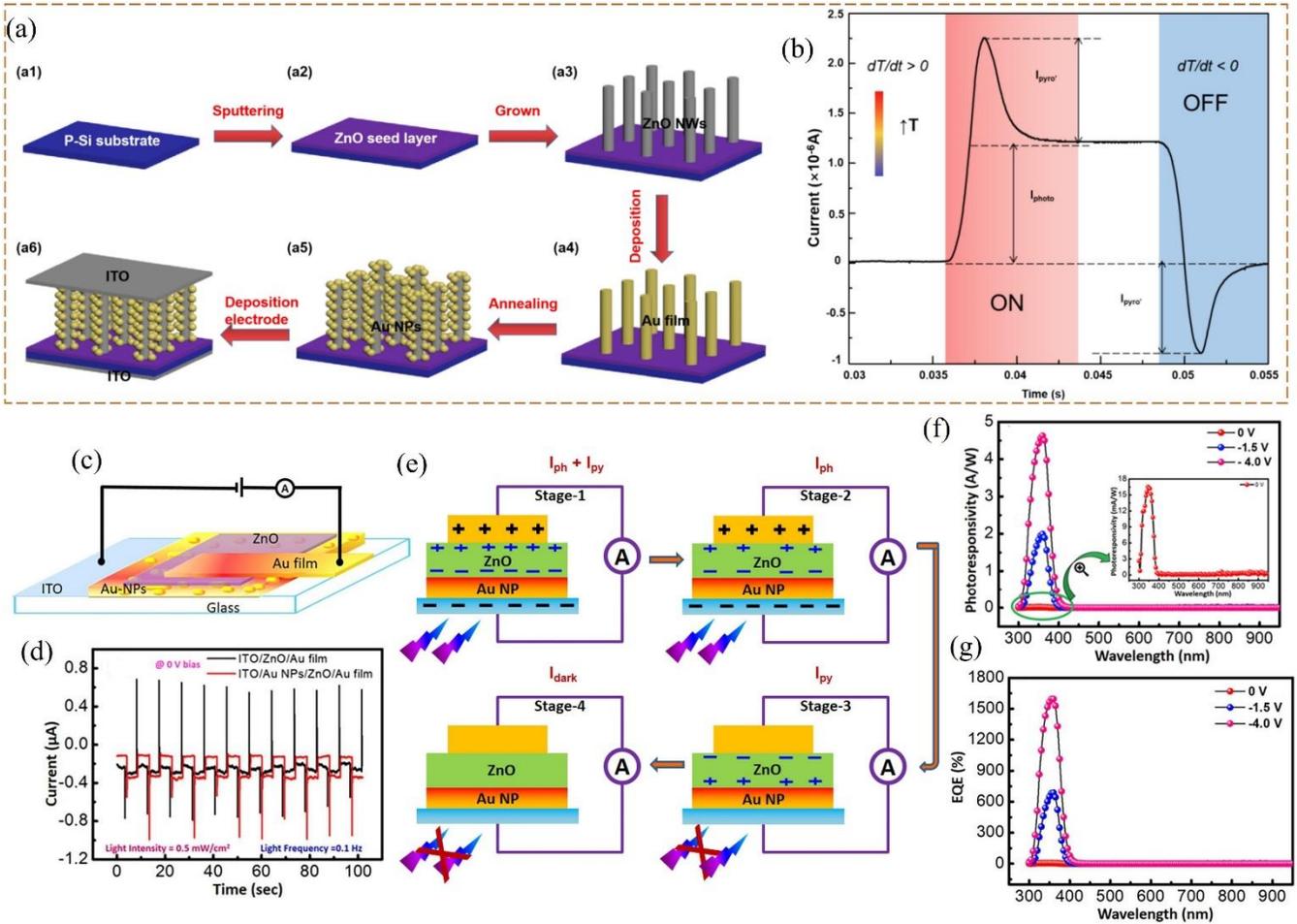

Fig.22: <u>(a-b)</u>: Au NPs/*n*-ZnO NWs/*p*-Si NIR PD. (a) Fabrication of Au NPs/*n*-ZnO NWs/*p*-Si NIR PD. (a1) *p*-Si substrate cleaning, (a2) ZnO seed layer sputtering, (a3) growth of ZnO NWs array; (a4) Au film deposition; (a5) formation of Au NPs via annealing; and (a6) electrode deposition; (b) $I_{pyro}$, generation in ZnO NWs/Si PD in a single cycle (continuous illumination without bias voltage. Adapted with permission.[319] Copyright 2021 ACS. <u>(c-g)</u>: ITO/Au-NPs/ZnO/Au PD. (c) Layout of the ITO/Au-NPs/ZnO/Au device. Active materials (Au NPs and ZnO thin film) are incorporated between ITO and Au film electrodes; (d) *I-t* characteristics of Device-1 (with Au NPs) and pristine one (without Au NPs). In pristine device, the pyroelectric effect is dominant, but in Device-1, both the pyroelectric and photovoltaic effects persist*;* (e) Illustration of the current generation because of the light-induced transient heating effect on pyroelectric polarization; (f) *R* of Device-1 at different *λ*s and biased condition. Inset shows magnified view of *R* at 0 V; (g) *EQE* of Device-1. The device performed better in UV-A region. Adapted with permission.[321] Copyright 2020 Elsevier B.V.

The fabrication of a UV-PD having a photo-induced pyroelectric effect or pyro-phototronic effect and a photovoltaic effect in combination has been reported.[321] The incorporation of sputter-deposited Au NPs into the device design capitalized on UV absorption in Au NPs via interband transition as well as exciton generation in ZnO, leading to greater $I_{phot}$ generation in the UV region (365 nm). Furthermore, the intrinsic pyroelectric feature of the synthesized, *c*-axis-oriented ZnO contributed to the increased speed of the device. In biased conditions, the maximal photoresponsivity and $D^*$ of the constructed device were 4.68 A/W and $8.18 \times 10^{11}$ Jones, respectively. In self-powered mode, the device had an ultrafast $\tau_{res}$ of 15 μ seconds. In this study, two device geometries were executed. Device-1 featured the configuration ITO/Au NPs/ZnO/Au film (Fig.22c), in which one electrode was a patterned ITO-coated glass substrate, the counter electrode was Au film, and the photoactive material was an Au NPs/ZnO bilayer inserted between these electrodes. To comprehend the contribution of Au NP incorporation, one pristine device was also built with only a ZnO layer with ITO/ZnO/Au film (configuration).

In the self-powered mode, Device-1 had a maximum photoresponsivity of 17 mA/W at a $\lambda = 365$ nm (Fig.22f inset). The $I$-$t$ characteristics of both devices (Fig.22d) showed that the rate of $I_{phot}$ generation in the Au-incorporated device was significantly faster than in the pristine device. Since the pristine device solely comprised pyroelectric ZnO crystal, the $I_{phot}$ was primarily generated by the pyroelectric effect. Photovoltaic part is nonexistent here. In Device-1, on the other hand, an energy barrier was formed between Au NPs and ZnO. This barrier produced an electric field that drives the charge carriers and generates photovoltage. Aside from the photovoltaic effect, ZnO's intrinsic pyroelectricity helped to generate $I_{phot}$. At a $\lambda = 365$ nm, Device-1's maximum $IQE$ was measured to be 2120% and one of the primary reasons for Device-1's high performance is the additional charge carriers generated by Au NPs. Furthermore, these Au NPs change the energy level of the fabricated device, causing charge carriers to flow.

The four-stage pyroelectric current generation is depicted in Fig.22e. When the device is simply illuminated with proper UV light, charge carriers are produced, resulting in $I_{phot}$ generation in the device. When it is lit up, the non-centrosymmetric ZnO crystal also experiences the short-term heating effect. This produces pyroelectric polarization in the same direction as the photovoltaic current flow and this polarization causes an additional transient current, the pyroelectric current ($I_{py}$), to be generated in the device. Therefore, in stage 1, the total current flow = $I_{phot} + I_{py}$. The effect of transient heat disappears in stage 2, and as a result, pyroelectric current disappears; in this state, the current flow is primarily due to $I_{phot}$. When the light is turned off, the device experiences another transient heat fluctuation in the opposite direction. This fluctuation causes a pyroelectric polarization, but in the opposite direction, resulting in current direction reversed. Because the photovoltaic contribution is nil in the dark mode, the flowing current in stage 3 is completely attributable to the pyroelectric effect ($I_{py}$). In stage 4, the effect of transitory heat becomes steady again, and pyroelectric polarization fades. As a result, the current flowing through the device is effectively the $I_{dark}$ of the device.

In $p$-Si/ZnO NWs PDs with photothermal-pyroelectric-plasmonic effects have demonstrated that plasmonic NPs arrays of various metals have distinct LSPR spectral absorption peaks, and pyroelectric PDs with inserted layers of Au or Ag NPs ($p$-Si/Au NPs/$n$-ZnO and $p$-Si/Ag NPs/$n$-ZnO) have exhibited waveband-selective detection under 405 and/or 940 nm light.[322] The temperature-dependent pyroelectric potential charges generated within ZnO NWs were shown to efficiently promote charge-carrier separation and transport by interacting with the intrinsic built-in electric field of $p$-Si/ZnO NWs. In addition to this, the inserted plasmonic NPs array results in increased light absorption, hot $e^-/h^+$ excitation, and a decreased barrier height at the $p$-Si/ZnO interface. This results in an LSPR potential that is induced by enhanced incident light absorption at a specific resonant $\lambda$ range at $p$-Si/ZnO NWs junction. Furthermore, its $\lambda$ selection ratio was raised by an 80-fold.

The operating mechanism of the plasma interface-modified waveband-selective $p$-Si/metal NPs/ZnO NWs PDs is depicted in Fig.23a. When the light is turned on, both ends of the pyroelectric NWs are heated almost simultaneously by the metal NPs and the incident light due to the metal NPs' high light absorption efficiency, resistive heating effect, and rapid plasma attenuation exothermic reaction. $E_{py}$ increases as the d$T$/d$t$ of the ZnO NWs increases. The excited surface plasmon can, because of the LSPR phenomenon, produce vibrations of the metal NP electron cloud, which, in turn, leads to the buildup of hot $e^-$-$h^+$ pairs at both ends of the NP, which is driven by the electric field ($E_b$). Hence, the hot electrons with high energy that are created on the surface of metallic NPs can inject into ZnO NWs mounting the schottky barrier (0.4-0.7 eV), which contributes to a noticeable $I_{phot}$ and generates a field ELSPR in the same direction as $E_{py}$. With this configuration, the origin of $I_{phot}$ is no longer restricted to photon energy that are higher than the ZnO band gap; rather, it is confined to photon energies that are higher than the SBH. This means that the sensor can detect light below the bandgap of the semiconductor without having a bias voltage. The created hot electrons in Au NPs overcome the contact barrier and leave Au NPs positively charged under the plasmon resonance condition, disturbing the equilibrium of the Fermi level status in metal/semiconductor heterojunction. Consequently, to compensate for the Fermi level difference, low energy electrons will flow into Au NPs. Au NPs can get electrons from the perpendicular electric field at the metal/semiconductor boundary, whereas ZnO NWs receive hot electrons from plasmons. Due to interfacial plasma-induced hot electron injection and scattered electron-phonons releasing heat, $p$-Si/metal NPs/ZnO NWs PDs created a faster and larger current-response peak marked $I_{pyro+photo+LSPR}$ (Fig.23b). The finite element analysis simulated localised EM fields of the Au NPs array (average size of 50 nm and an average spacing of 40 nm) are shown in Fig.23c. In metal NPs, the LSPR evidently induces intraband and interband excitation of electrons known as hot carriers.

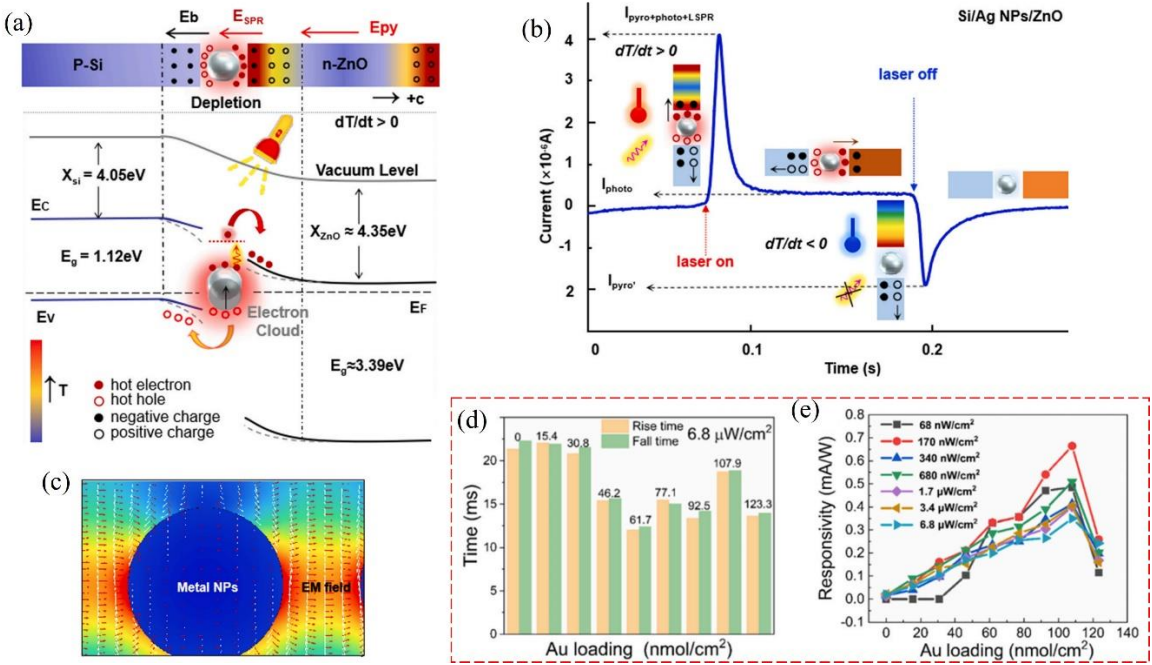

Fig.23: (a-c): *p*-Si/metal NPs/*n*-ZnO NWs pyroelectric PD, operational mechanism, and plasmonic NPs electromagnetic fields simulation. (a) Under illumination, potential distribution and band diagrams of the plasma interface modified pyroelectric PD; (b) Photoresponse behavior of pyroelectric PD (at $\lambda$ = 405 nm light, 5 mW/cm² power density, and 0 bias). Insets in (b): Depiction of the pyroelectric potential distribution; transport of photogenerated carriers and plasma induced hot carriers; (c) Demonstration of an optical near-field enhancement effect via simulation of plasmonic Au NPs' localized electromagnetic field interaction (at 940 nm). Adapted with permission.[322] Copyright 2021 Elsevier Ltd. (d-e): Schottky junction of Au NPs/ZnO NWs. (d) $\tau_r$ and $\tau_d$ (under $\lambda$ = 325 nm with 6.8 μW/cm² power density); (e) $R$ of devices before and after decorating Au NPs (in nmol/cm² loading amount) at different power densities of 325 nm light. Adapted with permission.[323] Copyright 2022 Elsevier Ltd.

Under UV illumination, the rate at which the temperature of ZnO NW changes affects the amount of pyroelectric current generation. To accelerate the rate of temperature change, two strategies have been devised: (a) UV light photothermal conversion with low power density is being enhanced.; (b) reducing the temperature of the substrate[324] on which the material is being deposited. The second strategy poses limitations on the application of a PD at RT or at high temperatures. In a study, the first approach was chosen to boost the pyroelectric current of the ZnO NW via fabricating a self-powered PD (Schottky junction of Au NPs/ZnO NWs), where the LSPR phenomenon was used.[323] When UV light illuminates a metal NP, LSPR can provide transitory thermal power, which can boost the light-induced heating effect and light absorption. The device was able to detect 325 nm light (68 nW/cm² power density) and under the same $\lambda$ illumination (with a power density of 170 nW/cm²), the $R$ and $D^*$ were dramatically improved by 3290% and 3298%, respectively. Because of the swift recovery, detection could be accomplished in 12 ms. After decorating Au of 107.9 nmol/cm², $R$ and $D^*$ reached 0.485 mA/W (Fig.23d-e) and $2.749 \times 10^{11}$ Jones for the self-powered PD under the illumination of 325 nm light (power density of 68 nW/cm²), whereas the device with 0 loading did not show response.

Akin to ZnO, TiO$_2$ NPs[325] and nanotubes,[326] have been investigated for UV photodetection because of their photoelectron transfer efficiency and high chemical stability. However, in comparison to ZnO nanostructured materials, TiO$_2$-based PDs cannot achieve high $R$ and a swift $\tau_{res}$ at the same time.[327] Maximizing the performance of self-powered UV PDs requires lowering the charge recombination rate and delayed carrier lifespan. TiO$_2$ is principally an *n*-type semiconductor, and building a *p-n* heterojunction necessitates the use of a suitable *p*-type material. *p*-type organic materials: P3HT,[328] PEDOT:PSS,[329] PANI,[330] while NiO, CuO, V$_2$O$_5$, etc. inorganic materials have been employed to build various PDs.[331]

The combined pyroelectric and photoelectric processes increase the intrinsic electric field, speed up charge transfer at the heterojunction interface, and delay the rate of $e^-\text{-}h^+$ pair recombination, allowing for high-performance UV photodetection.[332] TiO$_2$ nanorod arrays/polyterthiophene (TiO$_2$ NRs/PTTh) and TiO$_2$ nanorod arrays/Au NPs/polyterthiophene (TiO$_2$ NRs/Au/PTTh, Fig.24b) self-powered *p-n* heterojunctions, applicable to UV PDs were constructed.[332] In TiO$_2$ NRs/Au/PTTh PD, due to the pyro-phototronic effect between Au NPs and TiO$_2$ NRs, the detection performance was enhanced. The $R$ of TiO$_2$ NRs/PTTh and TiO$_2$ NRs/Au/PTTh UV PDs were respectively 0.0774 mA/W and 1.894 mA/W; $D^*$s were $2.155 \times 10^9$ Jones and $16.66 \times 10^9$ Jones, respectively. The $\tau_r$ and $\tau_d$ of the TiO$_2$ NRs/Au/PTTh UV PD were lowered when compared to the TiO$_2$ NRs/PTTh UV PD. In TiO$_2$ NRs/PTTh device, under zero bias, the $I_{phot}$ was ~9 nA. While the pyro-phototronic action causes the $I_{phot}$ in the TiO$_2$ NRs/Au/PTTh device to dramatically grow to ~150 nA. The transient current and the $I_{phot}$ are shown together in the observed peak of Fig.24d. The TiO$_2$ NRs/Au/PTTh device outperformed the TiO$_2$ NRs/PTTh device in terms of performance, due to both the pyro-phototronic that took place at the TiO$_2$ NRs/Au interface and the PVs impact at the *p-n* junction. The transient thermal effects of UV light irradiating the pyroelectric components generate hot electrons, causing the device to respond very fast. A built-in electric field is generated in the junction region after *p*-type PTTh/*n*-type TiO$_2$ NRs *p-n* junction is created (Fig.24a), generating a certain potential difference and this potential difference facilitates the segregation of photogenerated carriers and causes the electrons and holes to travel in opposing directions, contributing to the $I_{phot}$ in the external circuit. When photons strike TiO$_2$ NRs or PTTh, carriers are created if the photon energy is > or equivalent to the $E_g$ of the respective material. Holes will shift from the VB of TiO$_2$ NRs to the HOMO of PTTh since the HOMO of PTTh is lower than the VB of TiO$_2$ NRs. In parallel, electrons will travel from the polymer's LUMO to the CB of the TiO$_2$ NRs. The electrons in CB and the holes in HOMO are then transported, respectively, to the FTO electrode. The interfacial charge transfer was facilitated by the energy level matching of TiO$_2$ NRs and PTTh. The rate at which $e^-\text{-}h^+$ pairs recombine is lowered by this interfacial charge transfer. Following the addition of Au NPs (Fig.24a, right side image), in the TiO$_2$ NRs/Au/PTTh heterojunction, the transient thermoelectric potential will be produced during UV illumination as a result of the combination of Au NPs and TiO$_2$ NRs, which produces a transient photocurrent ($I_{photo}+I_{pyro}$). The holes will simultaneously go from the VB of TiO$_2$ NRs to the Fermi level of Au NPs and subsequently to the HOMO of PTTh because of the photoelectric effect. In the same way, the electrons in the PTTh will move from the LUMO to the Fermi level of the Au NPs. They will then move to the CB of the TiO$_2$ NRs, and both the holes and electrons will move to the FTO electrode. Transient photocurrent ($I_{pyro}$), which is produced in the opposite direction when the UV lillumination is stopped, is a manifestation of the transient thermoelectric potential. The current stabilizes and presents a steady-state current after the transient thermal effect vanishes ($I_{dark}$).

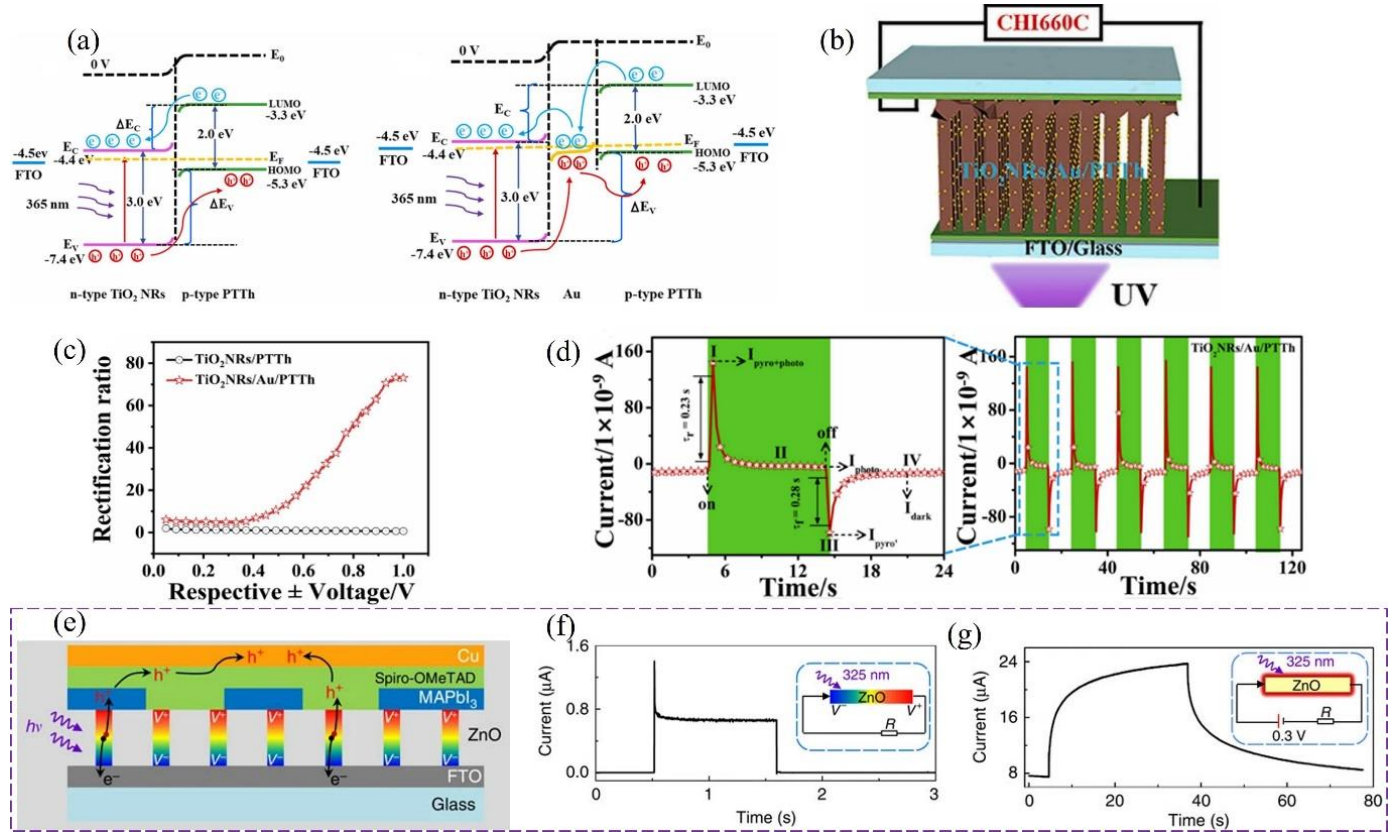

Fig.24: (a-d): TiO$_2$ NRs/PTTh and TiO$_2$ NRs/Au/PTTh UV PDs. (a) Energy band diagram of TiO$_2$ NRs/PTTh and TiO$_2$ NRs/Au/PTTh heterojunctions under self-powered model; (b) TiO$_2$ NRs/Au/PTTh PD; (c) rectification ratio of TiO$_2$ NRs/PTTh and TiO$_2$ NRs/Au/PTTh heterojunctions versus bias voltages under UV;(d) I-t curves of the TiO$_2$ NRs/PTTh and TiO$_2$ NRs/Au/PTTh heterojunctions.[332] (e-g): Self-powered ZPH PDs. (e) Schematic illustration of the working mechanism of ZPH PDs. Spiro-OMeTAD is HTM; (f-g) I–t curves of the ZPH PDs under different biases. (f) 0.0 V and (g) 0.3 V. Inset: Corresponding circuit diagram with current heating effect (red glow around ZnO) and pyroelectric effect of ZnO. Adapted with permission.[152] Copyright 2015, Wang et al.

Because of the binary photo-switching (ON/OFF = 1/0 states), PD is appropriate for light-based computing applications. However, such binary photo-switching PDs operate with or are governed by external bias, limiting their usefulness in intelligent devices. But, binary photo-switching can be made to work based on PPE in a self-powered PD. Recently, TiO$_2$/Si heterostructure-based PDs were shown to decode digital codes for practical applications.[333] SnS$_x$ features an energy band gap that is direct and controlled (1.3-1.7 eV), a high $\mu$ (10000-38000 cm$^2$/V-s),[318] and a high absorption coefficient ($> 5 \times 10^4$/cm). This material also exhibited a pyro-phototronic effect (PPE), which is significant in enhancing the performance of self-driven PDs at the $SnS_x/TiO_2$ NRs interface.[334] In tin sulfide ($SnS_x$)/$TiO_2$ NRs heterojunctions, a similar binary photoswitching functionality has been observed.[335]

To enhance the properties of ZnO-based PDs, the PPE of ZnO has been combined with the LSPR of Ag NPs.[336] The self-powered PD was composed of ZnO NWs and Ag NPs, and it can detect 325 nm UV light rapidly and effectively with a power density of 340 nW/ cm². The best $R$ and $D^*$ under the same UV light and power density were $8.82 \times 10^{-5}$ A/W and $4.9 \times 10^{10}$ Jones, respectively and the $\tau_{res}$ was reduced to 8.72 ms.

The fundamental challenges in designing high-performance ZnO-based UV PDs have been their relatively slow $\tau_{res}$ and performance variations, which are inevitable difficulties. Because the crystal structures of wurtzite ZnO are non-central symmetric, temperature changes in different parts of the material cause pyroelectric potentials.[337] Light irradiation induces a rapid rise in temperature within ZnO NWs, resulting in pyroelectric potential distributions along the crystal and polarization pyrocharges at both ends of the NW. In a study, a direct bandgap, large absorption coefficient, and high $\mu$ perovskite, $CH_3NH_3PbI_3$ (MAPbI₃), were used to construct a heterostructure with ZnO in order to minimize the $I_{dark}$ (0.1 nA) and maximize the pyroelectric effect improvements.[338] The ZnO/perovskite-heterostructured and self-powered photodetector (ZPH) PD's $\tau_{res}$ time was improved by five orders of magnitude, going from 5.4 seconds to 53 μs in the $\tau_r$ and from 8.9 seconds to 63 μs in the $\tau_d$. The $D^*$ and $R$ for the PD both increased by 322%.[152] The valence band and conduction band of the MAPbI₃ perovskite, measured with respect to vacuum level, are −5.4 and −3.9 eV, respectively.[338c] According to the absorption spectra of the ZnO-NW array and the ZnO/MAPbI₃ perovskite film, excitons are formed in the ZnO layer when exposed to UV light. Electrons are injected from ZnO layer and collected by the FTO electrode. Meanwhile, holes in the valence band reach the Cu electrode through the hole transport material (HTM). Based on the direction of the ZnO-NW array's *c*-axis, polarizations elicit pyroelectric potentials across the NWs, with positive potentials at the HTM/Cu electrode side and negative potentials at the FTO electrode side (Fig.24e). By aligning with the photovoltaic direction across the external circuit, this light-induced pyroelectric potential raises the transient short-circuit current and open-circuit voltage. Additionally, the pyroelectric potential of ZnO NWs may accelerate charge-carrier segregation at the interface by inhibiting photoexcited electrons in ZnO from recombining with holes at heterojunctions, improving $\tau_{res}$ even further. Larger bias voltages resulted in higher dark currents, which raised the device's background temperatures and, consequently, decreased the pyro-potential in ZnO NWs (insets of Fig.24f-g). When the bias voltage increased, there was a noticeable slow disappearance of the PPE-induced sharp peak (Fig.24g).

### 4.2.4. Other 2D and mixed-dimensional heterostructures for light detection

Heterostructures based on other materials: 1D SWCNT and nanowires, 3D topological insulators, Si nanostrips, metal phosphorous tri-chalcogenide, thermoelectric materials, and MXenes, etc., have also been explored.

Because of their broad-spectrum absorption and excellent photocarrier transfer with Gr, semiconducting single-wall nanotubes (s-SWCNTs) are a viable choice. Excellent optoelectronic capabilities have been observed when Gr is paired with a blend of dispersed s-SWCNTs and metallic SWCNTs.[339] The recent achievement of large-area s-SWCNTs thin film fabrication for transistors and PDs[373, 374] enables the fabrication of high-quality Gr/s-SWCNT heterostructures for optoelectronic devices. An efficient Gr/s-SWCNTs PD based on a double-layer heterostructure has shown photodetection capabilities.[340] The as-fabricated device exhibited enhanced performance over a broad range of $\lambda$s (UV - NIR), with a high $R$ of 78 A/W at a Vis wavelengths. The separation of the light-absorbing layer and the charge carrier transport channel led to a high photoconductive gain of $8 \times 10^4$; a fast $\tau_{res}$ of 80 µs, which is better than Gr-based heterostructured PD.

Thermoelectric materials such as $Bi_2Se_3$, $Bi_2Te_3$, and $Sb_2Te_3$ are examples of classical 3D topological insulators. These materials have a gapless metallic surface state but an insulating interior.[46] These materials display surface plasmons in the UV and Vis regions, which are explained by metallic surface states and nonequilibrium carriers caused by interband transitions. Mauser et al.[341] integrated these two characteristics to create a subwavelength thermoelectric *p-n* junction with intrinsic resonant absorption in the Vis region. The thermoelectric junction is made up of a periodic NW array, and each end is coupled to a wide pad made of the same material. (**Fig.**25a). The active materials were *n*-type $Bi_2Te_3$ (Seebeck coefficient of -84 µV/K and *p*-type $Sb_2Te_3$ (Seebeck coefficient of 242 µV/K). Even under spatially uniform irradiation, a substantial temperature gradient can be established due to the differential in absorptivity between the NW array and the wide pad. The thermoelectric junction was put on a suspended low-thermal-conductivity substrate, and the laser was shined on the *p–n* junction interface. The resulting on-resonance responsivity was ~38 V/W. The $\tau_{res}$ was ~ 0.34 ms due to the compact volume of the device that can promote heat dispersal.

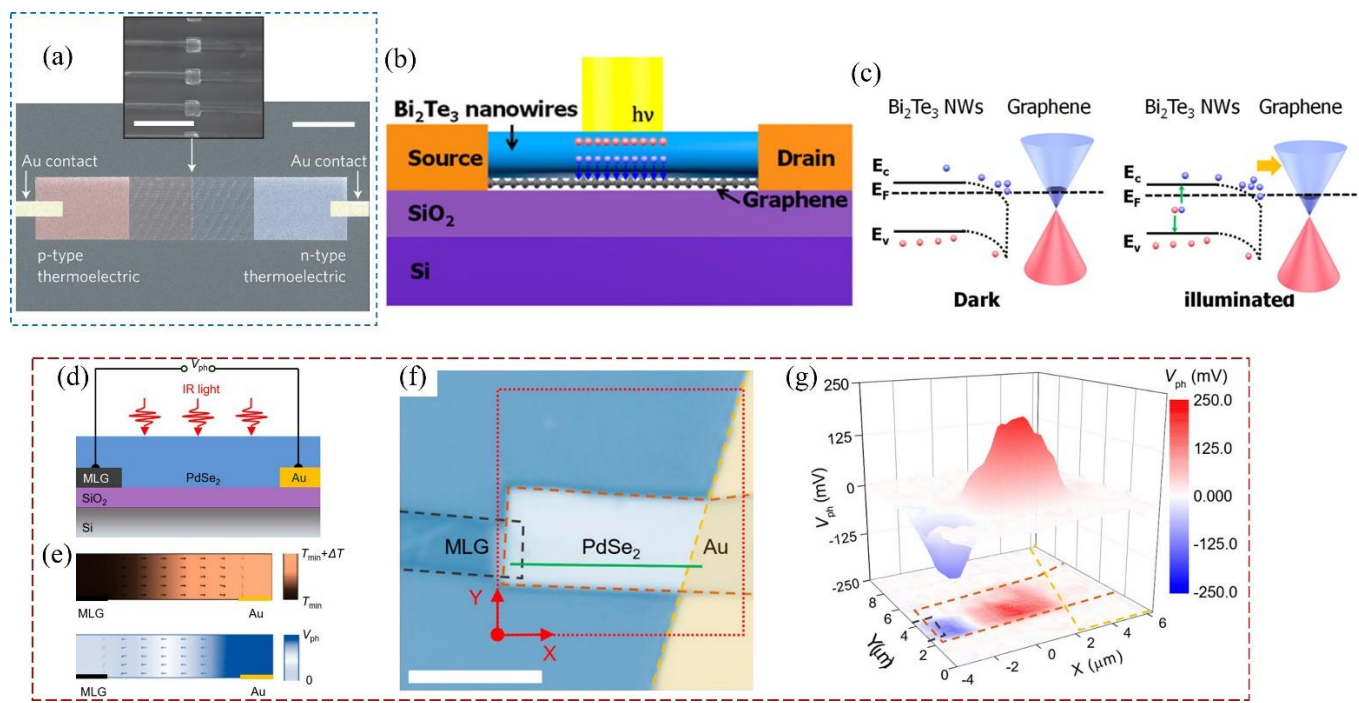

Fig.25: <u>(a)</u>: Bi₂Te₃/Sb₂Te₃ wire structure. False-color SEM of a fabricated *p/n* thermoelectric structure, with Au contacts. Scale bar, 20 µm. Inset: junction between Bi₂Te₃/Sb₂Te₃ wires. Scale bar, 1 µm.[341] <u>(b-c)</u>: Bi₂Te₃ NWs/Gr PD. (b) Illustration of generation of an electron (filled circle) and a hole (open circle) under the illuminated condition; (c-left side) dark, and (c-right side) illuminated status at the Bi₂Te₃ NWs/Gr junction.[342] <u>(d-g)</u>: PdSe₂ LWIR PD with asymmetric vdW contacts. (d) PdSe₂ PD under a global IR light illumination; (e) temperature and photovoltage distribution of the PdSe₂ detector under global IR light illumination. Black and blue arrows denote the gradient of temperature and electric field, respectively. (f) Image of the PdSe₂ PD with asymmetric vdW contacts. Scale bar: 5 µm; (g) The 532 nm laser-induced $V_{ph}$ mapping in a 3D plot exhibits an asymmetric distribution of photovoltage, indicating an asymmetric distribution of temperature in the channel. Adapted with permission.[343] Copyright 2022 ACS.

The photodetection characteristics of a PTE-based plasmon modulated Si nanostripe PD have been explored.[344] An open circuit photovoltage response of ~82 mV µW⁻¹ was recorded, for a 633 nm wavelength (0.09 µW illumination power) laser illuminating the Au plasmonic grating region (doped Si nanostripes with Seebeck coefficient of ~9.9 mV/K). The increased photovoltaic responsivity was caused by plasmon-enhanced optical absorption in the Si thin film as well as the substantial Seebeck coefficients of the Si nanofilm. Surface plasmon polaritons at the Au/Si interface and waveguide modes in the Si layer significantly improved optical absorbance. The Cr/Au coating on the Si nanostrips induced the sign of the Seebeck coefficient of Si to be inverted, as well as the sign of the photoresponse to be reversed.[344]

A drop-casting method was used to prepare Bi₂Te₃ NWs/Gr junction, which exhibited a high gain of ~ $3 \times 10^4$ (1.61 times larger than that of the only Gr PD) and wide bandwidth window (400–2200 nm).[342] Gr was *n*-doped after making contact with $Bi_2Te_3$ NWs. The photoresponsivity increased by 200% at 2200 nm (~0.09 mA/W) with narrow bandgap $Bi_2Te_3$ NWs and Gr junction. **Fig.25**b shows a schematic depiction of photogating effect in $Bi_2Te_3$/Gr PD. When photons strike the $Bi_2Te_3$ NWs/Gr PD, they create $e^-$-$h^+$ pairs in both the $Bi_2Te_3$ NWs and the Gr channel and drift through the channel. Because electrons and holes in $Bi_2Te_3$ NWs have different diffusion and drift time constants, holes remain in the $Bi_2Te_3$ NWs for longer than electrons, triggering the Dirac voltage shift after illumination (**Fig.25**b-c). If $e^-$-$h^+$ pairs were produced in the absence of $Bi_2Te_3$ NWs, recombination would occur rapidly.

As stated in Section 2.4, the greater the asymmetry of the Seebeck coefficient or the temperature gradient distribution, the greater the photovoltage generated, resulting in a higher photodetection responsivity. By fabricating a *p-n* junction with material doping or electrical gating, it is possible to establish an asymmetric Seebeck coefficient and modify the Seebeck coefficient. An asymmetric Seebeck coefficient can also be generated by an asymmetric metal contact with a large work function difference. A common approach for producing a high temperature gradient is localized illumination by focusing the light beam. Because of their distinct optical resonances, all of the techniques outlined above necessitate complex and/or advanced fabrication processes and high $\lambda$-selectivity designs.[46]

Pentagonal palladium selenide ($PdSe_2$) is a 2D layered thermoelectric material with a narrow $Eg$ (0-0.1 eV in a multilayer), high $\mu$ (216 cm$^2$/V-s), and an excellent air stability at ambient temperature.[345] $PdSe_2$ is a puckered pentagon with a low lattice thermal conductivity (4.5 W/m-K). $PdSe_2$ has a low electrical resistivity as well as a high Seebeck coefficient (300 $\mu$V/K).[346] Furthermore, the multilayer $PdSe_2$ demonstrates a 10% absorption in the LWIR regime.[345, 347] It has been shown that a PTE PD based on semimetal $PdSe_2$ nanoflakes has a strong thermoelectric property and a high anisotropic optical absorption.[343] The $PdSe_2$ nanoflake is asymmetric van der Waals (vdW) contacts with multilayer Gr (MLG) and Au, which act as an asymmetric heat sink (**Fig.25**d and f). A large temperature gradient was formed within a sub-wavelength-scale channel under LWIR global illumination owing to the significant difference in thermal conductivity between Au and MLG. This triggered the heated carrier to create a possible difference. Both the bottom Au and MLG van der Waals contacts are good ohmic contacts, which can help reduce contact resistance and consequently voltage noise level. Consequently, the PTE $PdSe_2$ detector demonstrated a RT *R* of more than 13 V/W and a *NEP* of < 7 nW/Hz$^{1/2}$ for the LWIR light. A swift response with a $\tau_{res}$ of ~50 $\mu$s was also observed. Furthermore, the photoresponse of $PdSe_2$ was highly anisotropic, resulting in a linear polarization angle-sensitive response with an anisotropy ratio of 2.06 at 4.6 $\mu$m and 1.21 at 10.5 $\mu$m. A MIR laser with $\lambda$ = 4.6 to 10.5 $\mu$m was employed to flood the $PdSe_2$ PTE detector to investigate the LWIR light detection characteristics. **Fig.25**e depicts the spatiotemporal distributions of temperature and potential under global IR illumination. Due to the greater cooling efficiency of the MLG electrode, the temperature localized at the Au contact was greater than that of the MLG contact under a global IR light irradiation. Consequently, a temperature gradient (black arrows) was generated from the MLG contact to the Au contact, inducing a PTE.

**Fig.25**g shows a 532 nm laser with a power of 1 mW producing thermoelectric photovoltage ($V_{ph}$) mapping in a 3D plot exhibiting the universal distribution of asymmetric $V_{ph}$ over the whole device channel. Peak $V_{ph}$ values can reach -117 and 193 mV at the MLG and Au contact sides, respectively.

Metal phosphorous tri-chalcogenides ($MPX_3$, $M$ = transition metal, $X$ = S, Se, Te) have a larger bandgap (1.2-3.5 eV)[348] compared to TMDs of 1.0–2.0 eV are potential candidates for UV (10 to 400 nm) detectors.[49] FePSe$_3$ ($p$-type 2D semiconductor) has an indirect $E_g$ of 1.3 eV (corresponding to NIR range).[349] Broadband photodetection from Vis-SWIR) was demonstrated using FePSe$_3$ thin flakes.[350] UV detectors fabricated from $MPX_3$ materials, on the other hand, have low photoresponsivity, moderate response speed, moderate $D^*$, and a narrow spectral response window. In order to widen the detection window, the bandgap must be tuned down so that the edge of the absorption band can be pushed into the IR spectrum.[351] An ultra-broadband FePSe$_3$/MoS$_2$ vdW heterodiode PD operationing in spectrum from SBUV-LWIR (0.265–10.6 µm) has been demonstrated (**Fig.26**a-d).[351] The device measured a high photoresponsivity of ~33600 A/W, very low *NEP* of ~7.1 × 10$^{-17}$ W/Hz$^{1/2}$ (**Fig.26**a and c), and a maximum $D^*$ of 1.51 × 10$^{13}$ cm Hz$^{1/2}$/W in the SBUV spectral range (200-280 nm). The device was also capable of sensing MWIR and LWIR at RT and exhibited a rectification performance with a rectification ratio ~10$^2$. A $D^*$ of 6.01 × 10$^8$ cm Hz$^{1/2}$/W in the MWIR regime and 4.92 × 10$^8$ cm Hz$^{1/2}$/W in LWIR (10.6 µm) was obtained in ambient air. The $\tau_r$ (from 10% to 90%) and $\tau_d$ (from 90% to 10%) were 0.32 and 0.36 ms, respectively (**Fig.26**b). **Fig.26**d, compares photoresponsivity of different UV photodetecting wide bandgap semiconductors versus response time.[351]

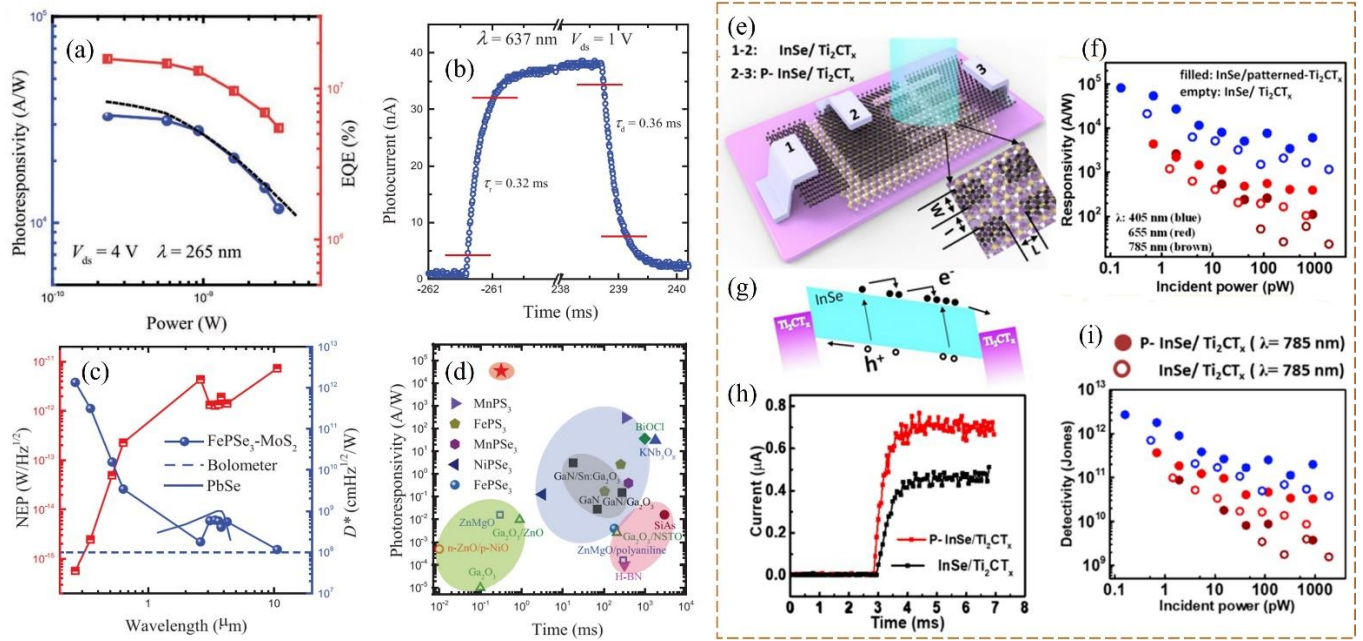

Fig.26: (a-d): FePSe₃/MoS₂ heterodiode. (a) Incident light power-dependent $R$ and $EQE$ of a device under SBUV 265 nm irradiation at a bias of $V_{ds} = 4$ V; (b) $\tau_{res}$ of heterodiode at a bias of 1 V; (c) $NEP$ versus $\lambda$ at a bias of 1 V and $\lambda$ dependence $D^*$ of the heterodiode compared with the commercially available PbSe and bolometer detectors; (d) comparison of photoresponsivity and $\tau_{res}$ of different UV photodetecting wide bandgap semiconductors. Red star refers to FePS₃-MoS₂. Adapted with permission.[351] Copyright 2022 Wiley-VCH GmbH. (e-i): InSe/Ti₂CT_x avalanche PDs. (e) Schematic of unpatterned and patterned avalanche PDs. A single InSe nanosheet was simultaneously used as the channel material in unpatterned and patterned InSe/Ti₂CT_x PDs, and a single Ti₂CT_x flake with constant width is used as the 2D electrode; (f) $R$, and (i) $D^*$ vs. incident light power for patterned and unpatterned avalanche PDs; (g) Illustration of avalanche effect in InSe/Ti₂CT_x PD; (h) time-resolved photoresponse measurement result ($\lambda = 405$ nm, $P = 625$ W/m² , $V_d = 3$ V). Adapted with permission.[352] Copyright 2019 ACS.

MXenes are a new family of 2D materials that have garnered a lot of interest since their discovery in 2011. MXenes are made by selectively etching away $A$ layers from their $MAX$ phases ($M_{n+1}AX_n$), where $M$ is an early transition metal, $A$ is an element from group IIIA -VIA, and $X$ is carbon and/or nitrogen. MXenes have high metallic conductivity and stability, and they are simple to synthesize in large quantities.[353] MXene (Ti₂CT_x) is reckoned as a suitable 2D electrode material for CMOS inverters due to its high work function (4.98 eV) and low resistance and has been used as an electrode in a InSe-based PD.{Yang, 2019 #313} In this study, two types of avalanche PDs (APDs) were fabricated; InSe/Ti₂CT_x and the other with the InSe/Ti₂CT_x electrode patterned into periodic nanoribbon arrays (i.e., a plasmonic grating structure) to enhance light absorption (denoted as P-InSe/InSe/Ti₂CT_x, **Fig.**26e). Electron beam lithography (EBL) method was used to pattern the electrode. Among these two APDs, P-InSe/InSe/Ti₂CT_x ultrathin PD outperformed the InSe/InSe/Ti$_2$CT$_x$ APD (i.e. nonpatterned InSe/Ti$_2$CT$_x$ electrode).[352] The PADs had a large Schottky barrier between InSe and Ti$_2$CT$_x$ due to the high work function of Ti$_2$CT$_x$. As a result, the PD had low $I_{dark}$s and can trigger the avalanche effect in InSe even at high drain voltages. This avalanche carrier multiplication helped improve the PD's performance.

The band structure of the InSe/Ti$_2$CT$_x$ PD when the avalanche event occurs is schematically illustrated in **Fig.26**g. The photoexcited carriers are accelerated to high energies by the strong electric field; this triggers carrier multiplication to generate more $e^-$-$h^+$ pairs, which increases the drain current.[56, 354] The photoresponse times of InSe/Ti$_2$CT$_x$ APD and P-InSe/Ti$_2$CT$_x$ were 0.8 and 0.5 ms, respectively (**Fig.26**h). Because of saturation of light absorption in both APD, the responsivity values decrease as incident power increases. Because of the higher $I_{phot}$, the P-InSe/Ti$_2$CT$_x$ APD yielded the maximum $R$ of $1 \times 10^5$ A/W under blue light irradiation (**Fig.26**f). The P-InSe/Ti$_2$CT$_x$ APD under 405 nm light illumination, has the greatest detectivity of $73 \times 10^{11}$ Jones, compared to $7.1 \times 10^{11}$ Jones for the nonpatterned InSe/Ti$_2$CT$_x$ APD (**Fig.26**i).

A PD comprising $p$-GaSe/$n$-WS$_2$ heterojunction that was inserted between the top and bottom Gr electrodes was constructed and evaluated for its photoresponse characteristics ranging from UV- Vis.[180] Fig.27a depicts a schematic of the constructed $p$–$n$ heterojunction PD on the SiO$_2$/Si$^{++}$ substrate. The rectification ratio (ratio of the forward/reverse current) touched $\approx 1.12 \times 10^5$ at $V_{ds} = +2/-2$ V. Under zero bias, the built-in electric field in the heterojunction sweeps photogenerated $e^-$-$h^+$ pairs in opposing directions across the junction and into the Gr electrodes (Fig.27b). The devices operated at zero bias with a $R$ of ~56 mA/W at $\lambda = 410$ nm, and an $EQE$ of ~16.97%. Fig.27c shows that the heterojunction has a high photoresponsivity from Vis - UV light. At $V_{ds} = 2$ V, the photoresponsivity remains relatively high over a wavelength range of $\lambda = 270$-740 nm. The device also exhibits substantial photoresponsivity at $V_{ds} = 0$ V throughout a wavelength range of = 270-680 nm due to the built-in electrical field between the $p$-GaSe and $n$-WS$_2$, indicating that it can operate in self-powered mode. The $\tau_r$ and $\tau_d$ were calculated to be 37 and 43 µs, respectively (Fig.27d).

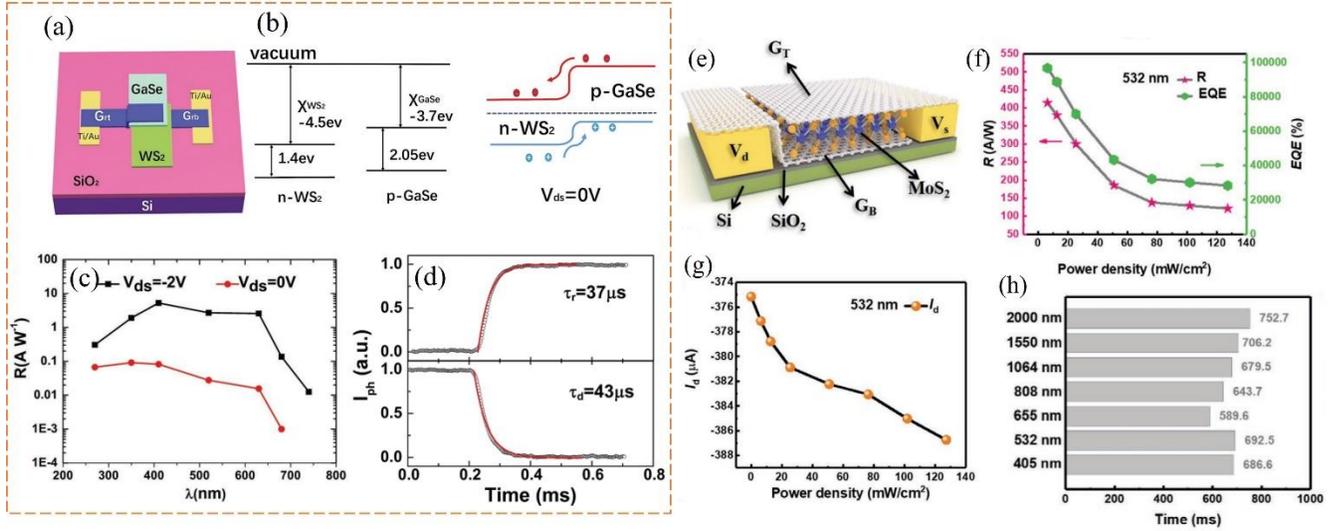

Fig.27: <u>(a-d)</u>: GaSe/WS₂ heterostructure. (a) Illustration of the Gr sandwiched GaSe/WS₂ heterostructure device; (b) Schematic band diagrams at the interface of the *p*-GaSe/n-WS₂ heterojunction. (b) Left image: Band structure of WS₂ and GaSe, electron affinities of WS₂ and GaSe are −4.5 and −3.7 eV, respectively. (b) Right image: Band alignment at the interface of the heterojunction at applied voltages $V_{ds}$ = 0 bias; (c) $R$ vs. λs at $V_{ds}$= ($V_{ds}$= 0, −2 V); (d) Normalized time response under LED illumination with a 5000 Hz light-switching frequency, red lines are fitting curves. Adapted with permission.[180] Copyright 2017 WILEY-VCH Verlag GmbH & Co. KGaA, Weinheim. <u>(e-h)</u>: Gr/MoS₂/Gr vertical heterostructured phototransistor; (e) Side view of the Gr/MoS₂/Gr vertical heterostructure; (f) $R$ and *EQE* as a function of incident power at $V_{ds}$ = −1 V for 532 nm; (g) $I_{ds}$-power relationship at $V_{ds}$ = −1 V, $V_g$ = 40 V for 532 nm; (h) $\tau_{res}$ of the device at $V_{ds}$ = −1 V, $V_g$ = 40 V. Adapted with permission.[355] Copyright 2020 Wiley-VCH GmbH.

A vertical Gr/MoS₂ (16.8 nm)/Gr vdWs heterostructure with Schottky barriers formed between MoS₂ and Gr were designed, fabricated, and tested for photoresponse from λ = 405 to 2000 nm.[355] The inclusion of Gr essentially broadened the device's working λ from Vis - IR. In the device configuration, the Schottky barrier height of top side graphene (G_T)/MoS₂ (ΔT) is higher than that of bottom side graphene (G_B)/MoS₂ (ΔB), and the device gets larger current when works in the reverse $V_{ds}$ bias. Varying gate voltages cause varied Fermi levels of G_B, altering the height of the Schottky barrier ΔB and the magnitude of the $I_{ds}$. Because of the shorter photogenerated carrier transmit distance and the optimized Fermi level of bottom side Gr in the vertical vdWs heterostructure, the PDs exhibited $R$ of ~414 A/W (at 532 nm laser), ~376 A/W (at 2000 nm laser); high *EQE* of ~96700% (at 532 nm laser) , ~23400% (at 2000 nm laser); $D^*$ of ~3.2 × 10¹⁰ Jones (at 532 nm laser), ~29 × 10⁹ Jones (at 2000 nm laser), and fast $\tau_{res}$ of ~692.5 ms (at 532 nm laser) and 752.7 ms (at 2000 nm laser). The variation in $I_{ds}$ with incident light power at $V_{ds}$ = 1 V is illustrated in Fig.27g. It is evident that the $I_{phot}$ in this Gr/MoS₂/Gr configuration would be enhanced by an increase in illumination power. Fig.27f shows the $R$ and electrical quantum efficiency associated with incident light power at gate voltages of $V_g$ = 40 V and $V_{ds}$ = 1 V, respectively. This suggests that these device photoresponsive parameters have decreased monotonously as the power increased, because of the decreased portion of photogenerated carriers that can be acquired at high light power. The high $R$ of device is related to: (a) the identical direction between the built-in potential and electron separation induced by the two Gr on both sides of $MoS_2$; and (b) the photogenerated carrier's short transmit distance, or the thickness of $MoS_2$ in the heterostructure device. Fig.27h presents the device's $\tau_{res}$ to different $\lambda$s with the best response to 655 nm light with a $\tau_{res}$ of 589.6 ms.

A vertical stack was farbicated, and its photoresponse properties were characterized by sandwitching a $MoS_2$ film between two conducting metal oxides, namely a transparent top indium tin oxide (ITO) layer and copper oxide ($Cu_2O$) (**Fig.28**a-d).[356] The ITO and $Cu_2O$ layers worked as highly doped $n$-type and $p$-type semiconductors, respectively, with high conductivities. The metal oxide layers protected the $MoS_2$ film from the environment and substrate, leading to reliable device characteristics. Further, they functioned as closely spaced carrier acquisition layers, with the distance between them determined by the thickness of the $MoS_2$ film. This structure exhibited $D^*$ of $32 \times 10^{13}$ Jones and $R$ of $57.7 \times 10^3$ A/W at an illumination power density ($P_{op}$) of 0.26 W/m² and external bias of -0.5 V. Futrther, ITO/$MoS_2$/$Cu_2O$ PD exhibited a small carrier transit time. The structure's short transit time and low $I_{dark}$ enabled for very high $D^*$ in excess of $10^{14}$ Jones at moderate optical power density, despite the absence of external bias due to the asymmetric PD's built-in field. The devices showed good photoresponse in $\lambda$ = 500-1100 nm range (**Fig.28**c). Because the active $MoS_2$ layer is buried under the ITO layer, the transparency of the top ITO layer was critical in determining the effectiveness of the device's photon absorption.

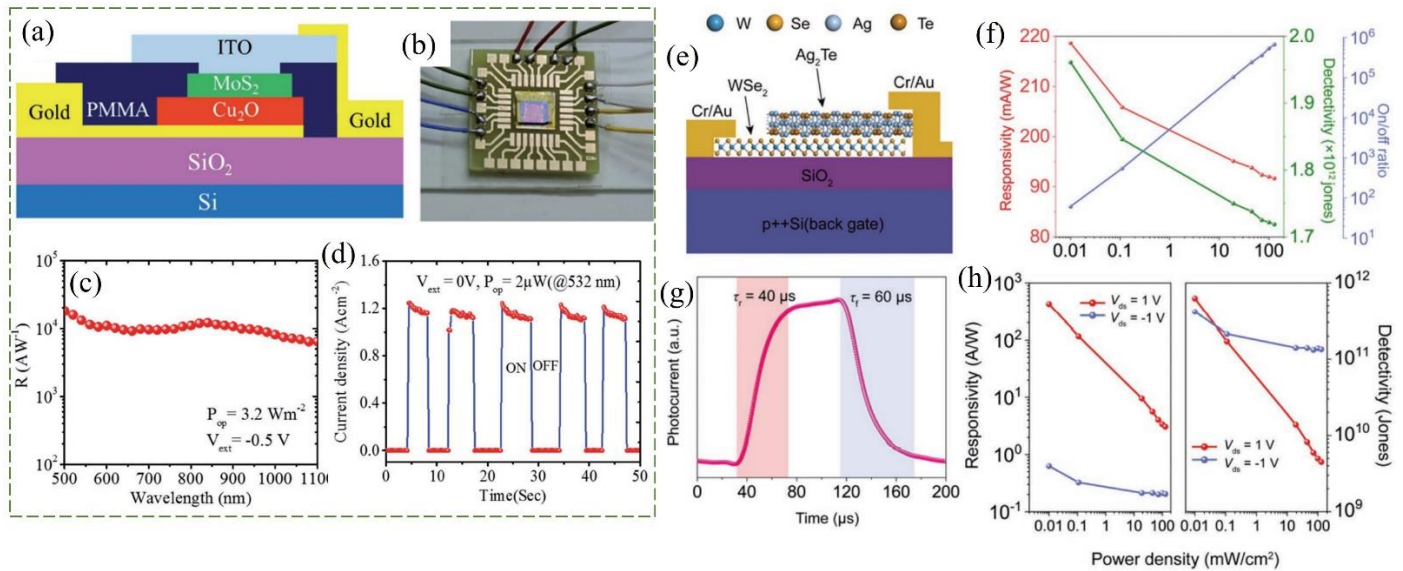

Fig.28: (a-d): ITO/$MoS_2$/$Cu_2O$ PD. (a) Cross-sectional view of the vertical device; (b) Optical image of wire bonded vertical PD chip; (c) $\lambda$ dependent $R$ indicating almost unchanged $R$ over a large range of $\lambda$s; (d) The device's transient response when a 532 nm laser is turned ON and The topological insulator -Ag$_2$Te was predicted to have a highly anisotropic surface Dirac cone.[358] Crystalline -Ag$_2$Te has extremely high $\mu$ and will most likely be used in future ballistic transport and topological electronics applications.[359] Ag$_2$Te is recognized as a good IR absorbing material because of its low $E_g$ semiconducting nature.[360] 0D Ag$_2$Te QDs[361] and 1D Ag$_2$Te nanotubes[360] have been reported to have NIR absorption and photoconductive response capabilities. The vdW epitaxy method was employed to synthesize ultrathin 2D $\beta$-Ag$_2$Te (with crystal thickness ~3.3 nm) on a mica substrate and subsequently used to fabricate Ag$_2$Te/WSe$_2$ heterojunction diodes via the mechanical stacking method[357] which showed ultrahigh rectification ratio ($2 \times 10^5$). 2D-Ag$_2$Te demonstrated $p$-type semiconducting nature ($\mu \approx 3336$ cm$^2$/V-s). The Ag$_2$Te/WSe$_2$ heterojunction PD was self-driven, yielding a maximum open circuit voltage of 0.41 V, and a detection range spanning Vis -NIR (405–940 nm). Under zero bias voltage, a high $R$ of 219 mA/W, a high $D^*$ of $1.96 \times 10^{12}$ jones, and a high light ON/OFF ratio of $6 \times 10^5$ were obtained. The PD was able to operate continuously for over 100 switching cycles while maintaining a fast $\tau_{res}$ (40 μs/60 μs) and no discernible $I_{phot}$ decline. Fig.28f shows the computed $R$, $D^*$, and ON/OFF ratios in photovoltaic mode for various power densities. The $R$ and $D^*$ of Ag$_2$Te/WSe$_2$ PD increase with decreasing light intensity and touch 219 mA/W and $1.97 \times 10^{12}$ jones, respectively, under 0.01 mW/cm$^2$, and stay at 192 mA/W and $1.72 \times 10^{12}$ jones even under a large power density of 129.03 mW/cm$^2$. Further, a high light ON/OFF ratio of $6 \times 10^5$ can be observed at 129.03 mW/cm$^2$ power density and it can reach 57 even under very low illumination (0.01 mW/cm$^2$). Fig.28g illustrates the response speed of PD, where the $\tau_{rise}$ of ~40 μs and $\tau_{fall}$ of ~60 μs were observed at a power density of 129.03 mW/cm$^2$. The variation of $R$ and $D^*$ with respect to various laser power densities is shown in Fig.28h. When $V_{ds} = +1$ V and the light power density is 0.01 mW/cm$^2$, the device's $R$ and $D^*$ can attain 428 A/W and 6.3 $10^{11}$ jones, respectively. When $V_{ds} = -1$ V, the device's $R$ and $D^*$ can reach 628 mA/W and $4.2 \times 10^{11}$ jones, respectively.

A metal/semiconductor/metal (MSM) WSe$_2$ PD with asymmetric contact geometry was fabricated via transferring exfoliated WSe$_2$ flakes to SiO$_2$/Si wafer by mechanical exfoliation method. The shapes of WSe$_2$ flakes were identified as triangular or flakes with sharp angles. In general, for an MSM device with symmetrical contact lengths (**Fig.29**b), where $W_L \approx W_R$, the net $I_{phot}$ is zero without bias. $W_L$ and $W_R$ are the contact lengths of the left and right metal– WSe$_2$ junctions, respectively. In contrast, if the contact lengths differ, a net $I_{phot}$ results. (**Fig.29**a).[362] A substantial contact length disparity yielded a high $R$ of 2.31 A/W and a huge open-circuit voltage of 0.42 V for MSM PD operating under 0 bias. The small $I_{dark}$ of ~1 fA observed in current MSM PD is much smaller than the high $I_{dark}$ common in conventional MSM PDs. A high $D^*$ of $9.16 \times 10^{11}$ Jones and high ON/OFF ratio were also observed. Further, the MSM PD also showed self-driven broadband photoresponse in 405–980 nm regime. In the present device, the electron mobility (26.3 cm²/V-s) is reported to be greater than the hole $\mu$ of ~7.4 cm²/V-s).

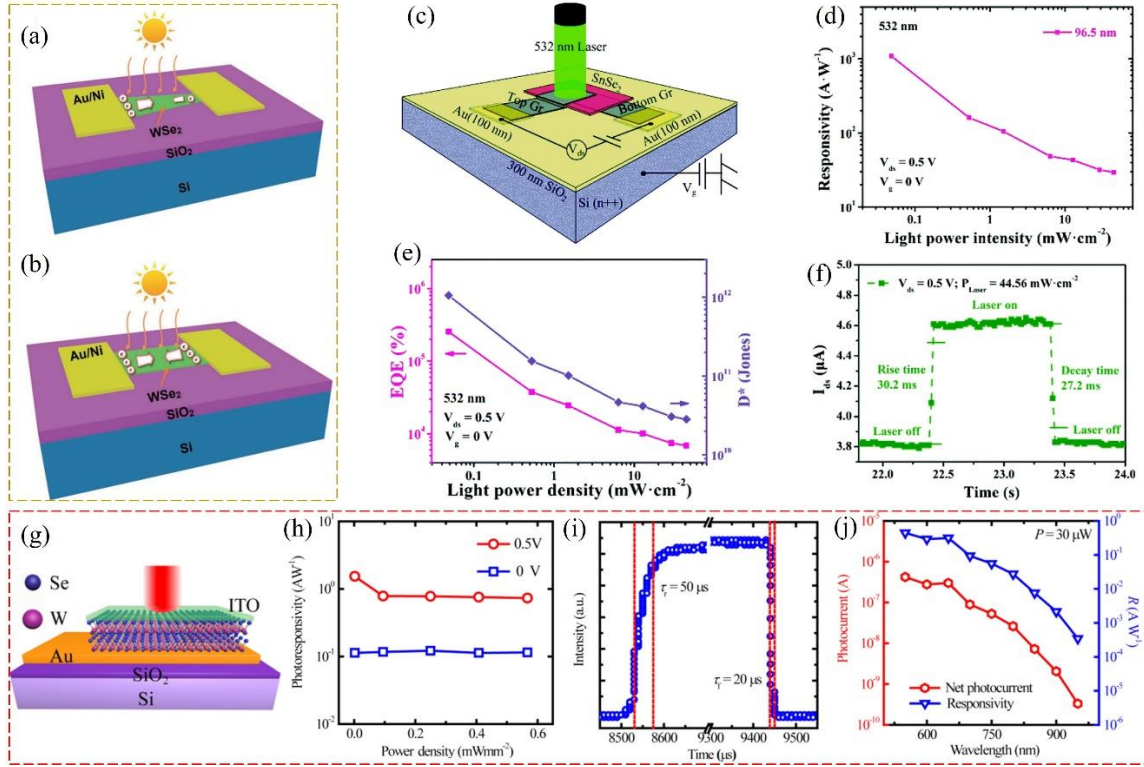

Fig.29: (a-b): Metal/WSe₂/Metal (MSM) *PD*s. 3D schematics of MSM PDs. (a) & (b) asymmetric contact (a net current is generated) and symmetric contact (no net current is generated) respectively when both device are illuminated by the light under zero bias. Adapted with permission.[362] Copyright 2018 WILEY-VCH Verlag GmbH & Co. KGaA, Weinheim. (c-f): Vertical Gr-SnSe₂-Gr sandwiched structures. (c) 3D illustration of a Gr–SnSe₂–Gr *PD*; (d) Variation of *R* with respect to light power density under $\lambda$ = 532 nm, $V_{ds}$ = 0.5 V, and $V_g$ = 0 V; (e) *EQE* and $D^*$ vs. light power density; (f) $\tau_r$ and $\tau_d$ of the Gr–SnSe₂–Gr under $\lambda$ = 532 nm illumination at $V_g$ = 0 V. Adapted with permission.[294] Copyright 2019, RSC. (g-j): Vertical Au–WSe₂–ITO PDs. (g) 3D illustration of Au–WSe₂–ITO Schottky barrier device; (h) *R* of the device vs. incident power density at $V_{bias}$ = 0 V and 0.5 V; (i) Time resolved photoresponse of the device at $V_{bias}$ = 0 V; (j) *R* and net photocurrent vs. $\lambda$s. Adapted with permission.[363] Copyright 2018, IOP Publishing Ltd.

Mechanical exfoliation and layer-by-layer stacking were used to fabricate several vertical Gr-SnSe₂-Gr sandwiched structures with varying SnSe₂ thicknesses (12.2 to 256 nm) (Fig.29c-f).[294] When the thickness of SnSe₂ was ~96.5 nm, a photoresponsivity of ~$1.3 \times 10^3$ A/W was observed under a 532 nm illumination. The PD also had a high *EQE*: $3 \times 10^5$ %, *D*\* of ~1.2 × $10^{12}$ Jones, and a fast $\tau_r = 30.2$ ms and $\tau_d = 27.2$ ms, ascribed to the low barrier height at the SnSe$_2$/Gr contact and a small photo-response distance.

By sandwiching a multilayer WSe$_2$ between a bottom Au electrode and a top indium tin oxide (ITO) electrode, a vertical Schottky junction PD was constructed (**Fig.29**g).[363] The device yielded a high photovoltaic responsivity of 0.1 A/W and a short $\tau_{res}$ of ~50 µs (**Fig.**29h-i). Furthermore, at zero bias, a high $I_{phot}$ to $I_{dark}$ ratio of ~$10^4$ and a very low $I_{dark}$ of $10^{-12}$ A were achieved. A very small vertical channel length for transit time and the built-in electric field, which accelerated photogenerated carrier separation contributed to the fast response.

### 4.2.5. Organic materials based photodetectors

Having considered the benefits of the inorganic component's inherent carrier mobility and the advantages of organic materials: the ease of synthesis, economic feasibility, physical flexibility, and non-toxic nature, developing an organic/inorganic hybrid heterojunction can be a very great option in the design and assembly of flexible high performing PDs. A PD containing plasmonic Au NPs and a crystalline rubrene has shown self-powered operation and broadband detection.[364] P3HT, or poly(3-hexylthiophene), is a well investigated polymer in organic PV systems[365] and has been tested in flexible inorganic/organic based thermoelectric films.[366] The effects of exteriorly applied strain on the PPE and photovoltaic effects were explored in a self-powered PEDOT: PSS/ZnO NWs heterojunction based PD.[367] Under 325 nm 2.30 mW/cm$^2$ UV light and a -0.45% compressive strain, the $I_{phot}$ of the PD was considerably increased from 14.5 to 103 nA by combining the pyro-phototronic and piezo-phototronic effects, resulting in a 600% enhancement. Also, it was shown that the PPE could be generated by applying a -0.45% compressive strain, which markedly improved the PD's response to 442 nm in relation to photocurrent $\tau_r$ and $\tau_d$.

A NIR photodiode with polymer CPDT-alt-BSe (conjugated polymer with $\lambda_{max} = 920$ nm) as the active layer and the inorganic material MoO$_3$ as the electron blocking layer (Fig.30a) has exhibited detection rate as high as $10^{12}$ Jones (at $\lambda = 1$ µm), a LDR of 86 dB, and was insensitive to temperature (Fig.30b).[368] MoO$_3$ at the anode coincides to the polymer's highest occupied molecular orbital (HOMO). The high switching speed has shown the potential application of this photodiode in detecting vascular pulses from a fingertip (Fig.30c). In this study, two photodiodes with different bulk heterojunction (BHJ) layer thicknesses were prepared. At zero bias, the thin 175 nm device had a greater $I_{phot}$ than the thick 310 nm device. However, the $I_{dark}$ was higher in the thin photodiode than in the thick one, showing a trade-off between $I_{dark}$ and $I_{phot}$ as the BHJ thickness was altered.

A heavy-metal-free flexible two terminal PD sensitive to $\lambda = 1.5$ µm photons is shown (Fig.30d).[369] This SWIR PD is based on the creation of a solution-processed hybrid comprised of an organic host conjugated diketopyrrolopyrrole-base polymer/PC$_{70}$BM bulk heterojunction and inorganic guest NaYF4:15%Er$^{3+}$ up-conversion nanoparticles (UCNPs). Under the $\lambda = 1.5$ µm SWIR photons illumination, the hybrid bulk-heterojunction (BHJ)/UCNP PDs demonstrated photoresponsivity of 0.73 and 0.44 mA/W for devices based on inflexible ITO/glass and flexible ITO/polyethylene terephthalate substrates, respectively. These hybrid PDs performed SWIR photodetection with a fast operation speed with a smaller $I_{phot}$ rise time of ~80 μs (Fig.*30*f). The *EQE*s of these hybrid PDs over the Vis-SWIR spectrum were ~10-2 % at $\lambda = 1.5$ μm (Fig.*30*g).

A dicyanovinyl-functionalized squaraine dye (SQ-H) was used as a donor material to fabricate an organic photodiode that showed high sensitivity to SWIR, $\lambda > 1000$ nm.[370] The energy levels of several layers of the device composed of polyethyleneimine (PEI) as the charge blocking layer in the device framework are shown in **Fig.30**h. Photoplethysmography for heart rate monitoring could be performed by the device built on the flexible substrate(**Fig.30**i). Clear and periodic systolic and diastolic signals for two wavelengths (1050 nm and 680 nm) were observed using this PD under ambient settings (**Fig.30**i-j).

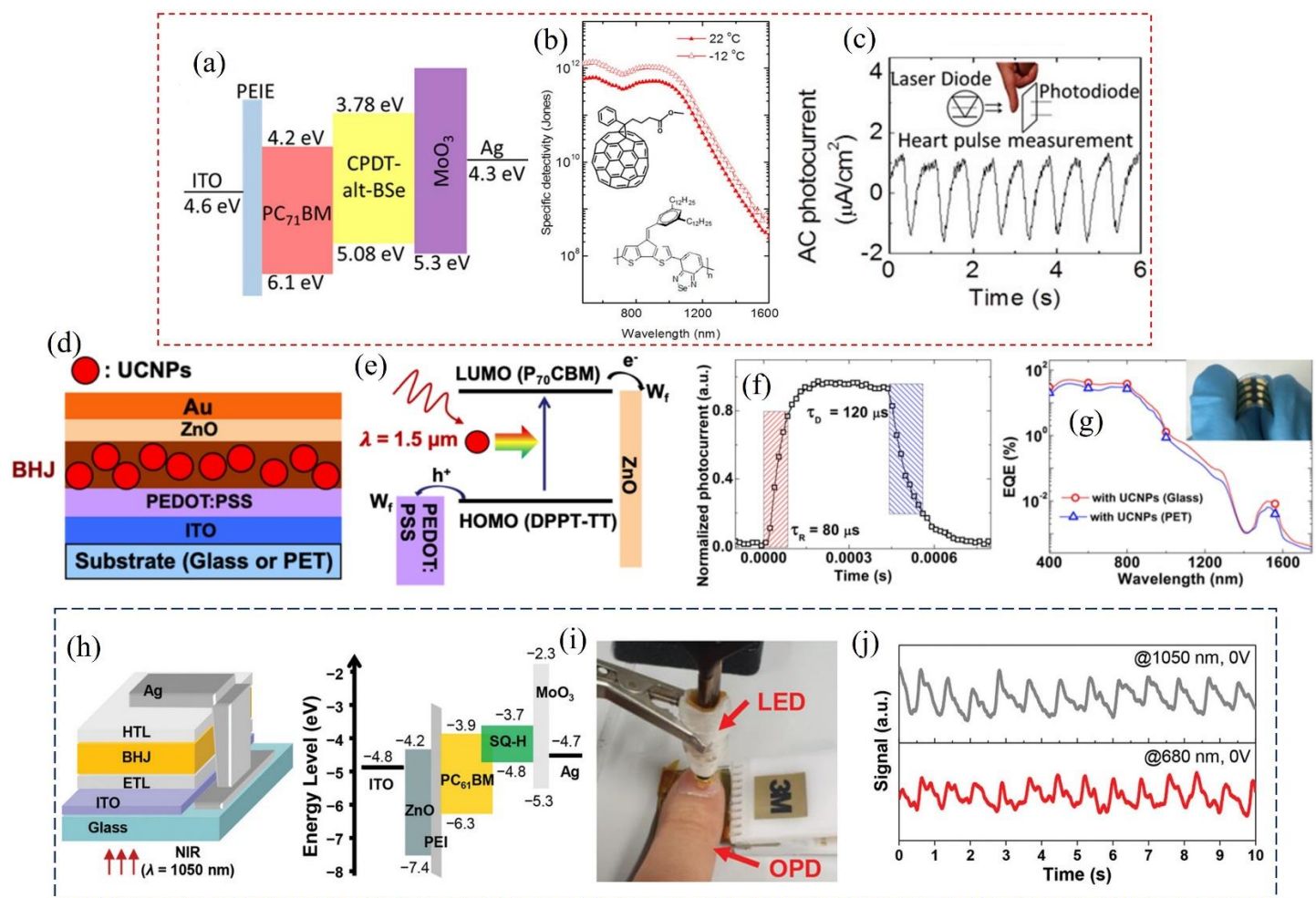

Fig.30: (a-c): Polymer CPDT-alt-BSe PD (a) Energy diagram of the photodiode; (b) $D^*$vs. $\lambda$s at 22 °C and 12 °C; (c) Heart rate measurement setup. A laser with a $\lambda$ of 904 nm and a power of 1 mW/cm² beamed NIR light through the finger, and the transmitted light was recorded by the organic photodiode, revealing signal modulation caused by arterial pulses.; Lower wave pattern in (c) shows change in $I_{phot}$ signals with person's heart rate. Adapted with permission.[368] Copyright 2017, ACS. (d-g): Hybrid BHJ(DPPTT-T/PC₇₀BM)/UCNP PD. (d) Hybrid BHJ(DPPTT-T/PC₇₀BM)/UCNP PD device architecture; (e) Operating principle of such hybrid

## 5 Summary, challenges, and opportunities

This review covered the recent progress in PDs based on single 2D materials and their 2D/*X*D (*X* = 0, 1, 2, and 3) heterostructures. An in-depth analysis of the photodetector's charge carrier generation, separation, and extraction mechanisms and the optimization strategies were given. Photodetection based on 2D materials has been demonstrated for a wide range of detection applications, such as broadband detection, high sensitivity, weak light detection, fast detection, polarization sensitive light detection, high data rate communication, spatially resolved photo image analysis, terahertz detection, and many others. The promising future of 2D material-based PDs can be attributed to their alluring physical features and superior device performance. Si has made a unique place for itself in Vis band photodetection by bringing processing technology to a high level of maturity. One example of this would be the cameras in smart phones.

Gr is a very appealing considering its salient features such as high $\mu$, wide spectral sensitivity and therefore can be used as high-speed charge carrier channel in heterostructures, thereby realizing high photogain. However, its applications as a standalone sensing element are not satisfactory because of its poor light absorption (~2%). Gr has demonstrated its usefulness as a superior material in IR PD design, particularly for high-speed data detection, by integrating an optical cavity/antenna, bandgap adjustment, and partnering with other light-absorbing materials. Unfortunately, such structures displayed dismal photoresponsivity and high $I_{dark}$, which hampered the device's commercialization. As already discussed, such problems can be minimized by inserting a thin insulating oxide layer at the interface of Gr-based Schottky PDs. For example, inserting GdIG between Gr and Si (i.e., Gr/GdIG/Si structure)[96] as it is already discussed. Other examples for heterostructures are inserting TiO$_2$ buffer layer between the MoS$_2$ channel and the HgTe QD layer[225] employing gate-tunable current-rectifying strategy as done in the case of vertical GaTe/InSe vdW *p-n* heterojunction.[293] Such strategies aid in achieving better quantum efficiency and a lower $I_{dark}$. Designing heterostructures such as Gr/2D semiconducting materials, where 2D semiconducting materials function as a light absorber and Gr acts as a high carrier transport pathway, is an effective way to improve device performance. A considerable number of 2D PDs exhibit excellent photoresponsivity, which can be primarily due to the prolonged carrier life. However, this benefit comes at the sacrifice of response speed. Such devices could be used to detect weak light; however, the high gain caused by the extended carrier life can also generate significant generation-recombination noise, resulting in a failure to increase $D^*$. The insufficient $\mu$ of some of the 2D TMDCs is a barrier to the device's performance, despite the strong light-matter coupling. Hence, a 2D PD that simultaneously possesses high $D^*$, broad bandwidth, and good photoresponsivity is generally desired.

BP has a high $\mu$, anisotropic transport, and wide tunable direct $E_g$, making it a particularly promising contender for high-performance optoelectronic applications, notably UV-IR and THz ranges. At RT, BP-based detector has shown $NEP$ of ~$10^{12}$ W Hz$^{1/2}$, which is in par with $NEP$s of commercial photodetectors. However, as mentioned earlier, BP deteriorate badly at RT. To make it more stable, researchers have used encapsulation, surface passivation, ion doping, and other techniques. Among them, building vdW heterostructures with Gr or h-BN as a protective layer may be the best option. The unstability of BP must be resolved via further investigation. The photodetection characteristics of certain transition metal trichalcogenides and metal halides have been studied. But, as with BP, the issue of their stability requires further examination and resolution.

Developing high-performance WG-type PDs on Si for the 2 μm or beyond wavelength range remains a significant problem. Most classic semiconductor materials, which function effectively at 1.3/1.55 μm, cannot be extended for longer $\lambda$s because the bandgap limits the cut-off wavelength for light absorption. To implement the photodetection of longer wavelengths, Si-based hybrid structures are needed.

To harness the synergy created by combining multiple layers of materials with compatible interband coupling, stacking order, etc., heterostructures based on the stack of individual 2D materials have been intensively studied for their IR properties. Gr has been proven to increase the detection range in Gr/TMD heterostructured PDs. Despite their numerous benefits, the ultrathin structure of 2D materials results in inadequate light absorption, low $EQE$, low $D^*$, and a failure to exploit the distinct advantages of 2D crystals. Commercialization possibilities will be determined not only by detector performance, but also by some particular advantages and capabilities in the ability to fabricate large-scale, high-quality 2D materials at a lower cost. It is of the utmost importance to work toward the establishment of large-scale integration of 2D crystals with already existing photonic and electronic platforms, like CMOS technologies, where lattice and thermal mismatch related problems may arise. Consequently, the photodetection technology that is based on 2D materials still needs significant development, particularly in the preparation of highly crystalline materials and films, the optimization of the transferring process, the design of novel device structures, the addition of supplementary regulation, and other aspects.

When compared to commercial optical materials, the absorption of 2D materials (TMDCs and Gr) with atomic-scale thickness is relatively lower. To strengthen the IR light and the active material interaction and coupling, multiple layers instead of a single layer are investigated. However, the contribution of a larger $I_{dark}$ is imminent, which makes multiplayer configurations unsuitable for IR detection. Therefore, in 2D material-based transistors, the active material of an optimum thickness is preferred to suppress the $I_{dark}$ via applied gate voltage. To tackle such problems, computational analysis may be beneficial. To minimize the $I_{dark}$ and other noises, the formation of a $p$–$n$ junction or Schottky barrier is important. Additional avenues to increase the amount of light absorption from atomically thin layered materials are the integration of plasma technologies, hybridizing with QDs, or developing heterostructures.

The first and most important step in the fabrication of 2D based PDs is to obtain high quality active material films. The preparation of 2D layered materials via CVD is wrought with many problems such as uneven distribution of 2D domains and non-uniformity in thickness. Such problems can be overcome by tuning the deposition parameters (reaction dynamics) during the film growth. The majority of reported high-performance photodetectors are based on mechanically exfoliated thin layers or heterostructures. The mechanical transfer of the epilayer technique has been widely employed for 2D material device production, but it invariably introduces impurities and defects in the layer or the interface, resulting in poor device performance, especially charge carrier transport at the interface. Such problems can be tackled via a careful synthesis method and surface treatment steps.

The design of cost-effective solution-based techniques for producing homogenous single-crystalline films on the scale of wafers remains an appealing prospect. To achieve this, 2D material growth and synthesis methods must first be thoroughly investigated and developed. Cosolvent route to obtain 2D structures for IR detection is less explored. This method creates soluble semiconductor precursors that, after deposition and mild treatment, can be transformed into crystalline semiconductors.[366] Hydrazine, a material that is well-suited for such applications, is highly toxic, explosive, and carcinogenic. Nine bulk $V_2VI_3$ semiconductors ($V$ = As, Sb, Bi; $VI$ = S, Se, Te) can be promptly dissolved at room temperature and pressure using a non-toxic binary solvent mixture made of 1,2-ethylenediamine (en) and 1,2- ethanedithiol (edtH$_2$).[371] Since then, this binary-solvent method has been used to make soluble precursors for many different metal chalcogenide semiconductors.[371-372] Similarly, an amine-thiol mixture can be used as a general solvent to dissolve a wide range of inorganic semiconductors, including Cu$_2$S, Cu$_2$Se, In$_2$S$_3$, In$_2$Se$_3$, CdS, SnSe, and others. This process produces high concentration semiconductor ink (> 200 mg/mL) at RT and pressure, which may then be used to make homogenous semiconductor thin films on SiO$_2$/Si wafers, glass, and plastic substrates.

Gr functionalization is a technique that can be used to alter the dynamics of charge carriers in Gr, which ultimately leads to improved photoresponse. It requires intense and thorough investigation.

The formation of van der Waals heterostructures is one of the primary strengths of 2D layered materials. This has led to the discovery of a wide variety of new characteristics and functionalities (example: self-driven vdWHs). The vertical vdWHs have a lot of potential to be used in the future to make new optoelectronic devices with high performance, such as broadband, high responsivity, and fast photodetection. In heterostructured PDs, the energy band structure of hybrid materials is crucial for photocarrier generation and transportation. Computational studies on the charge-transfer mechanism of band structures show the underlying photoelectric conversion process and provide theoretical support for device performance enhancement and future practical development of futuristic PDs. Because of their size modifiable semiconductor characteristics, narrow-bandgap QDs are ideal for band structure modification.

The incorporation of sensitizers like as plasmonic structures and organic dyes is one technique for increasing the light absorption of 2D PDs. None, however, offer the same degrees of spectrum tunability as QDs via composition and quantum confinement (modulating the QD size). Owing to their size-dependent band gap, exceptional light absorption coefficient, and low cost of synthesis by solution processing, QDs are very promising as active materials for photodetection.

The state-of-the-art IR PDs might use a synergistic combination of Gr field-effect transistors (GFET) for signal amplification and IR-sensitive colloidal QDs for photon absorption. These devices take advantage of low-cost solution-processed QD production, compatibility with complementary metal-oxide-semiconductors, and RT operation with decent photoelectric characteristics. Graphene/QD heterostructured PDs have shown profound performance applicable to UV, Vis, and IR photodetection primarily due to strong light absorption and spectral selectivity of QDs and high $\mu$ of Gr. The effective transfer of charge carriers from QDs to the Gr can be facilitated via the QD size effect, surface treatment, and chemical doping of QDs, which can improve the QDs light absorption and broaden the spectral response regimes of PDs. Despite the rapid progress of graphene/QD heterostructured PDs in recent times, there are still several challenges to be tackled. In graphene/QD heterostructured PDs, exceptionally high $R$ and $D^*$ were simultaneously reported; nonetheless, the response time often stays near to several seconds, necessitating speed improvements without compromising $R$ and $D^*$. Enhancing the stability of QD-based PDs is a second concern. In terms of photoelectric conversion efficiency, perovskite QDs have various benefits. Yet, its instability places constraints on device configuration. Effective strategies may involve surface engineering (surface passivation) or the deposition of specific passivation layers. However, compatibility between a passive layer and a QDs layer is essential for maintaining the long-term durability and reliability of PDs. Whilst Pb- and Hg-based QD PDs have shown good performance, their toxicity prevents them from being utilized further. Consequently, it is vital to synthesize nontoxic QDs and utilize them in the fabrication of nontoxic QD PDs.

Variations in the thicknesses of QD layer between different device prototypes may add extra limitations on the mass production of PDs. QDs have been deposited onto Gr by spin-coating, drop-casting, traditional inkjet printing, and contact printing. While these techniques are capable of coating large device surfaces, they have inherent constraints in terms of the spatial resolution and precision of the materials deposition, as well as the homogeneity of the QD layer (due to agglomeration of QDs). Contact printing is incompatible with various substrate topologies and necessitates homogenous QD deposition on a specified substrate. Conventional inkjet printing permits relatively limited spatial control over material deposition, with a feasible resolution of 100 μm for certain QD systems. Nanoprinting can be a suitable method for overcoming the aforementioned restrictions, as it provides a high spatial resolution of deposition on a variety of substrates and the capacity to deposit diverse materials on a small area for various PDs. Provided large area Gr is deposited on a suitable substrate, this can be subjected to hydrothermal environment, where Gr/substrate can be inserted in a hydro/solvothermal or solution-based synthesis environment to grow QDs uniformly on the surface of Gr. Such studies need further investigation. An in situ high-resolution TEM study or any other method where the charge transfer mechanism between the QDs and Gr can be studied at different illumination powers could give useful information about band alignment.

Photothermoelectric effect detectors are promising since they require little maintenance and can work in IR and THz regimes with cooling-free operation. With respect to thermoelectric properties of materials and thin-films, numerous classes of thermoelectric materials have shown excellent properties, especially, highly optimized Seebeck coefficient, thermal (low) and electrical (high) conductivities, and very high $\mu$ values. Further research is needed to determine how well-established thermoelectric materials may be used to generate the PTE effect, which can then be employed for detection of various $\lambda$s.[373] Another type of thermoelectric effect called Transverse Thermoelectric Effect (TTE) has recently been seen in crystalline material systems, which has sparked a lot of interest in its prospective applications as heat sensors, sub-nanosecond laser radiation detectors, and power producers.[366] Under the irradiation of a pulse laser with a $\lambda$ of 248 nm and a duration of 28 ns at RT, highly conducting $La_{0.5}Sr_{0.5}CoO_3$ (LSCO) epitaxial films showed an ultrafast transverse thermoelectric voltage of 7 ns rise time (i.e., TTE voltage response). Thermal co-evaporation was used to deposit mainstream $p$-$Sb_2Te_3$ and $n$-$Bi_2Te_3$ thermoelectric films on borosilicate glass and Kapton polyimide substrate.[374] Then, a flexible thermopile sensor based on $p$-$n$ films was constructed to demonstrate a potential thermal sensing component. $R$ and $D^*$ values for this device were 0.05 V/W and $1.6 \times 10^7$ cm $\sqrt{Hz}$/W, respectively. For additional discussion, readers are urged to refer to publication:[366]. Such investigations can be expanded to other materials. The author does not have any information on the progress of materials exhibiting TTE that can be employed commercially for photodetection. For M-G-M PDs, the resistance is typically $10^2$-$10^3$, and the $I_{dark}$ may reach mA even when the bias voltage is < 1 V. To lower the $I_{dark}$, it is preferable to construct PDs with the PTE effect that operate at near-zero bias.

PPE aided photodetection has made significant strides in recent years, particularly in the areas of ZnO-based materials, CdS, SnS, perovskites, and certain biomolecules with fast $\tau_r$, high $R$, and $D^*$ for different $\lambda$s. The PPE, when combined with ferroelectricity, LSPR, and piezoelectricity, can provide synergistic effects that improve photodetection performance. In this case, the deterministic effect that can be tuned to improve the photoelectric properties must be ascertained and implemented. Most pyro-phototronic photodetectors respond well to UV light; in the near future, devices for Vis and NIR light will be more attractive. Pyro-phototronic PDs have good photoresponsivity, a fast $\tau_r$, a very low $I_{dark}$, a low light ON/OFF ratio, a low light power detection capability, are self-powered, and can be made flexible, which is essential in flexible wearable applications. The pyroelectric and photovoltaic effects work together in pyrophototronic devices to strengthen the built-in electric field, accelerate charge transfer at the heterojunction interface, and lessen the recombination rate of $e^-$-$h^+$ pairs. A variety of ZnO nanostructures in combination with Ag, Au, PEDOT: PSS, P3HT, CdS have been studied in pyro-phototronic devices. Metallic nanostructures, particularly Au and Ag, can provide integrative effects like the plasmonic heat effect and LSPR to improve the $\tau_r$ and $D^*$ of self-powered PDs. Besides ZnO, $TiO_2$ NRs (noncentrosymmetric and wurtzite structures) are widely used in binary switching applications. Due to their features, they can promote PPEs. Recently, a few biocompatible or biosemiconductors (for example, bacteriorhodopsin, a notable photoreceptor protein) have come to the forefront for improving photoresponse, which can contribute to bioelectronic applications. Since the photoreceptor proteins are purely natural, genetic engineering of the host cell has the potential to change the very nature of the protein to make it more acceptable for light detection. However, how these proteins behave ex-situ (outside the cell in dry conditions) needs further study. The majority of self-powered PDs reported work solely in the UV region, with very few working in the broadband (UV-Vis-NIR) region. For example, the CdS nanorods/SnS nanoflakes heterojunction has broadened the detection range across the light spectrum. Therefore, more study is needed to optimize the properties and fabrication costs are needed to deploy in RT practical applications.

Metamaterial perfect absorbers (MMPAs) that operate at frequencies ranging from radio to optical have sparked a lot of interest in photodetection in recent years. MMPAs have been proven in applications such as photodetection,[375] and imaging device design,[376] etc. Fig.31 outlines some recent noteworthy publications in this topic. Despite its infancy, the discipline offers potential uses in the near future.

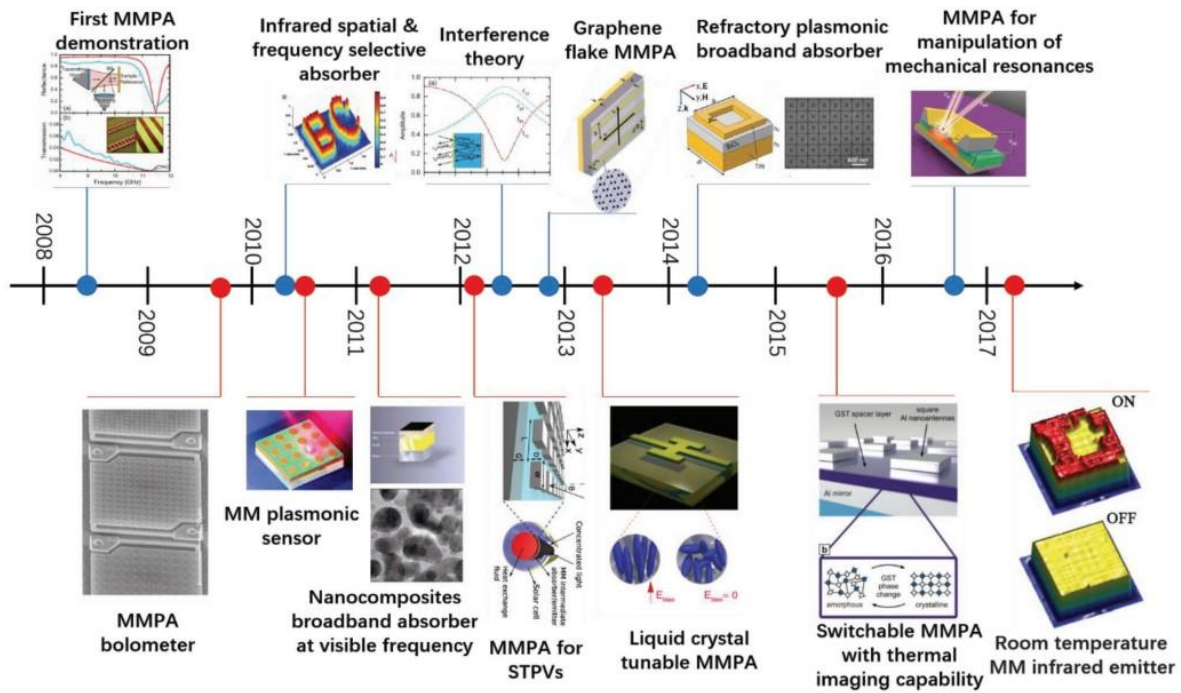

Fig.31: Selected representative MMPA publications. The publications indicated here are cited as references.[376b, 377] Adapted with permission.[376] Copyright 2010, AIP Publishing.

Finally, investigations on robustness of the PDs in terms of reliability, lifetime needs very in-depth analysis as changes in nanostructures can induce changes in *figures of merit* of the devices. Such issues may not be predominant in highly crystalline materials, but they do occur in QDs, plasmonic nanoparticle based heterstructures, where the problems related to random distribution of particles may compel two devices exhibiting different properties even both devices may have same fabrication process. In such cases, large tolerance akin to the random distribution of nanostructures must be practical.

**Conflicts of interest:** No competing interests.

**Data Availability Statement:** Not applicable.

**Acknowledgement:** The author would like to pay tribute to his supervisor, the late Prof. Vitalij Pecharsky, whose enduring influence and commitment to group members serve as a source of inspiration. Some of this research was performed during an appointment at the Ames Laboratory, which is operated for the U.S. DOE by Iowa State University under contract # DE-AC02-07CH11358.